\documentclass[aip,
amsmath,amssymb,
reprint,]{revtex4-1}

\usepackage[dvips]{graphicx}\usepackage{dcolumn}\usepackage{bm}\usepackage[utf8]{inputenc}
\usepackage{mathptmx}
\usepackage[T1]{fontenc} \usepackage{delarray}
\usepackage{amsmath,amssymb,array}
\usepackage{bm}
\usepackage[dvipsnames]{xcolor}
\usepackage{todonotes}

\usepackage{tikz}
\usetikzlibrary{arrows,automata}
\usetikzlibrary{shapes.geometric,positioning,calc}

 \newcommand{\ket}[1]{| { #1} \rangle}
 \newcommand{\bra}[1]{ \langle {#1} |}
\def \binom#1#2{{#1\choose #2}}

\usepackage{hyperref}
\begin{document}

\preprint{AIP/123-QED}

\title{Tools for quantum network design}

\author{Koji Azuma}
\email{koji.azuma.ez@hco.ntt.co.jp}
\affiliation{NTT Basic Research Laboratories, NTT Corporation, 3-1 Morinosato Wakamiya, Atsugi, Kanagawa 243-0198, Japan}
\affiliation{NTT Research Center for Theoretical Quantum Physics, NTT Corporation, 3-1 Morinosato-Wakamiya, Atsugi, Kanagawa 243-0198, Japan}
\author{Stefan B\"{a}uml}
\email{stefan.baeuml@icfo.eu}
 \affiliation{ICFO-Institut de Ciencies Fotoniques, The Barcelona Institute of Science and Technology, Av. Carl Friedrich Gauss 3, 08860 Castelldefels (Barcelona), Spain}
\author{Tim Coopmans}
\email{t.j.coopmans@tudelft.nl}

\author{David Elkouss}
\email{d.elkousscoronas@tudelft.nl}
\affiliation{QuTech, Delft University of Technology, Lorentzweg 1, 2628 CJ Delft, The Netherlands}

\author{Boxi Li}
\email{boxili@outlook.com}
\affiliation{QuTech, Delft University of Technology, Lorentzweg 1, 2628 CJ Delft, The Netherlands}\affiliation{ETH Z\"{u}rich, R\"{a}mistrasse 101, 8092,
Z\"{u}rich, Switzerland}
\affiliation{Peter Gr\"unberg Institute - Quantum Control (PGI-8), Forschungszentrum J\"ulich GmbH, D-52425 J\"ulich, Germany}
\date{\today}

\begin{abstract}
Quantum networks will enable the implementation of communication tasks with qualitative advantages with respect to the communication networks we know today. While it is expected that the first demonstrations of small scale quantum networks will take place in the near term, many challenges remain to scale them. To compare different solutions, optimize over parameter space and inform experiments, it is necessary to evaluate the performance of concrete quantum network scenarios. Here, we review the state of the art of tools for evaluating the performance of quantum networks. We present them from three different angles: information-theoretic benchmarks, analytical tools, and simulation.
\end{abstract}

\maketitle
\tableofcontents
\section{Introduction}
Communication networks are the basis of our connected society. In particular, the internet that we know today is a conglomerate of physical links of very different characteristics. Some are highly reliable and persistent, while some are very noisy and dynamic. From this basic resource, the network builds services, such as reliable remote transmission, multicast, and many others. 

Network engineering involves challenges of very different nature. At the hardware level, there might be the request of replacing a device element or of scaling an architecture from tens to thousands of nodes. At the network level, there might be the need to develop a new protocol. At the application level, new applications with different traffic patterns and requirements on the network might be deployed. To address these challenges and be able to take appropriate measures, it is necessary to evaluate solutions and then compare against alternatives and benchmarks in a quantitative way. 

Performance can be evaluated roughly in three different ways: experimentally, by means of simulation, and analytically.
Experimental methods evaluate the performance with hardware. They can range from prototyping to fully-fledged field test deployments. The advantage of experimental methods is that they give the most accurate answer to a concrete scenario, but they have several drawbacks. Notably, they can be extremely costly and offer little flexibility to change parameters. Moreover, they might be unavailable to evaluate directions of future research, i.e. if the hardware does not yet exist. Simulation and analytical methods require models of the different network elements. A first observation is that the quality of the evaluation depends on the accuracy with which the models capture the behavior of each element. 
The difference between simulation and analytical methods is that simulation methods imitate the behavior of the individual elements in the network and their interactions, while analytical methods compose analytical models representing some characteristic of interest to determine the aggregate performance without actually replicating the actions of the elements.
Both simulation and analysis can tackle a broad range of scenarios at the cost of a less accurate evaluation.

The choice of the evaluation method and the concrete tool depend on many factors. Some of them are cost, flexibility, accuracy, and validation. 

Moreover, the methods are not exclusive. For instance, a streamlined model for mathematical analysis can be validated by simulation or experiment. Or alternatively, one can use an analytical model to validate a new simulation approach. In general, cross-validation is a well known method for increasing the credibility and reliability of an evaluation tool \cite{law2019build}.

In principle, quantum networks enable the implementation of tasks beyond the reach of classical networks such as key distribution \cite{bennett1984quantum,ekert1991quantum}, clock synchronization \cite{komar2014quantum}, increasing the baseline of telescopes \cite{gottesman2012longer}, and many others \cite{wehner2018quantum}. However, while they share similar high-level challenges with their classical counterpart, there are noticeable differences. On the one hand, quantum information can not be copied, reducing the applicability of classical approaches for long-distance communication. On the other hand, entanglement between two parties is only useful if it is clear which particles at one location are entangled with which ones at the distant location. In turn, this places novel constraints to the network architecture. 

Motivated by recent experimental advances promising the deployment of the first quantum networks, the past years have seen a wealth of research dedicated to analytical and simulation tools for evaluating the performance of quantum networks. Our goal in this paper is to review them and make a unified description of all existing resources. We undertake this task in three steps. 

First, in Section \ref{sec:fundamental}, we review analytical information-theoretic tools. Given a model for communication channels and some precise definition of the available resources, information theory allows us to bound the optimal rate at which a communications task can be performed. This abstract optimal rate, called the channel capacity, typically can only be achieved in a very ideal setting, for instance as the number of channel uses tends to infinity. Hence, information-theoretical bounds represent a fundamental performance limit and can be used to benchmark concrete implementations. One important example of their role is the evaluation of quantum repeater implementations. Quantum repeaters enable, in principle, the transmission of quantum information over arbitrarily long distances with non-vanishing rates. While there has been strong experimental progress, it is unclear what performance is necessary to label an implementation as a quantum repeater. Information theory provides a clear tool for evaluating implementations. If two parties are connected by a quantum channel, their communication rate is bounded by the channel capacity. If the two parties place a device between them and, with the help of this device, they achieve a communication rate above capacity this certifies that the device is qualitatively improving the communication capabilities of the two parties.

Second, in Section \ref{sec:analytical}, we review analytical tools for evaluating the performance of concrete protocols. We focus our analysis in protocols for distributing entanglement as, once the entanglement has been distributed, it can be consumed for implementing different tasks. Given a protocol, the key performance parameters are the time it takes to distribute the entanglement and the quality of the entanglement. The main difficulty for this analysis is that many protocols rely on probabilistic elements and these elements can be stacked in non-trivial ways to construct the entanglement distribution protocol.  
In consequence, both the waiting time and the entanglement quality are random variables. In spite of this, as long as the physical models are simple enough, it is possible to approximate, and sometimes calculate, relevant parameters of these random variables for a large class of protocols including entanglement swapping, distillation, and cut-offs.

Meanwhile, simulation is the tool of choice for understanding different types of complex behavior in classical networks. Until recently, there existed very few options for quantum network simulations. In Section \ref{sec:simulation}, we discuss the role of simulation and the existing platforms. For this, first, we outline the methodology for evaluating the network performance based on simulation and then discuss the particularities of quantum networks together with a selection of existing tools for quantum networks.

We end the review with a summary of the tools presented and an outlook in Section \ref{sec:outlook}.

Let us clarify the scope of the review. We review tools for evaluating the performance of quantum networks as the key element that will enable the design of quantum networks. We leave the actual design and optimization of quantum networks\cite{aparicio2011protocol,dahlberg2019link,zhou2019security,kozlowski2020designing,huberman2020quantum,yu2019protocols,matsuo2019quantum,van2012quantum,van2013designing,huberman2020quantumB,pirker2018modular,pirker2018modular,rabbie2020designing,da2020optimizing,chakraborty2019distributed,kozlowski2020designing} beyond the scope of this review. 
We also omit the closely related topics of quantum networks for creating correlations \cite{aaberg2020semidefinite,kraft2020characterizing,wolfe2019quantum,renou2019genuine,renou2019limits} and of complex quantum networks \cite{brito2020statistical,perseguers2013distribution}.
Finally, there is a large literature regarding communication over direct links connecting two adjacent nodes. The fundamental limits over such a "network" for a task are given by the associated channel capacities, while the performance of a concrete protocol can be estimated in many different ways. There exist a number of references reviewing the fundamental and practical methods to evaluate a direct connection \cite{duer2007entanglement,holevo2012quantum,wilde2013quantum}. For this reason, we have chosen to restrict the scope to tools that are particular to networks beyond direct connections.
 \section{Fundamental limits for entanglement distribution over networks}
\label{sec:fundamental}
Even in the classical case, physical channels connecting two end points can have a very complicated behavior. However, for practical purposes, many communication channels can be abstracted into a stochastic map that takes a symbol at the input and places a symbol at the output following some probability distribution. One can then study the usefulness of this channel for a communication task. The usefulness is measured by the number of times that the task can be achieved divided by the number of uses of the channel. The optimal ratio maximized over all possible communication protocols without computational restrictions is called the capacity of the channel. While this is a very abstract notion, it is useful for benchmarking practical communication schemes. Moreover, the capacity of a classical channel can be easily approximated numerically \cite{arimoto1972algorithm,blahut1972computation}.
Quantum channels can also be abstracted by a mathematical object (see below~\ref{sec:ng}), and the capacity of a quantum channel for different communication tasks can be defined. However, it is presently unknown how to compute the capacity for a general channel and in many cases only bounds are available. Fortunately, for some very relevant channels such as the lossy bosonic channel, the capacity is known \cite{pirandola2017fundamental}. Building on the tools for point-to-point channels, a number of recent results allow to also bound the capacities for many communication tasks over networks, even including multi-user scenarios. In the rest of the section, we present a general description of such tasks, upper/lower bounds on their capacities, and efficient ways to evaluate the bounds.

\subsection{Networks as graphs}\label{sec:ng} 

A quantum network is composed of quantum information processing nodes and quantum channels. A quantum information processing node may represent a client, a quantum computer, or a quantum repeater node, but, in theory, it is regarded as an abstract node which can perform arbitrary local operations (LO) allowed by quantum mechanics \cite{nielsen2002quantum,wilde2013quantum}. 
This local operation is, in general, a stochastic process, described by a completely-positive (CP) map. More precisely, a local operation at node $X$ receives a quantum state $\hat{\rho}$ as an input, and returns a quantum state $\hat{\sigma}_k$ with probability $p_k$ as an output, where $\hat{\sigma}_k:=\hat{M}_k^X \hat{\rho} (\hat{M}_k^X)^\dag /p_k$ with Kraus operators $\{ \hat{M}_k^X\}_k$ satisfying $\sum_k (\hat{M}_k^X)^\dag \hat{M}_k^X =\hat{1}^X$ and $p_k={\rm Tr} [ (\hat{M}_k^X)^\dag \hat{M}_k^X \hat{\rho} ]$.
On the other hand, a quantum channel is a device to convey a quantum system from one node to another node, such as an optical fibre, a superconducting microwave transmission line, or an optical free space link.
In theory, any quantum channel ${\cal N}^{X \to Y}$ from a node $X$ to a node $Y$ is described by a completely-positive trace-preserving (CPTP) map. In particular, quantum channel ${\cal N}^{X \to Y}$ for an input state $\hat{\rho}$ is described by
${\cal N}^{X \to Y}(\hat{\rho}):={\rm Tr}_{E'} [\hat{U}^{XE\to YE'}( \hat{\rho} \otimes \ket{0}\bra{0}^E )(\hat{U}^{XE\to YE'})^\dag]$ with a unitary operator $\hat{U}^{XE\to YE'}$ on Hilbert space ${\cal H}^X \otimes {\cal H}^E={\cal H}^{Y} \otimes {\cal H}^{E'}$ and a state $\ket{0}^E$ of auxiliary system $E$.

\begin{figure*}
    \centering
    \includegraphics[width=150mm]{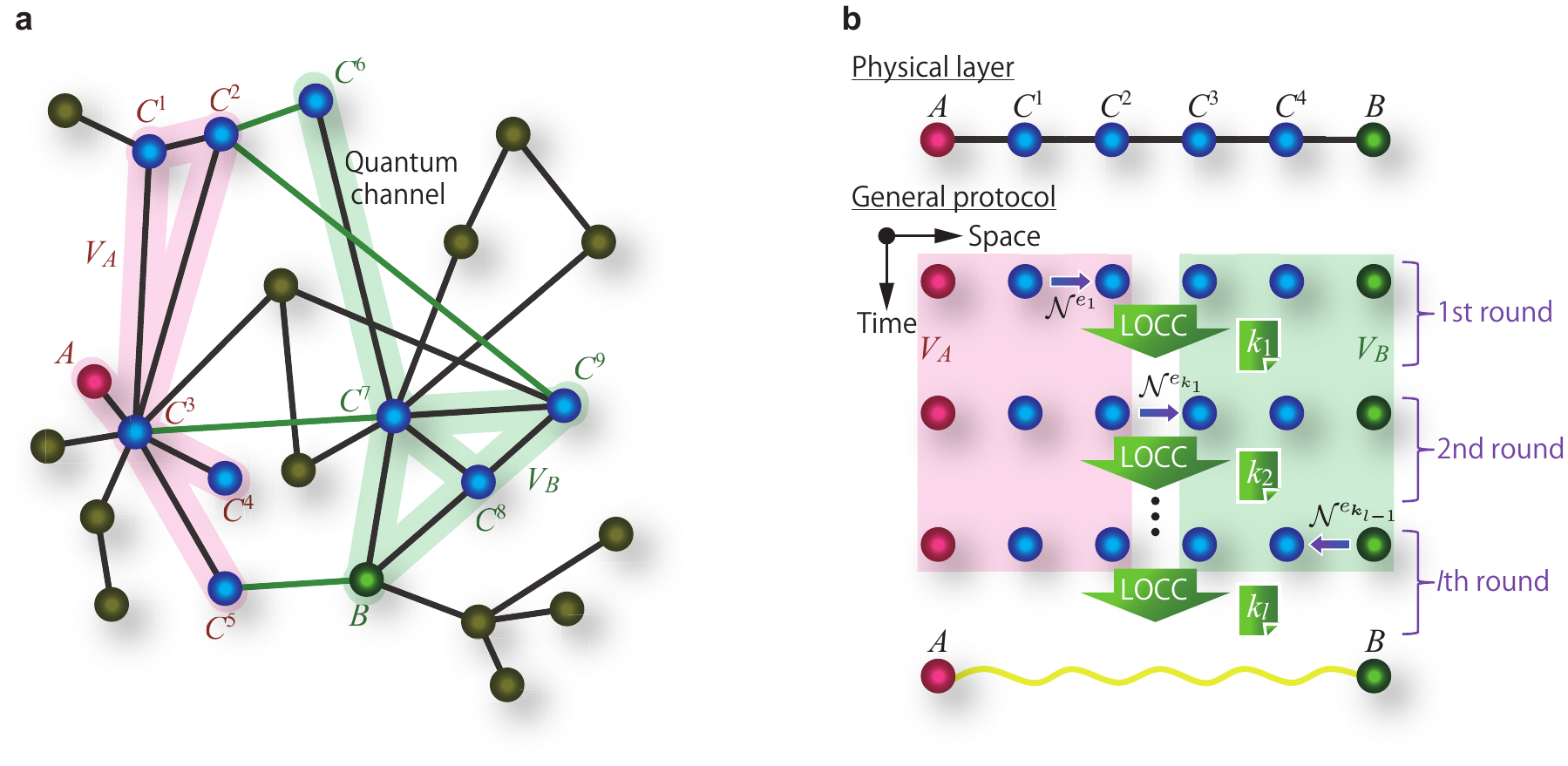}
   
    \caption{Quantum network and most general protocol. ({\bf a}) A subnetwork composed of nodes $V=\{A,B,C^1,C^2,\ldots,C^9\}$ is regarded as a quantum network associated with a graph $G=(V,E)$, where edge $e\in E$ is used to specify the existence of a quantum channel ${\cal N}^e$ in the network. A cut $\partial(V_A)$ is composed of green edges that separate the network into sets $V_A=\{A,C^1,C^2,\ldots,C^5\}$, shaded in red, and $V_B:=V\setminus V_A=\{B,C^6,C^7,\ldots,C^9\},$ shaded in green. ({\bf b}) Schematic description of the protocol performed at a physical layer of a linear network. In the first round channel $\mathcal{N}^{e_1}$ is used, followed by LOCC, providing output $k_1$. In the second round, depending on outcome $k_1$ in the previous round, channel $\mathcal{N}^{e_{k_1}}$ is used followed by LOCC providing output $k_2$ and so on. 
    Reprinted with permission from K. Azuma, A. Mizutani, and H.-K. Lo, Nat. Commun., vol. 7, no. 13523, (2016)\cite{AML16}, under a Creative Commons Attribution 4.0 International License.}
    
    \label{fig:network}
\end{figure*}

The structure of a quantum network, which might be a subnetwork of a larger quantum network in general, can be specified in terms of a graph (Fig.~\ref{fig:network}a). A quantum information processing node in a quantum network is regarded as a vertex $v$ in a graph, while a quantum channel ${\cal N}^{v\to v'}$ to send a quantum system from a node $v$ to another node $v'$ is associated with a directed edge $e=v\to v'$ (or, simply, $e=vv'$, where the former $v$ and latter $v'$ are the tail $t(e)$ and head $h(e)$ of the edge $e$, respectively) in the graph. 
With this rule, a given quantum network is specified by a graph $G=(V,E)$ with a set $V$ of vertices and a set $E$ of directed edges, in such a way that vertices in the set $V$ and directed edges in the set $E$ correspond to quantum information processing nodes and quantum channels in the quantum network, respectively.

In addition to a quantum network, we may use a classical communication network, such as a telephone network or the Internet. In general, this classical communication network may have a restriction, for instance, such that a quantum information processing node is not connected to another node with a classical communication channel, or such that classical communication between quantum information processing nodes is restricted to only one way. 
In practice, such a restriction about classical communication should be taken care of.
However, 
if we are interested in the ultimate performance of a quantum network, or if classical communication is supposed to be much cheaper than quantum communication in the era of quantum networks, then it may be reasonable to assume that classical communication between any nodes is free. Since LO are also free within every quantum information processing node, this leads to an assumption that local operations and classical communication (LOCC) among all quantum information processing nodes are free.

\subsection{Communication tasks}\label{sec:tasks}
A major goal of the quantum internet is to distribute entanglement among clients in a quantum network. This is because once clients share an entangled state, a communication task can be executed by consuming the entanglement, under LOCC, that is, without using the quantum network anymore.
What kind of entanglement is required depends on the task the clients want to perform. 
For example, if a client $A(\in V)$ wants to send an unknown quantum state with dimension $d$ to another client $B(\in V)$ via quantum teleportation\cite{bennett1993teleporting}, they need to share a bipartite maximally entangled state $\ket{\Phi_d}^{AB}:=\sum_{i=1}^d\ket{ii}^{AB}/\sqrt{d}$. Even if the client $A$ wants to send bits to the client $B$ secretly, since the maximally entangled state provides them $\log_2 d$ bits of secret key for the one-time pad, sharing the maximally entangled state is sufficient. Indeed, there is a quantum key distribution (QKD) scheme based on sharing the maximally entangled state. However, in general, a maximally entangled state is not necessary for QKD. In fact, there exists a larger class of so-called private states \cite{renner2005universally,horodecki2005secure} $\hat{\gamma}^{AB}_d$, from which clients $AB$ can obtain $\log_2 d$ secret bits. Such states are of the form $\hat{\gamma}^{AB}_d:=\hat{U}^{AB}_{\rm twist} (\ket{\Phi_d}\bra{\Phi_d}^{A'B'} \otimes \hat{\sigma}^{A''B''})(\hat{U}^{AB}_{\rm twist})^\dag$, with an arbitrary shared state $\hat{\sigma}^{A''B''}$ and an arbitrary controlled unitary operation in the form $\hat{U}^{AB}_{\rm twist}:=\sum_{i,j=1}^d \ket{ij}\bra{ij}^{A'B'} \otimes \hat{U}_{ij}^{A''B''}$. More generally, one could think of a group of clients $C:=C_1C_2\cdots C_n (\subset V)$ ($C_1,C_2,\ldots,C_n \in V$) who want to share a multipartite entangled state, such as a Greenberger–Horne–Zeilinger (GHZ) state\cite{greenberger1989going} 
$\ket{{\rm GHZ}_d}^{C}:=\sum_{i=1}^d\ket{ii\cdots i}^{C_1C_2\cdots C_n}/\sqrt{d}$, which can be used for secret sharing\cite{hillery1999quantum}, a clock synchronization protocol\cite{komar2014quantum} as well as quantum conference key agreement \cite{augusiak2009multipartite}. In the case of quantum conference key agreement there is, again, a larger class of so-called multipartite privates $\hat{\gamma}^{C}_d$, from which the group of clients can obtain $\log_2 d$ bits of conference key \cite{augusiak2009multipartite}. Such states are of the form $\hat{\gamma}^{C}_d:=\hat{U}^{C_1C_2 \cdots C_n}_{\rm twist} (\ket{{\rm GHZ}_d }\bra{{\rm GHZ}_d }^{C'_1C'_2\cdots C'_n} \otimes \hat{\sigma}^{C''_1C''_2\cdots C''_n})(\hat{U}^{C_1C_2 \cdots C_n}_{\rm twist})^\dag$, where $\hat{U}^{C_1C_2\cdots C_n}_{\rm twist}$ is an arbitrary controlled unitary operation in the form $\hat{U}^{C_1 C_2 \cdots C_n}_{\rm twist}:=\sum_{i_1,i_2,\ldots,i_n=1}^d \ket{i_1i_2\cdots i_n}\bra{i_1i_2\cdots i_n}^{C'_1 C'_2 \cdots C'_n} \otimes \hat{U}_{i_1i_2\cdots i_n}^{C''_1 C''_2 \cdots C''_n}$ and 
$\hat{\sigma}^{C''_1 C''_2 \cdots C''_n}$ is an arbitrary shared state.

Depending on the communication task that clients want to perform, clients $C$ may ask a provider to supply an resource entangled state $\hat{\tau}_d^C$ for the task. For example, $\hat{\tau}_d^C$ may be a bipartite maximally entangled state $\ket{\Phi_d}^C$, a private state $\hat{\gamma}^{AB}_d$, a GHZ state $\ket{{\rm GHZ}_d}^C$ or a multipartite private state $\hat{\gamma}^{C}_d$. The amount of the entangled resource present in $\hat{\tau}^C_d$ is quantified by the parameter $d$. Namely, $\log_2 d$ is the number of entangled qubits in a maximally entangled or GHZ state, or the number of secret bits available in a bi- or multipartite private state. The case of cluster states and general graph states will be discussed in Section \ref{sec:graph}. According to the request from the clients $C$, the provider prepares an available quantum network, associated with a graph $G=(V,E)$ such that $C\subset V$. This is because entanglement cannot be generated only with LOCC in general (or, precisely, entanglement is defined\cite{horodecki2009quantum} as a state which cannot be generated by LOCC, that is, which cannot be described by the form of fully separable states $\sum_i q_i \bigotimes_{v\in V} \hat{\rho}_i^v$, where $\{q_i\}_i$ is a probability distribution and $\hat{\rho}_i^v$ is a quantum state of node $v\in V$), and thus, the provider needs to use a quantum network.
Without loss of generality, a protocol of the provider which provides a final state $\hat{\rho}^V_{{\bm  k}_l}$ 
close to a target state $\hat{\tau}^C_{d_{{\bm k}_l}}$ 
with probability $p_{{\bm k}_l}$ is described as follows (see also Fig.~\ref{fig:network}b). A provider determines an available quantum network, associated with a graph $G=(V,E)$ such that $C\subset V$. 
The protocol starts by preparing the whole system in a fully separable state $\hat{\rho}_1^V$ and then by using a quantum channel ${\cal N}^{e_1}$ with $e_1\in E$. This is followed by arbitrary LOCC among all the nodes, which gives an outcome $k_1$ and a quantum state $\hat{\rho}_{k_1}^V$ with probability $p_{k_1}$ (because LOCC is based on local CP maps and sharing their outcomes by classical communication). 
In the second round, depending on the outcome $k_1$, a node may use a quantum channel ${\cal N}^{e_{k_1}}$ with $e_{k_1} \in E$, followed by LOCC among all the nodes. This LOCC gives an outcome $k_2$ and a quantum state $\hat{\rho}_{k_2 k_1}^V$ with probability $p_{k_2|k_1}$. Similarly, in the $i$-th round, according to the previous outcomes  ${\bm k}_{i-1}:=k_{i-1} \ldots k_2k_1$ (with ${\bm k}_0:=1$), the protocol may use a quantum channel ${\cal N}^{e_{{\bm k}_{i-1}}}$ with $e_{{\bm k}_{i-1}} \in E$, followed by LOCC providing a quantum state $\hat{\rho}_{{\bm k}_i}^V$ with a new outcome $k_i$ with probability $p_{k_i|{\bm k}_{i-1}}$. Eventually, after a finite number of rounds, say after an $l$-th round, the protocol presents clients $C$ with
$\hat{\rho}_{{\bm k}_l}^{C}={\rm Tr}_{V\setminus C} ( \hat{\rho}_{{\bm k}_l}^{V})$ close to a target state $\hat{\tau}_{d_{{\bm k}_l}}^{C}$ in the sense of $\|\hat{\rho}_{{\bm k}_l}^{C} -\hat{\tau}_{d_{{\bm k}_l}}^{C} \|_1 \le \epsilon$ for $\epsilon>0$, with probability $p_{{\bm k}_l}:=p_{k_l| {\bm k}_{l-1}}  \cdots p_{k_3| {\bm k}_2} p_{k_2|k_1} p_{k_1}$, where $\|\hat{X}\|_1$ is the trace norm defined by $\|\hat{X}\|_1 :={\rm Tr}\sqrt{\hat{X}^\dag \hat{X} } $.

Important quantities related to the evaluation of a protocol are (1) how many times quantum channels ${\cal N}^e$ are used in the protocol, (2) how valuable the state $\hat{\tau}_{d}^{C}$ given to clients $C$ is, and (3) how much error $\epsilon$ is allowed. As for (1), given that a general protocol could work in a probabilistic manner, $\bar{l}^e:=\sum_{i=0}^{l-1} \langle\delta_{e,e_{{\bm k}_i} }\rangle_{{\bm k}_i}=\sum_{i=0}^{l-1}\sum_{{\bm k}_i} p_{{\bm k}_i} \delta_{e,e_{{\bm k}_i}}$ with the Kronecker delta $\delta_{i,j}$ and $\bar{l}^E:=\sum_{e\in E} \bar{l}^e$ are important quantities, because $\bar{l}^e$ represents the average number of times quantum channel ${ \cal N }^e$ is used in the protocol and
$\bar{l}^E$ represents the average number of total channel uses.
Thus, $\{\bar{l}^e\}_{e \in E}$ can be a set to characterize the protocol.
If we introduce an average usage frequency $\bar{f}^e:=\bar{l}^e/\bar{l}^E (\ge 0)$ 
satisfying $\sum_{e \in E} \bar{f}^e=1$, a set $\{\bar{l}^E, \{\bar{f}^e \}_{e\in E} \}$ can be used to characterize the protocol. Or, by introducing an average usage rate $\bar{r}^e:=\bar{l}^e/l (\ge 0)$, a set $\{l, \{\bar{r}^e \}_{e\in E} \}$ can also be used to characterize the protocol, where $l$ is a parameter like time (because it has no explicit relation with $\bar{l}^e$ in contrast to $\bar{l}^E$).
As for (2), in general, it is not so obvious to define a quantity for evaluating the value of the target state $\hat{\tau}_{d_{{\bm k}_l}}^{C}$, because the value depends on the final application in general and may be determined by a complex function of the target state $\hat{\tau}_{d}^{C}$ or its parameter $d$. However, if the target state $\hat{\tau}_{d}^{C}$ is GHZ state $\ket{{\rm GHZ}_d}^C$ or private state $\gamma_d^C$, the value of the target state can be quantified by $\log_2 d$ showing how many qubits or private bits can be sent by consuming the target state under LOCC. As for (3), error $\epsilon$ in terms of the trace distance quantifies how far from a target state $\hat{\tau}_{{\bm k}_l}^{C}$ the final state $\hat{\rho}_{{\bm k}_l}^{C}$ is. In general, this trace distance $\| \hat{\rho}-\hat{\sigma}\|_1$ is related\cite{fuchs1999cryptographic} with another well known measure, the fidelity $F(\hat{\rho},\hat{\sigma}):=\| \sqrt{\hat{\rho}}\sqrt{\hat{\sigma}}\|_1$, as $2(1-\sqrt{F})\le \| \hat{\rho} -\hat{\sigma} \|_1 \le 2 \sqrt{1-F}$. Although the fidelity is a useful and intuitive tool, the trace distance is used often in theory for quantum communication, because the trace distance has universal composability\cite{ben2005universal}, in contrast to the fidelity.

\subsubsection{For two-party communication}

The upper bounds on how efficiently maximally entangled states and private states can be distributed for two clients have been derived by Pirandola\cite{P16,p19} and Azuma {\it et al}.\cite{AML16} In particular, in this case, the target state $\hat{\tau}^C_d$ is regarded as a single bipartite state $\ket{\Phi_d}^{AB}$ or $\hat{\gamma}_d^{AB}$, where $A,B\in V$. 
To obtain such upper bounds, Pirandola generalizes the Pirandola-Laurenza-Ottaviani-Banchi (PLOB) bound\cite{pirandola2017fundamental} on the private capacity for point-to-point quantum key distribution with a pure-loss bosonic channel, while Azuma {\it et al.} use the Takeoka-Guha-Wilde (TGW) bound\cite{takeoka2014fundamental} for that. The shared methodology is summarized by Rigovacca {\it et al.}\cite{rigovacca2018versatile}, which is described in what follows.

An entanglement measure\cite{horodecki2009quantum} ${\cal E}$ between parties $X$ and $Y$ does not increase on average under any LOCC between them and is zero for any separable state. In addition to this normal property as an entanglement measure, suppose that the entanglement measure ${\cal E}$ satisfies the following two properties: (i) If a state $\hat{\rho}^{XY}$ is close to a target state $\hat{\tau}_d^{XY}$ which is $\ket{\Phi_d}\bra{\Phi_d}^{XY}$ (or $\hat{\gamma}_d^{XY}$), that is, $\|\hat{\rho}^{XY}-\hat{\tau}_d^{XY} \|_1 \le \epsilon$, then there exist two real continuous functions $f_{\cal E}(\epsilon)$ and $g_{\cal E}(\epsilon)$ with $\lim_{\epsilon \to 0} f_{\cal E}( \epsilon)=0$ and $\lim_{\epsilon \to 0} g_{\cal E}( \epsilon)=1$, such that ${\cal E}(\hat{\rho}^{XY})\ge g_{\cal E}(\epsilon) \log_2 d - f_{\cal E}(\epsilon)$; (ii) we have ${\cal E}({\cal N}^{X'\to Y'} (\hat{\rho}^{XX'Y})) \le {\cal E}({\cal N}^{X'\to Y'}) +{\cal E}(\hat{\rho}^{XX'Y})$ for any state $\hat{\rho}^{XX'Y}$, where 
$
{\cal E}({\cal N}^{X \to Y}) :=\max_{\hat{\sigma}^{XX'}} {\cal E}({\cal N}^{X \to Y}(\hat{\sigma}^{XX'}))
$ is 
called the entanglement of channel\cite{takeoka2014fundamental,pirandola2017fundamental,christandl2017relative,wilde2017converse}. With an entanglement measure ${\cal E}$ satisfying Properties (i) and (ii), an upper bound is given in the following form:
For any protocol which provides clients $A(\in V)$ and $B(\in V)$ with maximally entangled state $\ket{\Phi_{d_{{\bm k}_l}}}^{AB}$ (or private state $\hat{\gamma}_{d_{{\bm k}_l }}^{AB}$) with probability $p_{{\bm k}_l}$ and error $\epsilon>0$, by using a quantum network associated with graph $G=(V,E)$, it holds
\begin{equation}
\langle \log_2 d_{{\bm k}_l} \rangle_{{\bm k}_l} \le \frac{1}{g_{\cal E}(\epsilon)} \sum_{e \in \partial(V_A)} \bar{l}^e {\cal E}({\cal N}^e) + \frac{f_{\cal E}(\epsilon)}{g_{\cal E} (\epsilon)} \label{eq:upper1}
\end{equation}
for any $V_A \subset V$ which is a set of nodes with $A \in V_A$ and $B \notin V_A$, where
$\langle x_{{\bm k}} \rangle_{{\bm k}} :=\sum_{{\bm k}}p_{{\bm k}} x_{{\bm k}}$ and $\partial(V_A) (\subset E)$ is the set of edges which connect a node in $V_A$ and a node in $V\setminus V_A$ (Fig.~\ref{fig:network}). Since Eq.~(\ref{eq:upper1}) holds for any choice of $V_A$, the minimization of the right-hand side of Eq.~(\ref{eq:upper1}) over the choice of $V_A$ gives the tightest bound on the left-hand side. In general, there might be cases where we have no computable expressions for point-to-point capacities, the best characterization which is simply the optimal rate achievable with a protocol over any number of channel uses. In consequence, it is unclear how to proceed to optimize additionally over the average use $ \bar{l}^e$ of each channel ${\cal N}^e$. In contrast, the upper bound (\ref{eq:upper1}) is always estimable, once we are given $\{{\cal E}({\cal N}^e)\}_{e\in E}$ or their upper bounds, because it depends only on a single use of channel ${\cal N}^e $, rather than multiple uses of it.

Examples of entanglement measure ${\cal E}$ satisfying Property (i)\cite{wilde2016squashed,christandl2017relative,rigovacca2018versatile} for either $\ket{\Phi_d}^{AB}$ or $\hat{\gamma}^{AB}_d$ and Property (ii)\cite{takeoka2014fundamental,takeoka2014squashed,christandl2017relative} are the squashed entanglement\cite{tucci2002entanglement,christandl2004squashed} $E_{\rm sq}$ and the max-relative entropy\cite{datta2009min} $E_{\rm max}$ of entanglement. The relative entropy\cite{vedral1997quantifying,vedral1998entanglement} $E_{\rm R}$ of entanglement satisfies\cite{horodecki2009general,pirandola2017fundamental,wilde2017converse} Property (i) for either $\ket{\Phi_d}^{AB}$ or $\hat{\gamma}^{AB}_d$ in general, but, at present, it is shown\cite{pirandola2017fundamental,wilde2017converse,P16} to satisfy Property (ii) only for some quantum channels, called teleportation simulable/stretchable channels or Choi simulable channels\cite{bennett1996mixed,gottesman1999demonstrating,horodecki1999general,knill2001scheme,wolf2007quantum,niset2009no,muller2012transposition,pirandola2017fundamental}. 
A channel ${\cal N}^{A\to B}$ is called teleportation simulable if there is a deterministic LOCC operation ${\cal L}^{A'A'':B}$ between systems $A'A''$ and $B$ such that
\begin{equation}
{\cal N}^{A\to B} (\hat{\rho}^A)={\cal L}^{A'A'':B} (  {\cal N}^{A\to B} (\ket{\Phi_d} \bra{\Phi_d}^{AA'})\otimes \hat{\rho}^{A''})
\end{equation}
for any state $\hat{\rho}^A$, where $d=\dim {\cal H}^A$. Examples of these channels are ones which ``commute'' with the unitary correction in the quantum teleportation. In particular, the quantum teleportation\cite{bennett1993teleporting} from system $A''$ in an unknown state $\hat{\rho}^{A''}$ to system $A$ is implemented by generalized Bell measurement $\{ \hat{M}^{A'A''}_i \}_{i=1,2,\ldots,d^2}$ on systems $A'A''$ in state $\ket{\Phi_d}\bra{\Phi_d}^{AA'}  \otimes \hat{\rho}^{A''}$, followed by a unitary operation $U^A_i$ depending on the random outcome $i$ of the Bell measurement. This means $\hat{\rho}^A=d^2 (\hat{U}_i^A \otimes \hat{M}_i^{A'A''})(\ket{\Phi_d}\bra{\Phi_d}^{AA'} \otimes \hat{\rho}^{A''})(\hat{U}_i^A \otimes \hat{M}_i^{A'A''})^\dag$. Therefore, if  every correction unitary $\hat{U}_i^A$ ``commutes'' with ${\cal N}^{A\to B}$, more precisely, if there is a unitary operation $\hat{V}_i^B$ with ${\cal N}^{A\to B}(\hat{U}_i^A \hat{\sigma}^A (\hat{U}^A_i)^\dag) =\hat{V}_i^B{\cal N}^{A\to B}(\hat{\sigma}^A ) (\hat{V}_i^B)^\dag$ for any state $\hat{\sigma}^A$ and any outcome $i$, 
then the channel ${\cal N}^{A\to B}$ is teleportation simulable. For example, depolarizing channels, phase-flip channels, and lossy bosonic channels are teleportation simulable, while amplitude damping channels are not teleportation simulable.

In contrast to cases of point-to-point quantum communication (that is, cases corresponding to a graph $G$ with $V=\{A,B\}$ and $E=\{A\to B\}$), the definitions of quantum/private capacities in a quantum network themselves may have variety\cite{AK17,BAKE20}. 

A rather general definition is 
\begin{multline}
{ \cal C}(G,\{{\cal N}^e\}_{e \in E}, \{q^e\}_{e \in E} )
=\lim_{\epsilon \to 0} \lim_{n \to \infty}
\\ \sup_{\Lambda(n, \{ q^e \}_{e \in E},\epsilon)} \left\{  \frac{\langle \log_2 d_{{\bm k}_l} \rangle_{{\bm k}_l}}{n} :\|\hat{\rho}_{{\bm k}_l}^{AB} -\hat{\tau}_{d_{{\bm k}_l}}^{AB} \|_1 \le \epsilon \right\} \label{eq:single-capa-1}
\end{multline}
for given $\{q^e\}_{e \in E}$ with $q^e \ge 0$,
where $n\ge  0$, $\Lambda(n, \{ q^e \}_{e \in E},\epsilon)$ is a set of protocols
which provide clients $A(\in V)$ and $B(\in V)$ with maximally entangled state $\ket{\Phi_{d_{{\bm k}_l}}}^{AB}$ (or private state $\hat{\gamma}_{d_{{\bm k}_l }}^{AB}$) with probability $p_{{\bm k}_l}$ and error $\epsilon>0$ by using quantum channels ${\cal N}^e$ $n q^e$ times on average at most (i.e., $\bar{l}^e \le n q^e$), 
and 
${\cal C}$ represents a network quantum capacity ${\cal Q}$ (private capacity ${\cal P}$) per time. Notice that any protocol with $l \bar{r}^e \le n q^e$ belongs to the set  $\Lambda(n, \{ q^e \}_{e \in E},\epsilon)$ of protocols. 

We can capture alternative definitions of capacity by putting additional constraints on $\{q^e\}_{e \in E}$. Let us introduce two particularly relevant alternative capacity definitions.

First, we can consider a scenario where the users can optimize over the usage frequency of individual channels and define the capacity as the optimal rate normalized by the total number of channel uses: 
\begin{multline}
{ \cal C}(G,\{{\cal N}^e\}_{e \in E}) \\
=\max_{q^e \ge 0, \;\sum_{e \in E} q^e =1}{ \cal C}(G,\{{\cal N}^e\}_{e \in E}, \{q^e\}_{e \in E} ),\label{eq:single-capa-2}
\end{multline}
where
${\cal C}$ represents a network quantum capacity ${\cal Q}$ (private capacity ${\cal P}$) per the average total number of channel uses and we have added the additional constraint:  $\sum_{e\in E}q^e =1 $.
 Notice that any protocol with $\bar{l}^E \bar{f}^e \le n q^e$ belongs to the set $\Lambda(n, \{ q^e \}_{e \in E},\epsilon)$ of protocols with $\sum_{e\in E} q^e=1 $. 
 
Second, we can consider the usage of the whole network as the unit resource. This network use metric can be captured by setting $q^e=1$ for all $e\in E$. This quantity is equivalent to the one introduced by Pirandola \cite{p19} for multi-path routing protocols with a so-called flooding strategy. 

Finally, we note that Pirandola also introduced the single path per network use capacity \cite{p19}. This quantity is not known to correspond to a restriction on $\{ q^e \}_{e \in E}$. For additional details on this quantity we refer to the original reference \cite{p19} and to a reference\cite{BAKE20} for a comparison with the channel use metric.

For all of these capacities except for the single path per network use, Eq.~(\ref{eq:upper1}) directly gives upper bounds,
\begin{align}
& { \cal C}(G,\{{\cal N}^e\}_{e \in E}, \{q^e\}_{e \in E} ) \le \min_{V_A} \sum_{e \in \partial(V_A)} q^e {\cal E}({\cal N}^e), \label{eq:upper2'} \\
&{ \cal C}(G,\{{\cal N}^e\}_{e \in E}) \le \max_{q^e \ge 0,\;\sum_{e \in E}q^e =1} \min_{V_A} \sum_{e \in \partial(V_A)} q^e {\cal E}({\cal N}^e), \label{eq:upper3'}
\end{align}
for ${\cal C}={\cal Q}, {\cal P}$. Also note that 
\begin{align}
& { \cal Q}(G,\{{\cal N}^e\}_{e \in E}, \{q^e\}_{e \in E} ) \le { \cal P}(G,\{{\cal N}^e\}_{e \in E}, \{q^e\}_{e \in E} ) , \\
&{ \cal Q}(G,\{{\cal N}^e\}_{e \in E}) \le { \cal P}(G,\{{\cal N}^e\}_{e \in E}) 
\end{align}
holds, because the maximally entangled state $\ket{\Phi_d}^{AB}$ is a special example of the private states $\hat{\gamma_d}^{AB}$.

An upper bound in the form (\ref{eq:upper1}) was originally derived for point-to-point quantum communication, that is, for cases corresponding to a graph $G$ with $V=\{A,B\}$ and $E=\{A\to B\}$.
The first useful upper bound on the private/quantum capacity is derived by Takeoka {\it et al.}\cite{takeoka2014fundamental}, with the use of the squashed entanglement $E_{\rm sq}$, which applies to any point-to-point quantum channel ${\cal N}^{A\to B}$. 
This TGW bound has no scaling gap with secure key rates (per pulse) of existing point-to-point optical QKD protocols, showing that there remains not much room to improve further the protocols in terms of performance. 
An alternative bound is derived by Pirandola {\it et al.}\cite{pirandola2017fundamental}, with the use of the relative entropy $E_{\rm R}$ of entanglement, to obtain tighter bounds for teleportation simulable channels, able to
close the gap with previously-derived lower bounds\cite{pirandola2009direct}. Indeed, this PLOB bound coincides with the performance of achievable point-to-point entanglement generation protocols
over erasure channels, dephasing channels, bosonic quantum amplifier channels, and lossy bosonic channels, implying that it represents the quantum/private capacities, i.e.,
\begin{equation}
{\cal Q}({\cal N}^e) ={\cal P}({\cal N}^e)= E_{\rm R}({\cal N}^e) \label{eq:PLOB}
\end{equation}
for those channels. For instance, in the case where the channel ${\cal N}^{A \to B}$ is a single-mode lossy bosonic channel with a transmittance $\eta$ ($0 \le \eta \le 1$), $E_{\rm R}({\cal N}^{A \to B}) = -\log_2(1-\eta)$ while we only know $E_{\rm sq}({\cal N}^{A \to B})\le \log_2[(1+\eta)/(1-\eta)]$. However, the PLOB bound can be applied only to teleportation simulable channels, in contrast to the TGW bound.
As a general upper bound applicable to arbitrary channels like the TGW bound, Christandl and M\"{u}ller-Hermes derive an upper bound\cite{christandl2017relative}, with the use of the max relative entropy $E_{\rm max}$ of entanglement, which is also shown to be a strong converse bound, i.e. the error tends to one exponentially once the rate exceeds the bound.
Further detail on the comparison between the upper bounds can be found in the review paper by Pirandola {\it et al.}~\cite{pirandola2019advances}.

Then, those upper bounds for point-to-point quantum communication were lifted up to ones for general quantum networks associated with a graph $G=(V,E)$, by Pirandola\cite{P16,p19}, Azuma {\it  et al.}\cite{AML16} and Rigovacca {\it et al}.\cite{rigovacca2018versatile}. The main idea here is to regard a quantum internet protocol as a point-to-point quantum communication protocol between a party having the full control over $V_A(\subset  V)$ and a party having the full control over $V \setminus V_A$.
In particular, Pirandola derives\cite{P16,p19} an upper bound (\ref{eq:upper1}) with ${\cal E}=E_{\rm R}$ for any quantum network composed of teleportation simulable channels, with the use of the relative entropy $E_{\rm R}$ of entanglement, by following the PLOB bound\cite{pirandola2017fundamental}, while Azuma {\it et al.} provide\cite{AML16} an upper bound (\ref{eq:upper1}) with ${\cal E}=E_{\rm sq}$ for any quantum network composed of arbitrary quantum channels, with the use of the squashed entanglement $E_{\rm sq}$, by following the TGW bound\cite{takeoka2014fundamental}. To make Pirandola's relative entropy bound applicable to arbitrary quantum networks like the upper bound of Azuma {\it et al.}, Rigovacca {\it et al.} present\cite{rigovacca2018versatile} a `hybrid' relative entropy bound for any quantum network composed of arbitrary quantum channels, by using an inequality introduced by Christandl and M\"{u}ller-Hermes\cite{christandl2017relative}. This bound corresponds to the case where ${\cal E}({\cal N}^e)$ in Eq.~(\ref{eq:upper1}) is assumed to be
\begin{equation}
{\cal E} ({\cal N}^e)=
\begin{cases}
E_{\rm R} ({\cal N}^e) & ( {\cal N}^e \in {\cal S}),\\
E_{\rm max}({\cal N}^e) & ( {\cal N}^e \notin {\cal S}),
\end{cases}
\end{equation}
where ${\cal N}^e \in {\cal S}$ indicates that quantum channel ${\cal N}^e$ is teleportation simulable. 

Pirandola gives\cite{P16,p19} an achievable protocol that works over a quantum network composed of teleportation simulable channels, while Azuma {\it et al.} present\cite{AK17} an abstract but general achievable protocol working over any quantum network. Here we focus on the latter protocol. In particular, Azuma {\it et al.} give an achievable protocol over a general quantum network associated with a graph $G=(V,E)$, by assuming that we have an achievable entanglement generation protocol over every channel ${\cal N}^e$. In particular, they assume that there is an entanglement generation protocol (including entanglement purification) over quantum channel ${\cal N}^e$ which provides a state $\hat{\rho}^e$ $\delta$-close to $ l^e R_\delta^e$ copies of a qubit maximally entangled state $\ket{\Phi_2}^{AB}$, called a Bell pair, by using quantum channel $({\cal N}^e)^{\otimes l^e}$, that is, $\|\hat{\rho}^e - (\ket{\Phi_2}\bra{\Phi_2}^e)^{\otimes l^e R_\delta^e} \|_1 \le \delta$ with $\delta \ge 0$ (for large $l^e$). If this entanglement generation protocol runs over every channel ${\cal N}^e$, the network can share the state $\bigotimes_{e \in E} \hat{\rho}^e$ with
\begin{equation}
\left\| \bigotimes_{e \in E} \hat{\rho}^e - \bigotimes_{e \in E} (\ket{\Phi_2}\bra{\Phi_2}^e)^{\otimes l^e R_\delta^e}   \right\|_1 \le |E| \delta. \label{eq:agg}
\end{equation}
If we regard each Bell state $\ket{\Phi_2}\bra{\Phi_2}^e$ of the Bell-pair network $\bigotimes_{e \in E} (\ket{\Phi_2}\bra{\Phi_2}^e)^{\otimes l^e R_\delta^e} $ as an undirected edge $e''$ with the same two ends of $e$, we can make a multigraph $G''=(V,E'')$ with the set of $E''$ of such undirected edges $e''$. In this multigraph $G''$, if we find a single path between nodes $A$ and $B$, called $AB$-path, then we can give a Bell pair to the nodes $A$ and $B$ by performing the entanglement swapping at the vertices along the $AB$-path. Hence, the number of Bell pairs which are provided to the nodes $AB$ equals to the number of edge-disjoint $AB$-paths. Menger's theorem in graph theory\cite{menger1927allgemeinen,bondy2008graph} tells us that the number $M$ of edge-disjoint $AB$ paths in a multigraph is equivalent to the minimum number of edges in an $AB$ cut (see Section~\ref{se:single-commodity} for this theorem). Thus, $M=\min_{V_A} \sum_{e \in \partial(V_A)} l^e R_{\delta}^e$ for the multigraph $G''$. If we perform the entanglement swapping along $M$ edge-disjoint $AB$ paths, we have 
\begin{equation}
\left\| \hat{\rho}^{AB} - (\ket{\Phi_2}\bra{\Phi_2}^e)^{M}   \right\|_1 \le |E| \delta
\end{equation}
from Eq.~(\ref{eq:agg}). Therefore, this protocol, called the aggregated quantum repeater protocol, supplies nodes $AB$ with $\min_{V_A} \sum_{e \in \partial(V_A)} l^e R_{\delta}^e$ Bell pairs with error $|E|\delta$. For fixed $q^e=l^e/l(>0)$ (or $q^e=l^e/l^E(>0)$ with $l^E=\sum_{e \in E} l^e$), if the assumed entanglement generation protocol over every channel ${\cal N}^e$ can achieve quantum capacity, that is, $\delta \to 0$ and $R_\delta \to {\cal Q}({\cal N}^e)$, by regarding $l$ ($l^E$) as $n$ and then taking $n \to \infty$, the aggregated quantum repeater protocol gives lower bounds on the network quantum capacities of protocols $\Lambda(n,\{q^e\}_{e\in E},\epsilon)$ as
\begin{align}
&\min_{V_A} \sum_{e \in \partial(V_A)} q^e {\cal Q}({\cal N}^e) \le {\cal Q}(G,\{{\cal N}^e\}_{e\in E},q^e\}_{e \in E}), \label{eq:lower1}\\
&\max_{q^e \ge 0,\;\sum_{e\in E} q^e=1}\min_{V_A} \sum_{e \in \partial(V_A)} q^e {\cal Q}({\cal N}^e) \le {\cal} {\cal Q}(G,\{{\cal N}^e\}_{e\in E}). \label{eq:lower2}
\end{align}

If a quantum network is composed only of quantum channels ${\cal N}^e$ that satisfy Eq.~(\ref{eq:PLOB}), as Pirandola has considered\cite{P16,p19}, lower bounds in Eqs.~(\ref{eq:lower1}) and (\ref{eq:lower2}) equal to upper bounds in Eqs.~(\ref{eq:upper2'}) and (\ref{eq:upper3'}) with ${\cal E}=E_{\rm R}$, respectively. For instance, in the case of a quantum  network composed only of lossy bosonic channels, the aggregated quantum repeater protocol achieves the private/quantum capacities of the network\cite{AK17}.

\subsubsection{For multiple-pair communication}\label{se:bound-multiple}

In general, a provider may be asked to concurrently distribute multiple pairs of bipartite maximally entangled states or bipartite private states, i.e., $\hat{\tau}_{\bm d}^C:=\bigotimes_i \ket{\Phi_{d^{(i)}}}\bra{\Phi_{d^{(i)}}}^{A_iB_i}$ or $\hat{\tau}_{\bm d}^C:=\bigotimes_i \hat{\gamma}_{d^{(i)}}^{A_iB_i}$, where $A_i, B_i \in V$, $C=\bigotimes_i A_iB_i$ and ${\bm d}:=(d^{(1)},d^{(2)},\cdots)$. The upper bounds for this problem are given\cite{P16,P19b} by Pirandola for any quantum network composed of teleportation simulable channels, with the use of the relative entropy $E_{\rm R}$ of entanglement, while they are given by B\"{a}uml {\it et al.}\cite{BA17} and Azuma {\it et al.}\cite{AML16} for any quantum network composed of arbitrary quantum channels, with the use of the squashed entanglement $E_{\rm sq}$. The works of Pirandola and Azuma {\it et al.} still use bipartite entanglement measures, while B\"{a}uml {\it et al.} enable us to use multipartite squashed entanglement\cite{yang2009squashed,avis2008distributed}. The upper bounds are described as follows, if we use the bipartite entanglement measure ${\cal E}$ holding the bound (\ref{eq:upper1}):
Any protocol which provides every pair $A_i B_i$ of clients with maximally entangled state $\ket{\Phi_{d^{(i)}_{{\bm k}_l}}}^{A_iB_i}$ (or private state $\hat{\gamma}_{d^{(i)}_{{\bm k}_l }}^{A_iB_i}$) with probability $p_{{\bm k}_l}$ and error $\epsilon>0$ (i.e., $\|\hat{\rho}^C_{{\bm k}_l} -\hat{\tau}_{{\bm d}_{{\bm k}_l}}^C \|_1 \le \epsilon$), 
by using a quantum network associated with graph $G=(V,E)$, follows
\begin{equation}
\sum_{i \in I(V')} \langle \log_2 d^{(i)}_{{\bm k}_l} \rangle_{{\bm k}_l} \le \frac{1}{g_{\cal E}(\epsilon)}  \sum_{e \in \partial(V')} \bar{l}^e {\cal E}({\cal N}^e) + \frac{f_{\cal E}(\epsilon)}{g_{\cal E} (\epsilon)} \label{eq:upper2}
\end{equation}
for any $V' \subset V$, where
$\langle x_{{\bm k}} \rangle_{{\bm k}} :=\sum_{{\bm k}}p_{{\bm k}} x_{{\bm k}}$, $\partial(V') (\subset E)$ is the set of edges which connect a node in $V'$ and a node in $V\setminus V'$, and
$I(V_A)$ is the set of indices $i$ whose corresponding pair $A_iB_i$ satisfies $A_i \in V$ and $B_i \in V\setminus V'$, or $A_i \in V\setminus V'$ and $B_i \in V'$. 

Capacities for multiple-pair communication could have more variety than those for two-party communication, because there are many pairs of clients\cite{BAKE20}. A possible definition is 
\begin{multline}
{ \cal C}^{\rm worst}(G,\{{\cal N}^e\}_{e \in E}, \{q^e\}_{e \in E} )
=\lim_{\epsilon \to 0} \lim_{n \to \infty}
\\ \sup_{\Lambda(n, \{ q^e \}_{e \in E},\epsilon)} \min_i \left\{  \frac{\langle \log_2 d_{{\bm k}_l}^{(i)} \rangle_{{\bm k}_l}}{n} :\|\hat{\rho}^C_{{\bm k}_l} -\hat{\tau}_{{\bm d}_{{\bm k}_l}}^C \|_1 \le \epsilon \right\} \label{eq:c1}
\end{multline}
for given $\{q^e\}_{e \in E}$ with $q^e \ge 0$,
where $n\ge  0$, $\Lambda(n, \{ q^e \}_{e \in E},\epsilon)$ is a set of protocols
which provide clients $A_i(\in V)$ and $B_i(\in V)$ with maximally entangled state $\ket{\Phi_{d^{(i)}_{{\bm k}_l}}}^{A_iB_i}$ (or private state $\hat{\gamma}_{d^{(i)}_{{\bm k}_l }}^{AB}$) with probability $p_{{\bm k}_l}$ and error $\epsilon>0$ (i.e., $\|\hat{\rho}^C_{{\bm k}_l} -\hat{\tau}_{{\bm d}_{{\bm k}_l}}^C \|_1 \le \epsilon$), by using quantum channels ${\cal N}^e$ $n q^e$ times on average at most (i.e., $\bar{l}^e \le   nq^e$), 
and 
${\cal C}^{\rm worst}$ represents a worst-case network quantum capacity ${\cal Q}^{\rm worst}$ (private capacity ${\cal P}^{\rm worst}$) per time.
Notice that any protocol with $l \bar{r}^e \le n q^e$ belongs to the set $\Lambda(n, \{ q^e \}_{e \in E},\epsilon)$ of protocols. 
The use of objective function $\min_i \{\langle \log_2 d_{{\bm k}_l}^{(i)} \rangle_{{\bm k}_l}/n\}_i$ means that a protocol is evaluated with the least achievable rate that is guaranteed for any client pair.
By putting an additional constraint $\sum_{e\in E}q^e =1 $ on the set $\{q^e\}_{e \in E}$, the capacity per time is transformed into one per the average total number of channel uses: 
\begin{multline}
{ \cal C}^{\rm worst}(G,\{{\cal N}^e\}_{e \in E}) \\
=\max_{q^e \ge 0, \;\sum_{e \in E} q^e =1}{ \cal C}^{\rm worst}(G,\{{\cal N}^e\}_{e \in E}, \{q^e\}_{e \in E} ). \label{eq:c2}
\end{multline}
Notice that any protocol with $\bar{l}^E \bar{f}^e \le n q^e$ belongs to the set $\Lambda(n, \{ q^e \}_{e \in E},\epsilon)$ of protocols with $\sum_{e\in E} q^e=1 $. 
Another possible definition is
\begin{multline}
{ \cal C}^{\rm weight} (G,\{{\cal N}^e\}_{e \in E}, \{q^e\}_{e \in E};\{s_i\}_i)
=\lim_{\epsilon \to 0} \lim_{n \to \infty}
\\ \sup_{\Lambda(n, \{ q^e \}_{e \in E},\epsilon)} \left\{  \frac{\sum_i s_i \langle  \log_2 d_{{\bm k}_l}^{(i)} \rangle_{{\bm k}_l}}{n} :\|\hat{\rho}^C_{{\bm k}_l} -\hat{\tau}_{{\bm d}_{{\bm k}_l}}^C \|_1 \le \epsilon \right\} \label{eq:c3}
\end{multline}
for given $\{q^e\}_{e \in E}$ with $ q^e \ge 0$ and given $\{s_i\}_i$ with $s_i\ge 0$, 
where 
${\cal C}^{\rm weight} $ represents a weighted multi-pair network quantum capacity ${\cal Q}^{\rm weight} $ (private capacity ${\cal P}^{\rm weight} $) per time.
The use of objective function $\sum_i s_i \langle \log_2 d_{{\bm k}_l}^{(i)} \rangle_{{\bm k}_l}/n$ means that a protocol is evaluated with the average of rates for client pairs $A_iB_i$.
By putting an additional constraint $\sum_{e\in E}q^e =1 $ on the set $\{q^e\}_{e \in E}$,
another possible definition is
\begin{multline}
{ \cal C}^{\rm weight} (G,\{{\cal N}^e\}_{e \in E};\{s_i\}_i) \\
=\max_{q^e \ge 0, \;\sum_{e \in E} q^e =1}{ \cal C}^{\rm weight} (G,\{{\cal N}^e\}_{e \in E}, \{q^e\}_{e \in E} ;\{s_i\}_i), \label{eq:c4}
\end{multline}
where ${\cal C}^{\rm weight} $ represents a weighted multi-pair network quantum capacity ${\cal Q}^{\rm weight} $ (private capacity ${\cal P}^{\rm weight} $) per the average total number of channel uses. As an important special case of ${\cal C}^{\rm weight} $, we may also consider total throughput ${\cal C}^{\rm total}$ defined as
\begin{multline}
{\cal C}^{\rm total} (G,\{{\cal N}^e\}_{e \in E}, \{q^e\}_{e \in E})  \\
={\cal C}^{\rm weight} (G,\{{\cal N}^e\}_{e \in E}, \{q^e\}_{e \in E};\{s_i=1\}_i), 
\end{multline}
and
\begin{equation}
{\cal C}^{\rm total} (G,\{{\cal N}^e\}_{e \in E}) 
={\cal C}^{\rm weight} (G,\{{\cal N}^e\}_{e \in E};\{s_i=1\}_i).
\end{equation}

Eq.~(\ref{eq:upper2}) gives upper bounds on the capacities. Let $R_i:=\langle \log_2 d_{{\bm k}_l}^{(i)} \rangle_{{\bm k}_l}/n$.
If we multiply Eq.~(\ref{eq:upper2}) by $n^{-1}$ with $n\ge 0$ and take the limit of $n\to \infty$ and $\epsilon\to 0$, it gives upper bounds on $R_i$ for every $i$, on $R_i + R_j$ for every $i \neq j$, on $R_i + R_j + R_k$ for every different $i,j,k$, $\cdots$, and on $\sum_i R_i$, by changing the choice of $V'$. For given $\{q^e\}_{e\in E}$, if we minimize every upper bound over possible choices of $V'$ for it, the minimized upper bounds give a minimal polytope in the Cartesian coordinates ${\bm R}:=(R_1,R_2,\cdots)$ which includes all possibly achievable points of vector ${\bm R}$. 
Since objective functions in the definitions of ${\cal C}^{\rm worst}(G,\{{\cal N}^e\}_{e\in E}, \{q^e\}_{e \in E})$ and ${\cal C}^{\rm weight}(G,\{{\cal N}^e\}_{e\in E}, \{q^e\}_{e \in E};\{s_i\}_i)$ are concave, maximization of the objective functions over vector ${\bm R} $ in the minimal polytope is convex optimization, which provides upper bounds on ${\cal C}^{\rm worst}(G,\{{\cal N}^e\}_{e\in E}, \{q^e\}_{e \in E})$ and ${\cal C}^{\rm weight}(G,\{{\cal N}^e\}_{e\in E}, \{q^e\}_{e \in E};\{s_i\}_i)$. If we further change $\{q^e\}_{e \in E}$ to obtain upper bounds on ${\cal C}^{\rm worst}(G,\{{\cal N}^e\}_{e\in E})$ and ${\cal C}^{\rm weight}(G,\{{\cal N}^e\}_{e\in E};\{s_i\}_i)$, we need to start again from finding the minimal polytope and then to maximize the objective function.
Instead of these recipes including not only minimization but also maximization, B\"{a}uml {\it et al.} provide\cite{BAKE20} a computationally efficient method to obtain upper bounds, which are explained in Section~\ref{sec:bounds}.

To find lower bounds on the capacities, we need to construct a protocol. A simple way to make such a protocol is to invoke the idea of the aggregated quantum repeater protocol\cite{AK17}. In particular, by running entanglement generation over every quantum channel ${\cal N}^e$, the network shares a quantum state $\bigotimes_{e\in E}\hat{\rho}^e$ $\epsilon$-close to $\bigotimes_{e\in E} (\ket{\Phi_2}\bra{\Phi_2}^e)^{\otimes l^e R_\delta^e }$ associated with an undirected multigraph $G''=(V,E'')$, where each undirected edge $e''\in E''$ with the same two ends of $e$ corresponds to a Bell pair $\ket{\Phi_2}^e$ between the ends of $e''$. 
Then, if we find a path between $A_i$ and $B_i$ in the undirected multigraph $G''$, we can provide a state close to a Bell pair $\ket{\Phi_2}^{A_iB_i}$ by performing entanglement swapping along the $A_iB_i$-path. 
Thus, depending on an objective function in a definition of a capacity, by finding out edge-disjoint paths between pairs $A_iB_i$ in the undirected mutligraph $G''$ properly, we can give a state close to $\hat{\tau}_{\bm d}^C:=\bigotimes_i \ket{\Phi_{d^{(i)}}}\bra{\Phi_{d^{(i)}}}^{A_iB_i}$. Along this recipe,
B\"{a}uml {\it et al.} provide\cite{BAKE20} a computationally efficient way to obtain lower bounds on the capacities, which are explained in Section~\ref{sec:bounds}.

\subsubsection{For multipartite communication}

The upper bound\cite{BA17} of B\"{a}uml {\it et al.} with the use of multipartite squashed entanglement\cite{yang2009squashed,avis2008distributed} works for broader classes of networks, by generalizing the idea of Seshadresan {\it et el.}\cite{seshadreesan2016bounds}.
Indeed, they have considered a quantum broadcast network which might include quantum broadcast channels as well. A quantum broadcast channel is a device to distribute a quantum system from a single node to multiple nodes, and thus, 
the edge $e$ of a quantum broadcast channel ${\cal N}^e$ is a directed hyperedge which has a single tail $t(e) \in V$ and multiple heads $h(e) (\subset V$). This implies that a quantum broadcast network is associated with a directed hypergraph $G=(V,E)$ with a set $V$ of vertices and a set $E$ of directed hyperedges. 
Besides, the goal of a quantum internet protocol working over such a quantum broadcast network is to distribute multipartite GHZ states $\hat{\tau}_{\bm d}^C:=\bigotimes_i \ket{{\rm GHZ}_{d^{(i)}}}\bra{{\rm GHZ}_{d^{(i)}}}^{C_i}$ or multipartite private states $\hat{\tau}_{\bm d}^C:=\bigotimes_i \hat{\gamma}_{d^{(i)}}^{C_i}$, where $C_i \subset V$.
Then, the upper bound is described as follows, with the use of multipartite squashed entanglement $E_{\rm sq}^{P}$:
Any protocol which provides every set $C_i$ of clients with GHZ state $\ket{{\rm GHZ}_{d^{(i)}_{{\bm k}_l}}}^{C_i}$ (or private state $\hat{\gamma}_{d^{(i)}_{{\bm k}_l }}^{C_i}$) with probability $p_{{\bm k}_l}$ and error $\epsilon>0$, 
by using a quantum broadcast network associated with a directed hypergraph $G=(V,E)$, follows
\begin{align}
\sum_{i} n^{C_i|P} \langle  \log_2 d^{(i)}_{{\bm k}_l} \rangle_{{\bm k}_l} \le & \frac{1}{g_{E_{\rm sq}}(\epsilon)}  \sum_{e \in E\setminus E_{\rm tri}^P} \bar{l}^e E_{\rm sq}^{P}({\cal N}^e) \nonumber \\ 
&+ \frac{f_{E_{\rm sq}}(\epsilon)}{g_{E_{\rm sq}} (\epsilon)} \label{eq:upper3}
\end{align}
for any partition $P=P_1:P_2:\cdots:P_k$ of the set $V$, where $\langle x_{{\bm k}} \rangle_{{\bm k}} :=\sum_{{\bm k}}p_{{\bm k}} x_{{\bm k}}$, $E_{\rm tri}^{P}(\subset E)$ is the set of hyperedges whose tail $t(e)$ and heads $h(e)$ all belong to one set of $P$, $n^{C_i|P}=0$ for the case of $|\{l \in \{1,\ldots,k \} : P_l \cap C_i \neq \emptyset \} |<2 $ and 
$n^{C_i|P}=|\{l \in \{1,2,\ldots,k \} : P_l \cap C_i \neq \emptyset \} |$ for the case of $|\{l \in \{1,2,\ldots,k \} : P_l \cap C_i \neq \emptyset \} | \ge 2 $. 
The multipartite squashed entanglement of broadcast channel ${\cal N}^e$ is defined by 
$
E_{\rm sq}^{P}({\cal N}^e):=\max_{\ket{\psi}^{t(e)}} E_{\rm sq}^{{P}_1 \cap (t(e)h(e) ) :\cdots:{P}_k \cap (t(e)h(e))  } ({\cal N}^e(\ket{\psi}\bra{\psi}^{t(e)}))
$, where if $P_j \cap (t(e)h(e))= \emptyset$, we strip $P_j \cap (t(e)h(e))$ from the partition of $E_{\rm sq}^{{P}_1 \cap (t(e)h(e) ) :\cdots:{P}_k \cap (t(e)h(e))  }$ in the right-hand side, and
$\ket{\psi}^{t(e)}$ is an arbitrary pure state which can be prepared at vertex $t(e)$. 

Capacities in this scenario are defined, similarly to Eqs.~(\ref{eq:single-capa-1}), (\ref{eq:single-capa-2}), and
(\ref{eq:c1})-(\ref{eq:c4}). In particular, we just need to regard $\langle \log_2 d_{{\bm k}_l}^{(i)} \rangle_{{\bm k}_l}$ corresponding to the size of $\ket{\Phi_{d^{(i)}}}^{A_iB_i}$  or $\hat{\gamma}_{d^{(i)}}^{A_iB_i}$ 
in the definitions of Eqs.~(\ref{eq:single-capa-1}), (\ref{eq:single-capa-2}), and
(\ref{eq:c1})-(\ref{eq:c4}) as one associated with $\ket{{\rm GHZ}_{d^{(i)}}}^{C_i}$  or $\hat{\gamma}_{d^{(i)}}^{C_i}$ in the current scenario. With this rephrasing, we can define various capacities
for the multipartite communication. 

As Eq.~(\ref{eq:upper2}) gives upper bounds on the capacities in the scenarios for multiple-pair communication, Eq.~(\ref{eq:upper3}) provides upper bounds on the capacities in the multipartite communication, through convex optimization (see Section~\ref{se:bound-multiple}). Lower bounds on the capacities can be obtained\cite{BA17} by borrowing the idea of the aggregated quantum repeaters\cite{AK17}, similarly to the scenarios for multiple-pair communication. In this case, by running entanglement generation over every quantum broadcast channel ${\cal N}^e$, the network shares a quantum state $\bigotimes_{e\in E}\hat{\rho}^e$ $\epsilon$-close to $\bigotimes_{e\in E} (\ket{{\rm GHZ}_2}\bra{{\rm GHZ}_2}^e)^{\otimes l^e R_\delta^e }$ associated with an undirected multi-hypergraph $G''=(V,E'')$, where each undirected hyperedge $e''\in E''$ with the same ends of $e$ corresponds to a GHZ state $\ket{{\rm GHZ}_2}^e$ among ends of $e''$. 
Then, if we find a Steiner tree spanning clients $C_i$, which is an acyclic sub-hypergraph connecting all vertices of $C_i$, in the undirected multi-hypergraph $G''$, we can provide a state close to a qubit GHZ state $\ket{{\rm GHZ}_2}^{C_i}$ by performing generalized entanglement swapping\cite{wallnofer20162d} at vertices composing the Steiner tree.
Hence, depending on an objective function in a definition of a capacity, by finding out edge-disjoint Steiner trees spanning clients $C_i$ in the undirected mutli-hypergraph $G''$, we can give a state close to $\hat{\tau}_{\bm d}^C:=\bigotimes_i \ket{{\rm GHZ}_{d^{(i)}}}\bra{{\rm GHZ}_{d^{(i)}}}^{C_i}$. However, this problem of finding the number of edge-disjoint Steiner trees, referred to as Steiner tree packing, is known to be an NP complete problem, even in the case of graph theory\cite{kaski2004packing}. 
Focusing on the scenario to distribute a single GHZ state or a single multipartite private state over a usual quantum network, B\"{a}uml {\it et al.} provide\cite{BAKE20} a computationally efficient method to obtain upper and lower bounds, which are explained in Section~\ref{sec:bounds}.

Recently, an alternative bound on the conference key and GHZ rates is provided by Das {\it et al.}\cite{das2019universal}. In this work the entire network is described by a so-called quantum multiplex channel, i.e. a quantum channel with an arbitrary number of in- and output parties and the protocol consists of many uses of the quantum multiplex channels interleaved by LOCC.

\subsection{Flow problems in optimisation theory}\label{sec:flows}
A number of results provide upper and lower bounds on the rates at which entangled resources for the tasks described in Section \ref{sec:tasks} can be distributed in a quantum network given by an arbitrary graph\cite{P16,p19,P19b,AML16,AK17,BA17,rigovacca2018versatile,BAKE20}. Besides, the bounds are closely related to standard problems in graph theory.
Here we provide some background information on the graph and network theoretic tools, before seeing such a relation in detail (which are reviewed in Section~\ref{sec:bounds}). In particular, we review concepts such as network flows, the max-flow min-cut theorem, multicommodity flows, Steiner trees, as well as the complexity of the various optimisation problems occurring in this context. The discussion in this section applies to general flow networks in an abstract sense. Applications to quantum networks will follow in Section \ref{sec:bounds}. 

In the following we consider an undirected, weighted graph $G'=(V,E')$, i.e. a graph consisting of vertices $v\in V$, undirected edges $\{vw\}\in E'$ where each edge is assigned a weight $c_{\{vw\}}\geq 0$. Note that in graph theoretic literature the weights are also referred to as capacities of the edges. Here we use the term weight when talking about abstract graphs to avoid confusion with information theoretic capacities in Section \ref{sec:bounds}. 

\subsubsection{Single commodity flows}\label{se:single-commodity}

We begin with the simplest scenario of a network flow, in which a single abstract commodity is assumed to flow through a network. For every edge $\{vw\}\in E'$ we can define edge-flows $f_{vw}\geq 0$ and $f_{wv} \geq 0$ such that
\begin{equation}\label{eq:capConstr}
f_{vw}+f_{wv}\leq c_{\{vw\}}.
\end{equation}
The quantities $f_{vw}$ and $f_{wv}$ could be interpreted as the amount of the abstract commodity flowing from vertex $v$ to vertex $w$ and from vertex $w$ to vertex $v$, respectively, while being constraint by a capacity $c_{\{vw\}}$. Note that an alternative definition of flows which is commonly used is based on directed edges each of which is assigned a weight. Whereas the results we are going to present can be formulated in both scenarios, in our case it is convenient to use the version based on undirected graphs. 

Further, we choose two vertices $s,t\in V$, which we call the source and target, respectively. We can define a flow from $s$ to $t$ as 
\begin{equation}\label{eq:stflow}
f^{s\to t}(G'):=\sum_{v:\{vs\}\in E'}\left(f_{sv}-f_{vs}\right),
\end{equation}
while requiring that for every $w\in V$ such that $w \neq s$ and $w \neq t$ it holds
\begin{equation}\label{eq:flowcons}
\sum_{v:\{vw\}\in E'}\left(f_{wv}-f_{vw}\right)=0.
\end{equation}
Eq. (\ref{eq:stflow}) can be regarded as the net flow leaving the source, whereas Eq. (\ref{eq:flowcons}) can be regarded as flow conservation in every vertex except the source and target. By the flow conservation, it holds
\begin{equation}
\sum_{v:\{vs\}\in E'}\left(f_{sv}-f_{vs}\right)=\sum_{v:\{vt\}\in E'}\left(f_{vt}-f_{tv}\right),
\end{equation}
justifying the interpretation as a flow from $s$ to $t$. The obvious task now is to maximise the flow $f^{s\to t}$ defined by Eq.~(\ref{eq:stflow}) with respect to the capacity constraint given by Eq.~(\ref{eq:capConstr}) and the flow conservation constraint given by Eq.~(\ref{eq:flowcons}), which is a linear program (LP). Namely

\begin{align}\label{LP1}
f^{s\to t}_{\max}(G')=\textrm{max}& \sum_{v:\{vs\}\in E'} (f_{sv}-f_{vs})\\
&\forall \{vw\}\in E':\ f_{wv}+f_{vw}\leq c_{\{wv\}}\nonumber\\
&\forall w\in V: w\neq s,t,\ \sum_{v:\{vw\}\in E'} (f_{wv} -f_{vw})=0, \nonumber
\end{align}
where the maximisation is over $2|E'|$ edge flows $f_{vw}\geq0$ and $f_{wv}\geq0$. The LP given by Eq.~(\ref{LP1}) further has $|E'|$ inequality constraints and $|V|-2$ equality constraints. Using $|E'|$ slack variables the $|E'|$ inequality constraints can be transformed into equality constraints, resulting in an LP in a standard form with $N=3|E'|$ nonnegative variables and $M=|E'|+|V|-2$ equality constraints. Using interior point methods \cite{ye1991n3l}, such an LP can be computed using $\mathcal{O}(\sqrt{N}L)$ iterations and $\mathcal{O}(N^3L)$ arithmetic operations, where $L$ scales as $\mathcal{O}(MN+M+N)$ \cite{wright1997primal}.

Next, let us introduce the concept of cuts. Given a subset $W\subset V$ of vertices, we define a $W$-cut as the set of edges
\begin{equation}\label{def:cut}
\partial (W)=\left\{\{wv\}\in E':w\in W, v\in V\setminus W\right\},
\end{equation}
i.e. as a set of edges, the removal of which disconnects $W$ and $V\setminus W$. In the case where $s\in W$ and $t\in V\setminus W$, we call Eq.~(\ref{def:cut}) an $st$-cut and use the notation $\partial (W_{s:t})$. Further, we define the weight of a cut as
\begin{equation}
c_{\partial (W)}=\sum_{\{vw\}\in\partial (W)}c_{\{vw\}}.
\end{equation}
For a given source and target, we can now define the minimum $st$-cut as
\begin{equation}
c^{s:t}_{\min}(G'):=\min_{W_{s:t}}c_{\partial (W_{s:t})}.
\end{equation}
We are now ready to introduce the max-flow min-cut theorem \cite{ford1956maximal,EFS96}, which states that for all $s,t\in V$ it holds
\begin{equation}\label{theo:maxFlowMinCut}
f^{s\to t}_{\max}(G')=c^{s:t}_{\min}(G').
\end{equation}
Let us note that so far we have only made the assumption that all edge flows are nonnegative real numbers. Depending on the network scenario considered, however, it might be necessary to require the values $f_{vw}$ and $f_{wv}$ to be integers for all $\{vw\}\in E'$. In that case, the weights $c_{\{vw\}}$ can, without loss of generality, be taken to be integers as well. A graph with integer weights can equivalently be written as a multigraph, where each edge $\{vw\}$ is replaced by $c_{\{vw\}}$ parallel edges, each with weight equal to one. The task of finding the maximum integer flow from $s$ to $t$ is then equivalent to finding the number of edge disjoint paths from $s$ to $t$. It has been shown by Menger \cite{menger1927allgemeinen}, that the number of edge disjoint paths is equal to the minimum $st$-cut. Finally, let us note that adding an integer constraint turns the optimisation problem given by Eq.~(\ref{LP1}) into an integer linear program. Integer Linear programming has been shown to be an NP-complete problem \cite{karp1972reducibility}. 

\subsubsection{Multi commodity flows}\label{sec:multiflows}
The theory of network flows can be generalised to scenarios with $k$ source-target pairs $(s_1,t_1),...,(s_k,t_k)$, and $k$ different commodities flowing through a network concurrently. To describe such a scenario, for every edge, we define $2k$ edge flows $f^{(i)}_{vw}\geq 0$ and $f^{(i)}_{wv} \geq 0$ for $i=1,...,k$. We then require that 
\begin{equation}\label{eq:MultiCapConstr}
\sum_{i=1}^k\left(f^{(i)}_{vw}+f^{(i)}_{wv}\right)\leq c_{\{vw\}}.
\end{equation}
We can now define the flow of commodity $i$ from $s_i$ to $t_i$ as
\begin{equation}
f^{s_i\to t_i}(G'):=\sum_{v:\{vs_i\}\in E'}\left(f^{(i)}_{s_iv}-f^{(i)}_{vs_i}\right).
\end{equation}
Further we require that every commodity $i$ is conserved in every vertex except its source $s_i$ and target $t_i$, i.e. for all $w\in V$ such that $w \neq s_i$ and $w \neq t_i$ it holds
\begin{equation}\label{eq:fc1}\
\sum_{v:\{vw\}\in E'}\left(f^{(i)}_{wv}-f^{(i)}_{vw}\right)=0.
\end{equation}
When maximising the flows we now have a number of possibilities for figures of merit. Here we focus on two commonly used figures of merit. The first option is to maximise the sum over the $k$ flows, resulting in the maximum total concurrent flow
\begin{equation}\label{eq:total}
f^\text{total}_{\max}(G')=\max\sum_{i=1}^kf^{s_i\to t_i}(G'),
\end{equation}
where the maximisation is under the constraints that Eq.~(\ref{eq:MultiCapConstr}) holds for all edges $\{vw\}\in E'$ and Eq.~(\ref{eq:fc1}) holds for all commodities $i\in\{1,...,k\}$ and for all vertices $w\in V$ such that $w \neq s_i$ and $w \neq t_i$. Eq.~(\ref{eq:total}) is a linear program, which in a standard form has $N=(2k+1)|E'|$  nonnegative variables and $M=|E'|+k(|V|-2)$ equality constraints, and hence scales polynomially in the network parameters.

Whereas there might be situations where the total concurrent flow is a suitable figure of merit, we note that, depending on the network structure and weights, Eq.~(\ref{eq:total}) could be maximised by a flow instance where the $f^{s_i\to t_i}$ vary greatly. The second figure of merit we consider here is fairer in the sense that it maximises the least flow which can be achieved for any of the commodities concurrently. Namely, we have the worst case concurrent flow 
\begin{equation}\label{eq:wc}
f^\text{worst}_{\max}(G')=\max\min_{i\in\{1,...,k\}}f^{s_i\to t_i}(G'),
\end{equation}
where again the maximisation is under the constraints that Eq.~(\ref{eq:MultiCapConstr}) holds for all edges $\{vw\}\in E'$ and Eq.~(\ref{eq:fc1}) holds for all commodities $i\in\{1,...,k\}$ and for all vertices $w\in V$ such that $w \neq s_i$ and $w \neq t_i$. Adding a slack variable $f^\text{worst}\geq0$, Eq.~(\ref{eq:wc}) can be reformulated as the maximisation of $f^\text{worst}$ such that, in addition to the above constraints, for all $i\in\{1,...,k\}$ it holds $f^\text{worst}\leq f^{s_i\to t_i}$, which is a linear program. This in a standard form has $N=(2k+1)|E'|+1+k$ nonnegative variables and $M=|E'|+k(|V|-2)+k$ equality constraints, again scaling polynomially in the network parameters.

Let us note that both multicommodity flow maximisations, given by Eqs.~(\ref{eq:total}) and (\ref{eq:wc}) reduce to the flow maximisation given by Eq.~(\ref{LP1}) in the case of a single commodity, i.e. $k=1$. An obvious question arising now is whether the max-flow min-cut theorem, Eq.~(\ref{theo:maxFlowMinCut}) can be extended to the case of multicommodity flows. Again, there are different quantities to consider. The first way to generalise the minimum $st$-cut to multiple source-target pairs is to consider multicuts. For $k$ commodities, $(s_1:t_1),...,(s_k:t_k)$-multicut is a set of edges the removal of which disconnects $s_i$ from $t_i$ for all $i\in\{1,...,k\}$. As in the case of cuts, the weight of a multicut is defined as the sum of weights of the edges it contains. Minimising over all multicuts, we obtain the minimum multicut $c^\text{multicut}_{\min}$. In the case where $k=1$, the problem of computing the minimum multicut reduces to the minimum cut, and hence it can be solved in polynomial time via the max-flow min-cut theorem. Surprisingly, even for $k=2$, the problem can be reduced to a single-commodity flow \cite{hu1963multi} and can be computed in polynomial time. However, in general, the problem of finding the minimum multicut is known to be NP-hard \cite{garg1997primal}. 

A second way in which the minimum $st$-cut can be generalised to multiple source-target pairs is the min-cut ratio, which is defined as
\begin{equation}
c^\text{cut ratio}_{\min}(G')=\min_{W\subset V}\frac{\sum_{\{vw\}\in\partial (W)}c_{\{vw\}}}{d(\partial (W))},
\end{equation}
where the minimisation is over arbitrary subsets $W\subset V$ and $d(\partial (W))$ is the number of source-target pairs separated by $\partial (W)$. Computation of the min-cut ratio has been shown to be NP-hard in general, as well \cite{aumann1998log}.

Whereas it is clear from a complexity point-of-view that, for general $k$, there cannot be an equality relation between $c^\text{multicut}_{\min}$ or $c^\text{cut ratio}_{\min}$ to a multi-commodity flow instance that can be computed in polynomial time, there are relations up to a factor of $\mathcal{O}(\log k)$. Namely, it has been shown by Garg {\it et al.}\cite{garg1996approximate} that
\begin{equation}\label{theo:Garg}
f^\text{total}_{\max}(G')\leq c^\text{multicut}_{\min}(G') \leq g_1(k) f^\text{total}_{\max},
\end{equation}
where $g_1(k)$ is of $\mathcal{O}(\log k)$. Similarly, it has shown by Aumann {\it et al.}\cite{aumann1998log} that
\begin{equation}\label{theo:Auman}
f^\text{worst}_{\max}(G')\leq c^\text{cut ratio}_{\min}(G') \leq g_2(k) f^\text{worst}_{\max}(G'),
\end{equation}
where $ g_2(k)$ is of $\mathcal{O}(\log k)$. 

\subsubsection{Steiner trees}
When considering scenarios involving groups consisting of more than two network users rather than pairs of users, we need to generalise the concept of a flow from a source to a target or, in the discrete case, the concept of edge-disjoint paths in a unit-capacity multigraph to a multipartite setting. In the discrete case, this leads to the concept of Steiner trees. Namely, given a unit-capacity multigraph $G''=(V,E'')$ and a set $S\subset V$ of vertices, a Steiner tree spanning $S$, or in short $S$-tree, is a connected, acyclic subgraph of $G''$ connecting all $s_i\in S$. By definition, in the $S$-tree, any pair of vertices $s_i,s_j\in S$ with $i\neq j$ is connected by exactly one path. In the special case where $S=V$, an $S$-tree is also a spanning tree. The relevant task now is to find the number of edge disjoint $S$-trees in $G''$. This problem, which is known as Steiner tree packing, is NP-complete in general \cite{cheriyan2006hardness}. Note however that in the case where $|S|=2$, the problem reduces to finding the number of edge-disjoint paths, which by Menger's theorem is equal to the min-cut.

In the general case, it is possible to upper bound the number of edge-disjoint $S$-trees by the $S$-connectivity of the graph $G''$: As in an $S$-tree any pair $s_i,s_j\in S$ with $i\neq j$ of vertices has to be connected by a path, the number $t_S(G'')$ of edge-disjoint $S$-trees in $G''$ cannot exceed the number of edge-disjoint paths between any pair $s_i,s_j\in S$ with $i\neq j$ in $G''$. Minimised over all $s_i,s_j\in S$ with $i\neq j$, this is known as the $S$-connectivity $\lambda_S(G'')$ of the graph. Application of Menger's theorem to each pair $s_i,s_j\in S$ with $i\neq j$ individually shows that $\lambda_S(G'')$ is equal to the minimum $S$-cut, which is defined as
\begin{equation}\label{eq:Scut}
\lambda_S(G'')=\min_{s_i,s_j\in S,i\neq j}c^{s_i:s_j}_{\min}(G'').
\end{equation}

Further, there have been a number of results \cite{kriesell2003edge,lau2004approximate,petingi2009packing} providing lower bounds on the number of edge disjoint $S$-trees $t_S(G'')$, which are of the form
\begin{equation}\label{eq:TreeConnect}
t_S(G'')\geq\lfloor g_3\lambda_S(G'')\rfloor-g_4.
\end{equation}
Kriesell \cite{kriesell2003edge} conjectured that $g_3=\frac{1}{2}$ and $g_4=0$. Lau \cite{lau2004approximate} has shown that Eq.~(\ref{eq:TreeConnect}) holds for $g_3=\frac{1}{26}$ and $g_4=0$. Finally, Petingi {\it et al.}\cite{petingi2009packing} show the relation for $g_3=\frac{|V\setminus S|}{2}$ and $g_4=1$.

\subsection{Efficient bounds for communication tasks}\label{sec:bounds}
As was shown by B{\"a}uml {\it et al.}\cite{BAKE20}, a combination of the results presented in Sections \ref{sec:tasks} and \ref{sec:flows} provides efficiently computable upper and lower bounds on the rates at which various resource states can be distributed in a network. The upper and lower bounds are in the form of single- and multi-commodity flow optimisations in an undirected version of the original network graph. The weights of the corresponding edges are in terms of the quantum capacities ${\cal Q}$ (for lower bounds) and the entanglement ${\cal E}$ of the channels (for upper bounds). Thus, given one knows these quantities for every channel constituting the network, one can obtain bounds on the network capacities by simply running linear programs. 

Given a quantum network described by a directed graph $G=(V,E)$ as defined in Section \ref{sec:tasks}, we define an undirected graph $G'=(V,E')$ as follows:
if there is only a single directed edge connecting two vertices $v,w \in V$, we regard the edge as an undirected edge of the graph $G'$ with the same two ends; if there exist two edges $vw\in E$ and $wv\in E$ for two vertices $v,w\in V$, we only add one undirected edge $\{v,w\}=\{w,v\}$ to $E'$. The reason we switch from the directed graphs to undirected ones here is that, whereas the original quantum channels $\mathcal{N}^e$ for $e\in E$ are directed, in protocols such as the aggregated repeater one presented in Section \ref{sec:tasks}, those channels are used to distribute Bell pairs, which are intrinsically undirected under unlimited LOCC, in the beginning. The remainder of the protocol is then concerned with transforming the resulting Bell-state network into the desired target state. 

Next, each edge $\{vw\}\in E'$ of $G'$ can be equipped with two weights $ {c}_{\mathcal{E}}(\{vw\})$ and $ {c}_{\mathcal{Q}}(\{vw\})$ to be used in the flow maximisations for the upper and lower bounds, respectively. For protocols in $\Lambda(n,\{q^e\}_{e\in E},\epsilon)$, we define 
\begin{align}
 {c}_{\mathcal{E}}(\{vw\})&=q^{vw}\mathcal{E}(\mathcal{N}^{vw})+q^{wv}\mathcal{E}(\mathcal{N}^{wv}),\label{c:upper}\\
 {c}_{\mathcal{Q}}(\{vw\})&=q^{vw}\mathcal{Q}(\mathcal{N}^{vw})+q^{wv}\mathcal{Q}(\mathcal{N}^{wv}),\label{c:lower}\
\end{align}
where we set $\mathcal{E}(\mathcal{N}^{vw})=\mathcal{Q}(\mathcal{N}^{vw})=0$ whenever $vw\notin E$.

\subsubsection{Bipartite communication}
We begin by considering the bipartite case where two vertices, $A$ and $B$, wish to obtain Bell states or bipartite private states. The corresponding capacities are given by Eqs.~(\ref{eq:single-capa-1}) and (\ref{eq:single-capa-2}). Looking at Eq.~(\ref{eq:upper2'}), we now see that the upper bound presented there is indeed equal to the min-cut in the undirected graph $G'$ with weights $ {c}_{\mathcal{E}}(\{vw\})$ for all $\{vw\}\in E'$. By the max-flow min-cut theorem, Eq.~(\ref{theo:maxFlowMinCut}), it can be expressed as a flow maximisation of the form Eq.~(\ref{LP1}), which can be computed by linear programming. As the constraints in the maximisation over usage frequencies $q^e$ in Eq.~(\ref{eq:upper3'}) are linear as well, we can now formulate the entire upper bound given by Eq.~(\ref{eq:upper3'}) as a linear program. 

Similarly, the lower bound on the capacity provided by Eq.~(\ref{eq:lower2}) can be transformed into a flow maximisation in the undirected graph $G'$ with weights $ {c}_{\mathcal{Q}}(\{vw\})$, and hence the optimisation in Eq.~(\ref{eq:lower2}) becomes a linear program. In summary, we have \cite{BAKE20}
\begin{align}
\max_{q^e}  {f}^{A \to B}_{\mathcal{Q}\max}(G')& \le {\cal Q}(G,\{{\cal N}^e\}_{e\in E})\nonumber\\
& \le  {\cal P}(G,\{{\cal N}^e\}_{e\in E})\le \max_{q^e}  {f}^{A \to B}_{\mathcal{E}\max}(G'),
\end{align}
where we have defined $ {f}^{A \to B}_{\mathcal{Q}\max}(G')$ and $ {f}^{A \to B}_{\mathcal{E}\max}(G')$ as maximum flow from $A$ to $B$ in $G'$ with edge weights given by Eqs.~(\ref{c:lower}) and (\ref{c:upper}), respectively. Further, the maximisation is over $q^e \ge 0$ such that $\sum_{e\in E} q^e=1$ on both sides. Both the upper and lower bounds are linear programs of the form (\ref{LP1}), with additional optimisation over usage frequencies, which adds $|E|$ non-negative variables and one equality constraint, and hence still computable in polynomial time. 

The reason we can drop the integer constraint in the lower bound is that we are looking at the asymptotic limit \cite{BAKE20}. Roughly speaking, assuming that the $ {f}^{A \to B}_{\mathcal{Q}\max}(G')$ is maximised by non-integer edge flows $f_{vw}$, we can always find a large enough number $N$ such that $Nf_{vw}$ are close enough to an integer for all edges. In particular, by using the corresponding channel a large enough number of times, we can obtain $\lfloor Nf_{vw}\rfloor$ Bell pairs for all edges $\{vw\}\in E'$, which can then be connected by entanglement swapping along paths linking Alice and Bob.

\subsubsection{Multi-pair communication}
Let us consider the case of multiple user pairs $(A_1,B_1),...,(A_k,B_k)$ wishing to obtain bipartite resources such as Bell pairs or private states between them concurrently. In Section \ref{sec:tasks} it was shown that there are a number of different ways to define the corresponding capacities. One is to maximise the sum of the rates achievable by the user pairs. The corresponding quantum and private multi-pair network capacities $\cal{Q}^{\text{total}}$ and $\cal{P}^{\text{total}}$, respectively, are defined in Eq.~(\ref{eq:c4}) with $s_i=1$ for all $i$. A second way is to maximise the worst case rate achievable between all user pairs, with corresponding network capacities $\cal{Q}^{\text{worst}}$ and $\cal{P}^{\text{worst}}$, defined in Eq.~(\ref{eq:c2}).

It follows as a special case from the results of B{\"a}uml and Azuma\cite{BA17} that the total private multi-pair capacity $\cal{P}^{\text{total}}$ is upper bounded by the minimum multicut $ {c}^\text{multicut}_{E_\text{sq}\min}(G')$. Similarly, it has been shown \cite{BA17,BAKE20} that the worst-case private multi-pair capacity $\cal{P}^{\text{total}}$ is upper bounded by the minimum cut ratio $ {c}^\text{cut ratio}_{\mathcal{E}\min}(G')$ with respect to weights defined by Eq.~(\ref{c:upper}) for all entanglement measures ${\cal E}$ holding the bound (\ref{eq:upper1}). Applying the generalisations of the max-flow min-cut theorem to multi-commodity flows, given by Eqs.~(\ref{theo:Garg}) and (\ref{theo:Auman}), respectively, we can upper bound both private multi-pair capacities in terms of multi-commodity flow optimisations. 

Using a generalisation of the aggregated repeater protocol\cite{AK17}, lower bounds on the multi-pair quantum capacities $\cal{Q}^{\text{total}}$ and $\cal{Q}^{\text{worst}}$ can be achieved by the corresponding multi-commodity flow maximisations $ {f}^\text{total}_{\mathcal{Q}\max}(G')$ and $ {f}^\text{worst}_{\mathcal{Q}\max}(G')$ with respect to weights defined by Eq.~(\ref{c:lower}) \cite{BAKE20}. Again we note that the integer constraint in the flow optimisation can be dropped in the asymptotic case. In summary, it holds for the total multi-pair network capacities
\begin{multline}
\max_{q^e} {f}^\text{total}_{\mathcal{Q}\max}(G')\leq\mathcal{Q}^{\text{total}}(G,\{\mathcal{ N}^e\}_{e\in E})\\
\leq\mathcal{P}^{\text{total}}(G,\{\mathcal{ N}^e\}_{e\in E})\leq g_1( k)\max_{q^e} {f}^\text{total}_{E_\text{sq}\max}(G'),
\end{multline}
where the maximisation is over all $q^e\geq0$ with $\sum_eq^e=1$. Both upper and lower bounds are linear programs. Similarly, for the worst-case multi-pair network capacities can be upper and lower bounded by polynomial-size linear programs, as
\begin{multline}
\max_{q^e} {f}^\text{worst}_{\mathcal{Q}\max}(G')\leq\mathcal{Q}^{\text{worst}}(G,\{\mathcal{ N}^e\}_{e\in E})\\
\leq\mathcal{P}^{\text{worst}}(G,\{\mathcal{ N}^e\}_{e\in E})\leq g_2( k)\max_{q^e} {f}^\text{worst}_{\mathcal{E}\max}(G'),
\end{multline}
where $\mathcal{E}$ can be any entanglement measure for which the bound (\ref{eq:upper1}) holds. Again, the upper and lower bounds are polynomial-size linear programs. As we discuss in Section \ref{sec:multiflows}, the gaps $g_1(k)$ and $g_2(k)$, both of which are of order $\mathcal{O}(\log k)$, leave the possibility for more efficient protocols, e.g., quantum network coding. 

Let us also note that the multi-commodity flow based approach to obtain achievable rates in multi-pair communication can be extended to realistic scenarios with imperfect swapping operations and without entanglement distillation. In such a case, the fidelity achievable between the user pairs drops with the number of swapping operations. Chakraborty {\it et al.}\cite{chakraborty2020entanglement} overcome this problem by adding an additional constraint on the lengths of paths to the optimisation of the total multi-commodity flow. They provide a linear program scaling polynomially with the graph parameters, which provides a maximal solution if it exists.

\subsubsection{Multipartite communication}
Finally we consider the case where a set of users $\mathcal{A}=\{A_1,...,A_k\}$ wishes to establish a $k$-partite GHZ or private state. It is straightforward to generalise the bipartite quantum and private network capacities given by Eq.~(\ref{eq:single-capa-2}) to the multipartite case. We denote the corresponding multipartite quantum and private network capacities $\mathcal{Q}^{\mathcal{A}}(G,\{\mathcal{ N}^e\}_{e\in E})$ and $\mathcal{P}^{\mathcal{A}}(G,\{\mathcal{ N}^e\}_{e\in E})$, respectively. From the results of B{\"a}uml and Azuma\cite{BA17} it follows that the multipartite private network capacity is upper bounded by the minimum $\mathcal{A}$-cut in $G'$ with weights $c_{E_\text{sq}}(\{vw\})$:
\begin{equation}
\mathcal{P}^{\mathcal{A}}(G,\{\mathcal{ N}^e\}_{e\in E})\leq\max_{q^e}\min_{A_i,A_j\in \mathcal{A},i\neq j}c_{E_\text{sq}\min}^{A_i:A_j}(G').
\end{equation}
By application of the max-flow min-cut theorem (for every pair $A_i,A_j\in \mathcal{A}$ with $i\neq j$), as well as introduction of a slack variable, the right-hand side can be expressed as a linear program \cite{BAKE20}. Namely we maximise $f\geq0$ such that for all pairs $A_i,A_j\in \mathcal{A}$, $i\neq j$, $f\leq f^{A_i\to A_j}$ and requiring that the capacity and flow conservation constraints, given by Eq.~(\ref{eq:capConstr}) and (\ref{eq:flowcons}), respectively, are fulfilled for every pair.

In order to lower bound $\mathcal{Q}^{\mathcal{A}}(G,\{\mathcal{ N}^e\}_{e\in E})$, we have to generalise the aggregated repeater protocol. The first part of the protocol, in which Bell states are distributed resulting in a Bell network described by a multigraph $G''$, remains the same. The second part, however, has to be generalised to the distribution of GHZ states among the parties in $\mathcal{A}$. Namely, instead of connecting paths of Bell pairs by entanglement swapping, it is possible to connect $\mathcal{A}$-trees of Bell pairs by means of a generalised entanglement swapping protocol \cite{wallnofer20162d}, resulting in GHZ states among $\mathcal{A}$\cite{BA17,yamasaki2017graph,BAKE20}. Thus the relevant figure of merit is the number of edge disjoint $\mathcal{A}$-trees in $G''$. 

Whereas the number of edge disjoint $\mathcal{A}$-trees is an NP-hard problem, we can make use of Eq.~(\ref{eq:TreeConnect}) and Menger's theorem to relax the problem an integer-flow optimisation, which in the asymptotic limit can be further relaxed to a flow optimisation problem of the form $\min_{A_i,A_j\in \mathcal{A},i\neq j}f^{A_i\to A_j}_{\mathcal{Q}\max}$, which is a linear program. In summary we have the following bounds:
\begin{align}
&\frac{1}{2}\max_{q^e} \min_{A_i,A_j\in \mathcal{A},i\neq j}f^{A_i\to A_j}_{\mathcal{Q}\max}(G')\leq\mathcal{Q}^{\mathcal{A}}(G,\{\mathcal{ N}^e\}_{e\in E})\nonumber\\
&\leq\mathcal{P}^{\mathcal{A}}(G,\{\mathcal{ N}^e\}_{e\in E})\leq \max_{q^e}\min_{A_i,A_j\in \mathcal{A},i\neq j}f^{A_i\to A_j}_{E_\text{sq}\max}(G'),
\end{align}
where we have set $g_3=\frac{1}{2}$. 
Both the upper and lower bounds are linear programs with $2 \binom{ |{\cal A}|}{2} |E'| +|E|+1$ variables and 
$\binom{ |{\cal A}|}{2} (|E'| +1)+|E|$ inequality constraints, i.e., of polynomial size.

\subsection{Beyond classical routing in quantum communication networks}\label{sec:beyond}

So far we have focused on the distribution of Bell, GHZ and private states by means of protocols based on the aggregated repeater protocol, which basically is a classical routing protocol applied on quantum networks. Other quantum network protocols based on classical routing have been introduced, among others, by Pant {\it et al.}\cite{pant2019routing}, Schoute {\it et al.}\cite{schoute2016shortcuts} and Chakraborty {\it et al.}\cite{chakraborty2019distributed}. We will now give a brief, and by no means exhaustive, overview of other quantum network protocols, which go beyond the classical routing paradigm.

\subsubsection{Quantum network coding}
The lower bounds presented in Section \ref{sec:bounds} are based on routing protocols, where Bell pairs are established between nodes connected by channels and then connected by means of entanglement swapping along paths connecting the users \cite{AK17}. Whereas, for a large class of quantum channels, in the single pair case, the upper bounds can be achieved in this way, the situation becomes more involved in the multi-pair case, where we have seen gaps between upper and lower bounds. Such gaps leave room for improvement in terms of more efficient protocols such as a quantum version of network coding\cite{ahlswede2000network}.

In classical network theory, network coding protocols differ from routing protocols in that bits of information are being acted upon by some operation before being sent via a channel, allowing the combination of several bits of information into one. Whereas routing is optimal in single-pair scenarios, coding can provide an advantage in multi-pair problems in directed networks, as can be seen in the famous butterfly graph. In undirected networks, the question of a coding advantage in the multiple unicast setting is still an open question \cite{li2004network}. In the case of directed classical multicast settings, where messages are sent from a source node to several destination nodes, it has been shown that coding can always achieve the capacity, which is equal to the so-called cutset bound, whereas routing cannot, which can, again, be demonstrated using a butterfly graph \cite{el2011network}.

Returning to the setting of quantum networks, a quantum version of classical network coding has been shown to outperform the simple routing approach presented above \cite{kobayashi2010perfect,kobayashi2011constructing}. Namely, assuming we have a butterfly network where each link consists of a single Bell pair, concurrent quantum routing between the two diagonal node pairs is not possible as any path connecting the first pair of diagonal nodes disconnects the second pair when being used up. However, using the Bell states as identity channels via teleportation, we can apply the network coding protocol presented by Kobayashi {\it et al.}\cite{kobayashi2010perfect,kobayashi2011constructing}, to distribute Bell pairs concurrently between both user pairs. The protocol mainly consists of creation of the state $|+\rangle=(|0\rangle+|1\rangle)/\sqrt2$ at each node followed by a translation of a classical network coding protocol into sequences of Clifford operations. The resulting encoded states are transmitted via the identity channels. A similar advantage in a butterfly network over the Steiner tree routing approach can be obtained when distributing GHZ states using quantum routing \cite{epping2017multi}.

Going to the asymptotic setting, on the other hand, routing in combination with \emph{time-sharing}, i.e. the distribution of entanglement between one pair in one round and another pair in another round and so on, is sufficient to achieve the cut based upper bounds in the multi-pair scenario\cite{leung2010quantum}. Note that there is a tradeoff between the increased need of quantum memory in time-sharing and the need to perform more complex operations in quantum network coding.

 \subsubsection{Graph state protocols}\label{sec:graph}
The aggregated repeater protocol discussed above consists of an initial stage distributing Bell states across all edges of the graph, resulting in a Bell state network involving all connected vertices in the graph. This Bell state network is then transformed into the desired target state among a single or multiple pairs or groups of users. Thus, the Bell state network can be seen as a universal resource state for a number of network tasks. A different strategy is to create a large multipartite entangled graph state among all, or a large subset of, the vertices, which serves as a universal resource state that can subsequently be transformed into the desired target states. In this class of protocols there are two tasks to be considered. The first is to use a network of quantum channels and free operations (such as LOCC) to create the graph state that serves as resource. The second is to transform the graph state into the desired target states by means of free operations. 

The first task has been considered by Epping {\it et al.}\cite{epping2016large}, where the goal is a multipartite entangled graph state among all vertices in the network, except the ones that merely serve as repeater stations to bridge long distances. This is achieved by protocol involving local creation of qubits in $\ket{+}$-states, controlled-$Z$ operations entangling two qubits, Pauli-$X$ measurements and the sending of qubits via channels, corresponding to directed edges. In a first step all vertices are connected to form a graph state, which is done as follows: In vertices with only $n$ outgoing edges, $n+1$ $\ket{+}$-states are created, all of which get entangled by means of controlled-$Z$ operations and $n$ qubits are sent via the outgoing edges. In vertices with incoming and outgoing edges, all incoming qubits get entangled among each other as well as with locally created qubits in $\ket{+}$-states which are then sent via the outgoing edges. In a second step the repeater stations are being removed from the graph state by Pauli-$X$ measurements. This protocol is then combined with stabiliser codes to compensate for losses in the transmission channels, noisy gates as well as errors in the preparation and measurement. In subsequent work \cite{epping2016robust} the technique is combined with quantum network coding and generalised to qudit graph states.

The second task has been considered by Hahn {\it et al.} \cite{hahn2019quantum}, which is concerned with the transformation of a graph state into Bell states between a single or multiple pairs of users, GHZ states among groups of users and, most generally, the transformation from one graph state to another graph state. The free operations are restricted to local Clifford operations and Pauli measurements. Whereas it is always possible to isolate the shortest path connecting two users and make the connection by measuring the intermediate vertices, Hahn {\it et al.} provide a protocol that requires the measurement of less nodes. Their approach is based on a graph transformation known as local complementation \cite{bouchet1988graphic} and the fact that a graph state can be transformed into another graph state by local Clifford gates whenever the corresponding graphs can be interconverted by means of local complementations \cite{van2004graphical}. In related work by Dahlberg {\it et al.}\cite{dahlberg2018transforming,dahlberg2018transform}, the question of interconvertability between graph states by means of single-qubit Clifford operations, single-qubit Pauli measurements and classical communication is studied. The authors show that the problem of deciding whether two graph states can be interconverted by this class of free operations is NP-complete and present an algorithm that provides the sequence of operations necessary to perform the task, if possible. Another approach is presented by Pirker {\it et al.}\cite{pirker2018modular}, where tensor products of GHZ states or fully connected decorated graph states are used as resource states, allowing for creation of arbitrary graph states among the network users.

\subsubsection{Coherent control}

The adaptive protocols described in Section \ref{sec:tasks}, while using quantum channels, quantum states, and quantum de- and encoding operations, are classical in the way they choose which channel is used in each round: One could think of the LOCC operations having a classical output register that is used as a  control register determining which channel is used next. There exist, however, scenarios in which the choice of channels is not determined by a classical register, but coherently using a quantum control register. For example, a `quantum switch' has been introduced, where a quantum control register determines the order of two channels ${\cal N}_1$ and ${\cal N}_2$. By entering a state in quantum superposition, say $\ket{+}_c$ into the control register, a superposition of the orders of ${\cal N}_1$ and ${\cal N}_2$ is achieved \cite{chiribella2013quantum}.

Surprisingly, it has been shown that such a superposition of orders can boost the rate of  classical and quantum communication beyond the limits of conventional quantum information theory\cite{ebler2018enhanced,chiribella2018indefinite,chiribella2019quantum}. In particular, it has been shown that two identical copies of a completely depolarising channel can become able to transmit classical information when used in such a superposition of orders. Similar effects were also observed when quantum control is used to create a superposition between single uses of two channels \cite{abbott2018communication}.

Miguel-Ramiro {\it et al.}\cite{miguel2020genuine} have introduced the concept of coherent quantum control into the context of quantum networks, allowing for coherent superpositions of network tasks. In particular they provide protocols where two nodes in a network are connected by a superposition of different paths. Other examples include the distribution of a quantum state to different destinations in superpositions and the distribution of a superposition of GHZ states among different user groups. We note that such protocols are not included in the paradigm of adaptive LOCC protocols used in Section \ref{sec:tasks}.

 \section{Waiting times and fidelity estimation over abstract quantum networks}
\label{sec:analytical}
\newcommand{\pgen}{p_{\rm{g}}}
\newcommand{\qgen}{q_{\rm{g}}}
\newcommand{\pswap}{p_{\rm{s}}}
\newcommand{\pdist}{p_{\rm{d}}}
\newcommand{\tcoh}{t_{\rm{coh}}}
\newcommand{\mean}[1]{\langle #1 \rangle}
\newcommand{\Conv}{\mathop{\scalebox{3.0}{\raisebox{-0.2ex}{$*$}}}}

In this section, we review analytical tools for characterizing the performance of quantum networks and the algorithms that immediately follow the analytical expressions.
In particular, we consider the literature that studies the time it takes to distribute remote entanglement over a quantum network, referred to as the waiting time, and the quality of the entanglement.
Due to their more modest quantum information processing requirements, we devote a large part of the section to quantum repeaters which are built from probabilistic schemes, i.e.\ the so-called first-generation repeater \cite{munro2015inside}. 
As a consequence of the probabilistic nature of such schemes, the waiting time is a random variable; thus, it is not represented by a single number but instead by a probability distribution.
Our presentation focuses on the fidelity with respect to the desired maximally entangled state as a measure of entanglement quality. However, many of the tools can directly be used for estimating other figures of merit such as the secret key rate.

This section is organized as follows.
We start in Section~\ref{sec:abstract models of quantum networks} with the modeling of a quantum network, which includes the mathematical abstraction of different components in a quantum network.
In Section~\ref{sec:analytical study on the rate and fidelity}, we discuss the analytical tools used in evaluating the performance of networks.
In some cases, those analytical tools yield closed-form expressions, of which the evaluation requires the assistance of numerical algorithms.
We discuss three such cases in Section~\ref{sec:numerical study on the rate and fidelity}: Markov chain methods, probability-tracking algorithms, and sampling with Monte Carlo methods.
Finally, we review in Section~\ref{sec:second- and third generation repeater protocols} the analysis of quantum repeater protocols that include quantum error correction. 
We have chosen to limit the scope of this section to discrete variable protocols. 

\subsection{Abstract models of quantum networks}
\label{sec:abstract models of quantum networks}
Here, we summarize common models of different network components, with an emphasis on how they contribute to the statistics of the waiting time and quality of the entangled state.
Throughout the section, we refer to a pair of entangled qubits shared by spatially-separated nodes, as a `link'.

\textit{Entanglement generation}. 
By entanglement generation, we refer to the delivery of a fresh Bell state between two nodes in the network which are directly connected through a communication channel, such as an optical fiber.
We refer to the generated entanglement as an elementary link.
There are several schemes for the generation of elementary links~\cite{munro2015inside}, and in each of them, the generation is performed in discrete attempts until the first successful attempt.
We assume that each attempt is of constant duration $\Delta$ and has constant success probability $\pgen$.
The attempt duration $\Delta$ is determined by the distance and speed of light in the medium; in the rest of this section, we set $\Delta=1$ for simplicity.
It is also commonly assumed that the distinct attempts are independent and thus that the state $\hat{\rho}$ that is produced is constant, i.e. it is independent of the number of attempts required to produce it.
The state $\hat{\rho}$ is a noisy Bell state which typically incorporates different sources of noise, photon loss, and detector inefficiency. 

\textit{Entanglement swapping}. 
The photon transmission probability in fiber decays exponentially and thus fundamentally limits that distance over which elementary link generation can be performed~\cite{takeoka2014fundamental}.
This limit can be overcome by adding additional nodes in between, so-called quantum repeaters, which perform entanglement swaps to connect two short-distance links into a single long-distance one \cite{zukowski1993}.
Typically, entanglement swaps are probabilistic, with a fixed success probability $\pswap$ which is normally independent of the states swapped but depends on the physical implementation.
For matter qubits that can be controlled directly, an entanglement swap can be implemented with deterministic quantum gates, i.e. $\pswap = 1$.
If entanglement swapping is implemented with optical components, the entanglement swapping becomes probabilistic, i.e., $\pswap<1$ and typically $\pswap\le 0.5$ \cite{calsamiglia2001maximum}.
There are also more sophisticated optical swapping schemes with a probability larger than one half \cite{grice2011arbitrarily,olivo2018ancilla,ewert20143}. 
In some models, where the memory decoherence to the vacuum state is considered, the success probability can also be a variable \cite{wu2020nearterm}.

\textit{Entanglement distillation}. 
Performing an entanglement swap on two imperfect Bell states yields long-distance entanglement of lower quality than the original two states individually had, and its quality is further decreased by imperfections in the quantum operations.
In principle, entanglement distillation can be used to improve the fidelity by probabilistically converting multiple low-quality entangled pairs of qubits into a single one of high quality \cite{bennett1996purification}.
In contrast with entanglement swapping, the success probability $\pdist$ depends on the entangled states that are distilled \cite{bennett1996purification,deutsch1996quantum}.

\textit{Entangled state representation}. 
Arguably, the simplest model of the fresh elementary link state is a Werner state \cite{werner1989quantum}, which characterizes the state with a single parameter $w$: 
$$\hat{\rho}=w \ket{\Phi_2} \bra{\Phi_2} +  (1-w) \hat{I} /4\ ,$$ 
where $\ket{\Phi_2}$ is the desired maximally-entangled two-qubit state and $\hat{I} /4$ the maximally mixed state on two qubits.
Although operations such as entanglement distillation do not always output a Werner state, any two-qubit state can be transformed into a Werner state with LOCC without changing the fidelity \cite{duer2007entanglement}.
A more general model is a probabilistic mixture of the four Bell states.
This state is also always achieved \cite{bennett1996mixed} by applying local Pauli gates randomly to an arbitrary state of a pair of qubits, and it is not less entangled than a Werner state. In principle, one could also track the full density matrix, though many studies choose the previous two representations to simplify the analysis.
Given the density matrix $\hat{\rho}$ of a state, its fidelity with a pure target state $\ket{\phi}$ is given by $\bra{\phi}\hat{\rho}\ket{\phi}$.
Throughout the section, the target state will be a Bell state.

\textit{Noise modeling}. 
Imperfections of the quantum devices, for example, operational noise and detector inefficiencies, are commonly modeled by depolarizing, dephasing, or amplitude damping channels.
The first two can be incorporated relatively simply into analytical derivations as they correspond to the random application of Pauli gates.
Amplitude damping requires tracking the full density matrix.
One could, however, replace an amplitude damping channel with the more pessimistic choice of a depolarization channel, which does not change the output state's fidelity with the target state, or alternatively twirl the damped state by applying random Pauli operations \cite{briegel1999quantumrepeaters}.

Particularly relevant in the context of entanglement generation using probabilistic components is the noise caused by time-dependent memory decoherence: in case multiple links are needed, the earliest link is generally generated before the others are ready and thus needs to be stored in a quantum memory. The storage time leads to a decrease in the quality of the entanglement, and the longer the qubit is stored, the more its quality degrades. Due to the interplay between waiting time and time-dependent decay of entanglement quality, memory noise is particularly hard to capture. Sometimes this problem is sidestepped by analyzing protocols with running time qualitatively shorter than the memory decoherence time.

\textit{Node model.}  
For simplicity, the network nodes can be modeled by a fully-connected quantum information processing device capable of generating entanglement in parallel with its neighbors. However, it is important to note that many platforms do not conform to this model. For instance, NV-centers in a single diamond have a single optical interface. Hence, if nodes hold only a single NV center, entanglement generation can only be attempted with one adjacent node at a time. Moreover, the connectivity between the qubits follows a star topology, i.e. direct two-qubit gates between arbitrary qubits are not possible.

\textit{Cut-off}. 
Due to memory decoherence, the quality of the stored entanglement decreases as the waiting time grows.
One common strategy to compensate for memory decoherence is cut-offs: if a link remains idle for too long, it is discarded.
By discarding entanglement whose storage time exceeds some pre-specified threshold, one improves the quality of the delivered entanglement at the cost of longer waiting time.

Additionally, it is possible to build on top of this idea a simplified model of memory decoherence: the quantum information is preserved perfectly for a fixed duration and then lost \cite{collins2007multiplexed,abruzzo2014measurement}.

Note that the inclusion of cut-offs in entanglement distribution schemes complicates their analysis because of the additional effect of waiting time on the state quality.

\begin{figure*}
    \includegraphics[width=\linewidth]{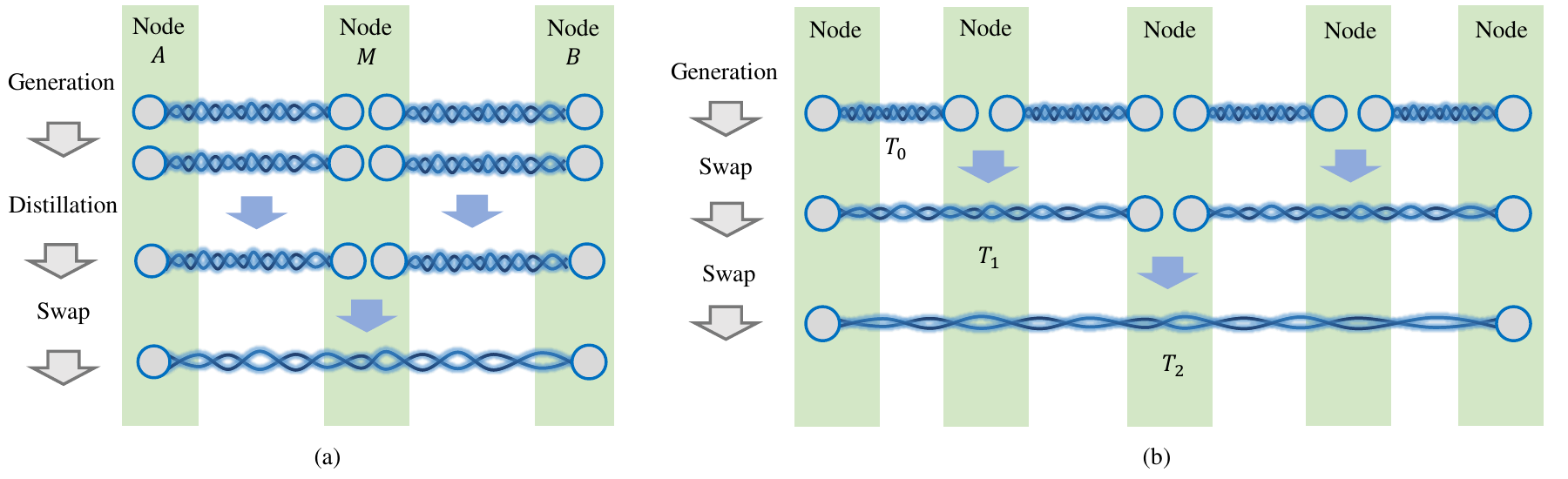}
    \caption{
    \label{fig:bdcz}
    Examples of a nested repeater protocol.
    {\bf (a)} An example with $n=1$ nesting level, on $2^n + 1 = 3$ nodes.
	Nodes $A$ and $M$ generate two entangled pairs in parallel, followed by performing entanglement distillation on the two pairs, and repeat this procedure until the distillation step has succeeded.
	Since the protocol is nested, nodes $M$ and $B$ do the same.
	Once both distillation steps on each side of $M$ succeed, $M$ performs an entanglement swap, which produces entanglement between $A$ and $B$.
	If the entanglement swap fails, then the protocol restarts, i.e. $A$-$M$ and $M$-$B$ start with entanglement generation again.
    {\bf (b)} An example with $n=2$ nesting levels ($5$ nodes) without entanglement distillation.
    At each nesting level, the distance that entanglement spans is doubled.
    By $T_n$, we denote the random variable describing the delivery time of entanglement at level $n$.
    In Section~\ref{sec:analytical}, we consider nested repeater protocols of entanglement generation and swapping such as in (b).
    Whenever the protocol includes entanglement distillation, as in (a), or cut-offs, it is mentioned explicitly.
    }
\end{figure*}

\textit{Nested protocols}. 
One particularly relevant network topology is the repeater chain, where all nodes are arranged in a line. 
Nested protocols offer a structured approach to distributing entanglement across a repeater chain~\cite{briegel1998quantum,duan2001long,kok2003construction,childress2005fault,van2006hybrid,munro2008high,azuma2012quantum,zwerger2012measurement}.
In this section, unless explicitly mentioned, we follow the BDCZ scheme\cite{briegel1998quantum}, with the restriction that each entanglement swap doubles the distance that an entangled pair spans.
In such a scheme, the number of repeater segments is $2^n$ ($2^n + 1$ nodes) where $n$ is the number of nesting levels at which an entanglement swap is performed.
We depict examples of BDCZ protocols in Fig.~\ref{fig:bdcz}.
We denote the waiting time random variable of a repeater scheme on $2^n$ segments as $T_n$.
Many of the tools for determining the waiting time statistics and quality of the produced entanglement discussed below also apply to other schemes than nested repeater protocols.

\subsection{Analytical study of the waiting time and fidelity}
\label{sec:analytical study on the rate and fidelity}

In this section, we present analytical tools to compute the waiting time and the fidelity of the entangled state produced between the end nodes of a repeater chain.

We consider the nested repeater chain protocol on $2^n$ segments (see Section~\ref{sec:abstract models of quantum networks}) with only entanglement swapping, i.e.\ no distillation or cut-offs unless explicitly mentioned. For simplicity, we assume that the generation probability $\pgen$ is the same for each pair of adjacent nodes and the swapping probability $\pswap$ is equal at each nesting level. We also assume that the nodes are capable of generating entanglement in parallel. 
Finally, we ignore the (constant) duration of local operations and classical communication for simplicity, although all of the tools mentioned are capable of incorporating these.

We first investigate methods that instead of tracking the full probability distribution, only track an approximation of the average waiting time and quantum state at each nesting level of the entanglement-distribution protocol.
To demonstrate the tools used in computing the distributions, we include an explicit calculation for a protocol on two nodes with a single repeater positioned in between.
As this exact calculation cannot be directly generalized to a higher nesting level in a nested protocol with more than a single repeater, we then consider the idea of approximating the waiting time by assuming it follows the statistics of elementary-link generation, where the mean waiting time is computed using the approximation from the previous level.
Finally, we review the mathematical tools and approximation methods used to analyze deterministic swapping protocols and distillation-based repeater schemes. 

\subsubsection{The mean-only approach}
\label{sec:mean-only}

In many early analyses of repeater protocols, only the mean waiting time is considered for each nesting level: it is assumed that the entanglement is delivered after a fixed time duration determined by the generation rate.
We refer to it as the mean-only approach.
In this approach, the mean waiting time is computed as the product of the mean waiting time at each nesting level ($1/\pgen$ at the bottom level, $1/\pswap$ at the higher levels), yielding $T_n=1/\pswap^n \pgen$ \cite{briegel1998quantum,duan2001long}. This approach can be refined by noting that at each nesting level the protocol proceeds only when two adjacent pairs are ready. Then, the mean waiting time can be approximated by $T_n=(3/2)^n/\pswap^n \pgen$ \cite{jiang2007fast,childress2005fault,azuma2012quantum,brask2008memory,sangouard2011quantum,loock2019extending,abruzzo2013quantum,asadi2018quantum}.
The factor $3/2$ comes from the fact that in the limit of very small success probability, the waiting time of preparing two links is approximately $3/2$ times that of one link.
We discuss the statistics behind this factor later in Section~\ref{sec:approximation with geometric distribution}.

The mean-only approach is a good approximation when $\pgen$ and $\pswap$ are much smaller than 1\cite{jiang2007fast,brask2008memory}.
However, it only approximates the mean, i.e. it does not provide the entire probability distribution of the waiting time. Hence, it is not suited for investigating time-dependent aspects such as memory noise or cut-offs.
With this method, memory decoherence is either approximated by an inefficiency constant \cite{abruzzo2013quantum} or studied only for the communication time \cite{hartmann2007role}.
To provide a better estimation, one needs to consider the waiting time distribution and the statistics it results in, which we discuss below.

\subsubsection{Single repeater swap protocol}
\label{sec:single repeater swap protocol}

Here, we explicitly compute the probability distribution of the waiting time for the simplest repeater chain: a single repeater between two end nodes.
We also derive an expression for the mean fidelity decay due to memory decoherence.
Many problems regarding single-repeater protocols have an analytical solution because the entanglement generation follows a known distribution.
By studying this simple scenario, we demonstrate the common concepts and methods used to describe and compute the statistics of waiting time and fidelity.
In Sections \ref{sec:approximation with geometric distribution} and \ref{sec:deterministic swap}, we use this calculation as a basis for the analysis of nested repeater protocols of more nodes.

We describe the waiting time of elementary-link generation as a random variable $T_0$, following a geometric distribution given by
\begin{equation}
    \Pr(T_0=t) = p(1-p)^{t-1},
	\label{eq:geometric-distribution}
\end{equation}
where $t \in \{1,2,3...\}$ and $p=\pgen$.
This distribution plays a central role in the statistics of entanglement distribution, as we see in the remaining part of this section.
In the limit of $\pgen \ll 1$, the geometric distribution can be approximated by an exponential distribution, 
\begin{equation}
	\label{eq:exponential-distribution}
\Pr(T_0 = t) = \pgen \cdot \exp(-\pgen t)
\end{equation}
	which is a continuous distribution with $t \geq 0$.
Note that we have set the attempt duration $\Delta$ of entanglement generation to 1 (Section \ref{sec:abstract models of quantum networks}).

To perform an entanglement swap, both elementary links have to be prepared first. We define the time used in preparing them as $M_0$:
\begin{equation}
    \label{eq:max of two T}
    M_0=\max(T_0, T_0'),
\end{equation}
where $T_0'$ is an independent copy of $T_0$.
The distribution of $M_0$ can be computed using the fact that
\begin{equation}
	\label{eq:cdf-max}
	\Pr(\max(X, Y) \leq t) = \Pr(X \leq t) \cdot \Pr(Y \leq t)
\end{equation}
	for any independent random variables $X$ and $Y$.
The mean of $M_0$, i.e.\ the waiting time until both elementary links have been prepared, is given by~\cite{sangouard2011quantum}:
\begin{equation}
    \label{eq:mean waiting two links}
    \mean{M_0} = \frac{3-2\pgen}{(2-\pgen)\pgen}
    .
\end{equation}
After two elementary links are prepared, the repeater node performs an entanglement swap, which is a probabilistic operation with success probability $\pswap$.
The total waiting time for generating the entanglement between two end nodes is therefore
\begin{equation}
    \label{eq:compound sum}
    T_1 = \sum_{k=1}^K M_0^{(k)}
    ,
\end{equation}
where $K$ represents the number of swap attempts until it succeeds and $M_0^{(k)}$ are independent copies of $M_0$.
Eq. (\ref{eq:compound sum}) is referred to as a compound distribution since the number of summands $K$ is also a random variable.
For an entanglement swap, the number of attempts $K$ also follows a geometric distribution (Eq.~\eqref{eq:geometric-distribution}) with success probability $p=\pswap$.
Because $K$ and $M_0$ are independent, the average waiting time is given by
\begin{equation}
    \label{eq:average waiting time t1}
	\mean{T_1} = \mean{M_0} \mean{K} = \frac{3-2\pgen}{(2-\pgen)\pgen} \cdot \frac{1}{\pswap}
    .
\end{equation}
The intuition behind Eq.~\eqref{eq:average waiting time t1} is that, on average, the repeater node requires $\mean{K}$ swap attempts until the first successful swap, and for each swap attempt, $\mean{M_0}$ attempts are needed to prepare the two elementary links.

Computing the fidelity of the two elementary links just before swapping can be done as follows.
If the generation of elementary links is not deterministic, i.e.\ if $\pgen < 1$, the two elementary links are in general not produced at the same time, requiring the earlier of the two to be stored in a quantum memory.
This storage time results in decoherence of the earlier link.
To estimate the fidelity decrease, we need to first compute the distribution of the memory storage time, i.e.\ the time difference between the generation of the earlier and the later link.
We define $\qgen=1-\pgen$.
The probability that one link is prepared $j$ steps before the other is given by \cite{schmidt2019memory,loock2019extending}
\begin{equation}
    \label{eq:memory storage time}
    \Pr(T_0-T_0'=j)
    =
    \sum_{t=1}^{\infty} \pgen^2 \qgen^{2(t-1)+j}
    =
    \frac{\pgen \qgen^j }{2-\pgen}
    .
\end{equation}
Here we assume that $T_0>T_0'$.
Modelling the fidelity decrease as exponential-decaying function of the storage time, the fidelity of the earlier link decays by a factor 
$\Gamma = \mean{
    \exp(
            -|T_0-T_0'|/\tcoh
    )
}
$,
where $\tcoh$ denotes the memory coherence time.
Plugging in Eq.~(\ref{eq:memory storage time}), we obtain
$$
    \Gamma = 
    \frac{\pgen}{2-\pgen}
    +
    2\sum_{j=1}^{\infty}
    \exp\left(
        -\frac{j}{\tcoh}
    \right)
    \cdot
    \frac{\pgen \qgen^j }{2-\pgen}\ .
$$
The factor 2 before the sum corresponds to the possibility that either link can be generated earlier than the other. Finally, the evaluation of the sum gives~\cite{schmidt2019memory,loock2019extending}
\begin{equation}
    \label{eq:single repeater decay}
    \Gamma
    =
    \frac{\pgen}{2-\pgen}
    \left(
        \frac{2}{1-\qgen\exp (-\frac{1}{\tcoh})}-1
    \right)
    .
\end{equation}

In addition to the single-repeater scenario considered above, analytical results for the memory decay have also been obtained for more advanced single repeater protocols such as a protocol with cut-offs \cite{collins2007multiplexed} or protocols where two elementary links have to be prepared sequentially \cite{rozpedek2018parameters}.

Unfortunately, for higher-level nested protocols, i.e.\ $n \ge 1$, there is no analytical expression known for the mean waiting time $\mean{T_n}$ with $\pswap<1$, because $T_i$ for $i>0$ does not follow a geometric distribution, in contrast to $T_0$.

\subsubsection{Approximation with geometric distribution at higher level}
\label{sec:approximation with geometric distribution}
Above, we computed the waiting time probability distribution in the single-repeater scenario.
This calculation explicitly relied on the fact that the waiting time distribution of elementary-link generation follows a simple distribution, the geometric distribution (Eq.~\eqref{eq:geometric-distribution}).
Unfortunately, for nested repeater chains with more than a single repeater, no exact expression for the waiting time distribution has been found.

\begin{figure}
    \centering
    \includegraphics[width=\linewidth]{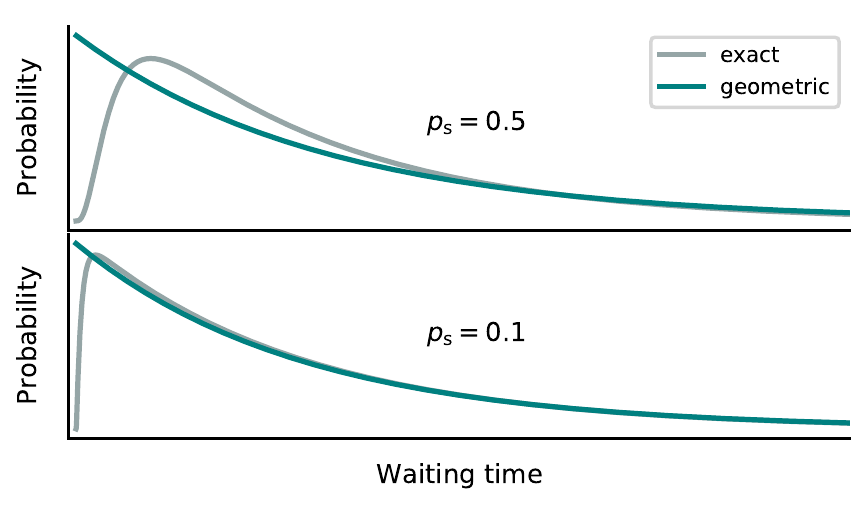}
	\caption{
The probability distribution of the exact waiting time $T_2$ of a nested swap protocol with 4 repeater segments (computed with the algorithm from Section~\ref{sec:probability-tracking algorithms}) and, as an approximation to the exact distribution, the geometric distribution from Eq.~(\ref{eq:geometric-distribution}) with the same mean, i.e. $p=1/\mean{T_2}$.
(Top) We see that the two distributions deviate most for short waiting times.
	This can be explained by noting that the exact probability that all entanglement generation steps and entanglement swaps succeed in the first few steps is very small.
This fact is not captured by the approximation.
(Bottom) We observe that for small swap success probabilities $\pswap$ (both axes are rescaled to compare only the shape of the distributions), the deviation becomes smaller.
}
\label{fig:compare distributions}
\end{figure}

However, the waiting time distribution at higher nesting levels can be approximated by assuming that, at a higher level, the waiting time distribution is still geometrically or exponentially distributed (Eq.~\eqref{eq:exponential-distribution}).
This approximation is usually used in an iterative manner.
One computes the average waiting time at the current level and uses it to define a geometric distribution with the same expectation value.
This new distribution is then used to study the next nesting level.
In Fig.~\ref{fig:compare distributions}, we compare the approximated distribution with the exact one.

\begin{figure*}
    \centering
    \includegraphics[width=\textwidth]{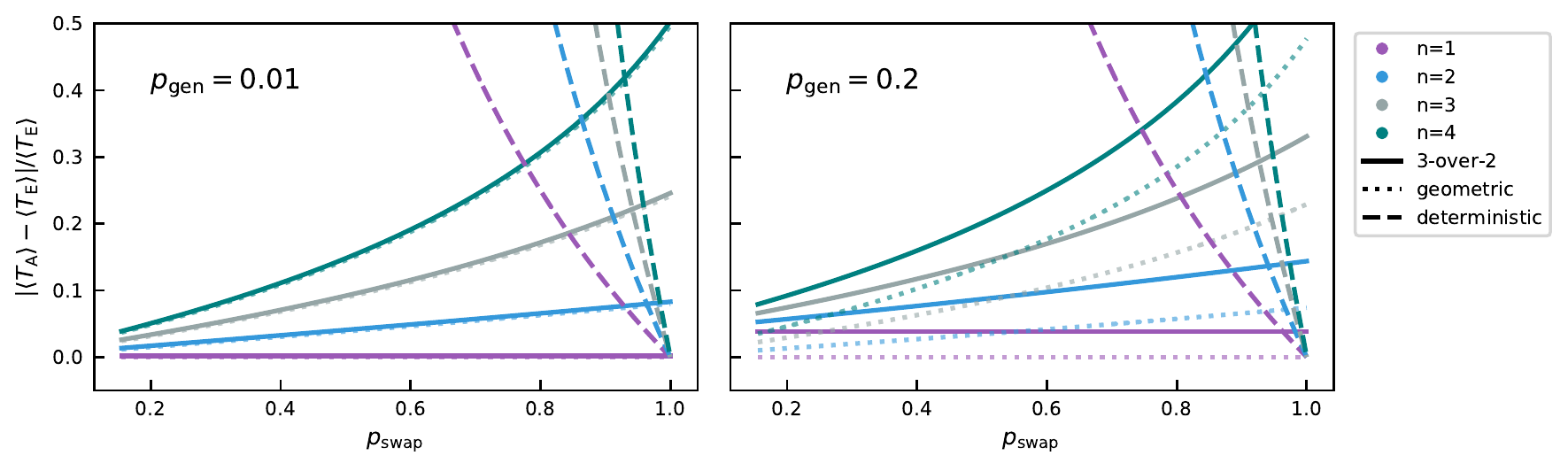}
    \caption{Comparing the relative error in the mean waiting time $|\mean{T_{\rm{A}}} - \mean{T_{\rm{E}}}| / \mean{T_{\rm{E}}}$ among different approximation methods for the mean waiting time of a nested repeater protocol on $n$ nesting levels, where $\mean{T_{\rm{A}}}$ and $\mean{T_{\rm{E}}}$ are the approximated and exact mean waiting time, respectively.
    We plot the relative difference between the approximated and the exact mean waiting time as a function of the swap success probability $\pswap$ for nested swap protocols up to level $n=4$.
    The three approximation methods are:
    1. The approximation with geometric distribution given by Eq.~(\ref{eq:geometric}).
    It is exact at level 1 and the deviation increases as the levels grow;
	2. The 3-over-2 approximation given by Eq.~(\ref{eq:3-over-2}), which is itself an approximation of Eq.~\eqref{eq:geometric}.
    In the limit of $p = \pgen \ll 1$, the difference between the two approximation vanishes;
	3. The deterministic swap approximation (Eq.~\eqref{eq:deterministic swap t}) assuming that the swap always succeeds ($\pswap=1$).
    Note that, in contrast to the former two, the latter approximation is a lower bound on the exact waiting time.
    The deviation of the deterministic swap approximation is almost independent of $\pgen$.
    The exact mean waiting time $\mean{T_{\rm{E}}}$ is computed using the algorithm in Section \ref{sec:probability-tracking algorithms}.
    Due to our implementation, we cannot reach the region $\pswap \rightarrow 0$.
	However, Shchukin \textit{et al.} numerically show that, in this limit, the deviation of the approximation with geometric distribution becomes negligible (described in Ref.~\cite{shchukin2019waiting}).
    }
    \label{fig:error of approxiamtion}
\end{figure*}

Let us give the explicit calculation under the approximation that the waiting time follows a geometric distribution at each nesting level.
We first calculate $\mean{T_{n-1}}$ and then approximate the distribution of $T_{n-1}$ with a geometric distribution (Eq.~\eqref{eq:geometric-distribution}) parameterized by $p=1/\mean{T_{n-1}}$.
Under this assumption, the mean waiting time $\mean{T_n}$ can be computed by a derivation analogous to the one leading to Eq.~\eqref{eq:average waiting time t1} in Section~\ref{sec:single repeater swap protocol} and is given by
\begin{equation}
    \label{eq:geometric}
    \mean{T_n} = \frac{3-2p_{n-1}}{(2-p_{n-1})p_{n-1}\pswap}
\end{equation}
with $p_{n-1}=1/\mean{T_{n-1}}$.
In the limit of $\pgen \rightarrow 0$ for the bottom level ($n=0$) and $\pswap \rightarrow 0$ for higher levels ($n>0$), the mean waiting time $\mean{T_{n-1}} \rightarrow \infty$ and thus $p_{n-1} \rightarrow 0$.
As a consequence, Eq.~\eqref{eq:geometric} can be approximated as
\begin{equation}
	\label{eq:3-over-2-intermediate}
    \mean{T_n} \approx \frac{3}{2p_{n-1}\pswap}
    .
\end{equation}
Effectively, it means that, on average, the waiting time of generating two links is approximately $3/2$ times that of a single link.
Applying Eq.~\eqref{eq:3-over-2-intermediate} iteratively over all nesting levels with $\mean{T_0} = 1/\pgen$ yields
\begin{equation}
    \label{eq:3-over-2}
    \mean{T_n} \approx \frac{3^n}{2^n \pswap^n \pgen}
    ,
\end{equation}
which is precisely the 3-over-2 approximation mentioned in Section \ref{sec:mean-only}.

The error introduced by the approximations Eq.~(\ref{eq:3-over-2}) and Eq.~(\ref{eq:geometric}) is shown in Fig.~\ref{fig:error of approxiamtion}.
As expected, the figure shows that the approximations behave well if the success probabilities of elementary-link generation and swapping are small, i.e. $\pgen \rightarrow 0$ and $\pswap \rightarrow 0$.
The figure also shows that the approximations are not so good for large success probabilities; the deviation from the exact mean waiting time increases as $\pswap$ grows, and the deviation is worse for Eq.~\eqref{eq:3-over-2} than for Eq.~\eqref{eq:geometric}.

To approximate the fidelity of the produced link between the end nodes of the repeater chain in the presence of memory decoherence, one can use Eq.~(\ref{eq:single repeater decay}) by replacing $\pgen$ with $1/\mean{T_{n-1}}$.
The approximation is computed analogously to the procedure described for the average waiting time; that is, for a given level, the average infidelity for the entanglement links just before a swap due to the memory decoherece is calculated and used to derive the initial infidelity for entangled links at the next level.
By assuming that the distribution at every level is given by the exponential distribution (Eq.~\eqref{eq:exponential-distribution}), Kuzmin {\it et al.} designed a semi-analytical method for computing fidelity with more sophisticated memory decay models \cite{kuzmin2019scalable, kuzmin2020diagrammatic} (see also Section \ref{sec:probability-tracking algorithms}). 

A different approach to keep the waiting time distribution geometric at a higher level is to design a special protocol.
For example, Santra {\it et al.}\cite{santra2018quantum} introduce a family of protocols with a memory buffer time. This buffer time is a threshold on the total waiting time for preparing the two links for the swap.
If the links are not ready before the buffer time is reached, the protocol aborts and restarts from entanglement generation. The buffer time is slightly different from the memory cut-off (see Section \ref{sec:abstract models of quantum networks}); with a memory cut-off the protocol aborts if the memory storage time of a single link (instead of both links) exceeds a threshold. 

The protocol is designed such that the buffer time at the current level becomes the time step at the next level.
As a consequence, the waiting time is geometrically distributed at each nesting level.
Note that the protocol results in avoidable additional memory decay as both links have to wait until the buffer time is reached even if they are prepared before that.
Despite this, by optimizing the buffer time, Santra {\it et al.} show that this protocol improves the final fidelity for some parameter regimes compared to the nested repeater protocol without buffer times.

An alternative approach was taken for optimizing repeater protocols including distillation, where the buffer time is chosen large enough so that the protocol becomes nearly deterministic \cite{goodenough2020}.
At a cost of longer waiting time and lower fidelity, the variance in the fidelity is kept small and the protocol can deliver entanglement at a pre-specified time with high probability.

\subsubsection{Deterministic entanglement swap}
\label{sec:deterministic swap}
So far we have focused on the regime where the success probability of entanglement swap is smaller than 1. 
In this section, we discuss nested repeater protocols with deterministic entanglement swapping ($\pswap = 1$) and without distillation.

First, let us compute the waiting time probability distribution in the case of deterministic swaps without distillation or cut-offs.
Recall that we assume that the time required to perform local operations is negligible so that the deterministic entanglement swap has no contribution to the waiting time.
For a repeater scheme with $n$ nesting levels, the total waiting time is the time until all $N=2^n$ elementary links are prepared, i.e. the maximum of $N$ independent copies of $T_0$:
\begin{equation}
    \label{eq:order statistics}
    T_N = \max(T_0^{(1)}, T_0^{(2)}, \dots, T_0^{(N)})
    .
\end{equation}
The cumulative distribution of $T_N$ from Eq.~(\ref{eq:order statistics}) is given by the general version of Eq.~\eqref{eq:cdf-max} for the maximum of $N$ independent and identically distributed random variables:
\begin{equation*}
    \Pr\left(T_N \leq t\right)
	=
    \Pr\left(\max(T_0^{(1)}, \dots, T_0^{(N)}) \leq t\right)
    =
\Pr(T_0 \le t)^N
\end{equation*}
from which the probability distribution of $T_N$ can be computed using
\begin{equation*}
    \Pr\left(T_N = t\right)
=
    \Pr\left(T_N \leq t\right)
	-
    \Pr\left(T_N \leq t-1\right)
	.
\end{equation*}

By $T_{N,k}$, we denote the random variable describing the time at which the first $k$ elementary links of the $N$ segments are generated.
We first give the expression for $\Pr(T_{N,k} \le t)$, the probability that at least $k$ links are generated before $t$.
This is equivalent to the probability that, at time $t$, the number of elementary links that have not yet been generated is $N-k$ or less \cite{vinay2017practical}:
\begin{equation*}
    \Pr(T_{N,k} \le t)
    = 
    \sum_{j=0}^{N-k} \genfrac{(}{)}{0pt}{}{N}{j} \left(1 - \Pr(T_0\leq t)\right)^{j} \Pr(T_0 \leq t)^{N-j}
\end{equation*}
where $\Pr(T_0 \leq t) = 1 - (1 - \pgen)^t$ since $T_0$ is geometrically distributed with success probability $\pgen$.
The probability that precisely $k$ of the $N$ segments are generated at time $t$ is
\begin{equation*}
    \Pr(T_{N,k}=t) = \Pr(T_{N,k} \le t) - \Pr(T_{N,k} \le t-1)
    ,
\end{equation*}
from which the mean waiting time is calculated as
\begin{equation}
	\label{eq:mean-TNk}
    \mean{T_{N,k}} = \sum_{t=1}^{\infty} t \cdot \Pr(T_{N,k}=t)
    .
\end{equation}

The mean waiting time $\mean{T_{N, k}}$ from Eq.~\eqref{eq:mean-TNk} can be computed by solving a recurrence formula where $\mean{T_{N,k}}$ is expressed as function of $\mean{T_{N,k-1}}$ \cite{bernardes2011rate,praxmeyer}.
For $k=N$, i.e.\ the waiting time that all elementary entanglement are prepared \cite{bernardes2011rate}, the solution reads
\begin{equation}
    \label{eq:deterministic swap t}
    \mean{T_{N,N}}(p)=\sum_{k=1}^N
    \genfrac{(}{)}{0pt}{}{N}{k}
    \frac{
        (-1)^{k+1}
    }
    {
        1-(1-\pgen)^k
    }
    .
\end{equation}
For $\pgen \ll 1$, this expression can be approximated by \cite{shchukin2019waiting,schmidt2019memory,vinay2017practical}
\begin{equation}
    \mean{T_{N,N}}(p)\approx
    \sum_{k=1}^N
    \genfrac{(}{)}{0pt}{}{N}{k}
    \frac{
        (-1)^{k+1}
    }
    {
        k\pgen
    }
    =    
    \sum_{k=1}^N
    \frac{
        1
    }
    {
        k\pgen
    }
    =
    \frac{H(N)}{\pgen}
\end{equation}
with
\begin{equation*}
    H(N)=
    \sum_{k=1}^N
    \frac{1}{k}
    \approx
    \gamma + ln(N) + \frac{1}{2N}
    +\mathcal{O}(\frac{1}{N^2})
    ,
\end{equation*}
where $H(N)$ is the $N$-th harmonic number and $\gamma\approx 0.57721$ is the Euler-Mascheroni constant.
In separate work, Praxmeyer included finite memory time with cut-off into the calculation and obtained \cite{praxmeyer}
\begin{equation}
    \label{eq:deterministic swap with cut-off}
    \mean{T_{N, N}} = 
    \frac{
        1 - (1-\qgen^\tau)^N
        + (1-\qgen^N)
        \left[
            \tau - \sum_{j=1}^{\tau-1}(1-\qgen^j)^N
        \right]
    }
    {
        (1-\qgen^{\tau+1})^N - \qgen^N(1-\qgen^\tau)^N
    }
    ,
\end{equation}
where $\tau$ is the cut-off threshold and $\qgen=1-\pgen$.

Similar derivations as the ones above can be used for the waiting time until the first, instead of the last, elementary link has been generated.
Those derivations are relevant for the analysis of multiplexed repeater protocols and the mean waiting time in those cases has been analyzed in \cite{collins2007multiplexed,vinay2017practical}.

To our knowledge, in contrast to the waiting time, there is no exact fidelity calculation with exponential memory decoherence for deterministic swap.
A lower bound on the fidelity can be obtained by assuming the worst case, i.e. the swap is performed only after all elementary links are generated \cite{bernardes2011rate,vinay2017practical}.

These expressions presented here for the deterministic-swap case also apply to repeater chains where the numbers of segments is not a power of 2, as well as to more general network topology \cite{khatri2019practicalPRR}.
The reason for this is that if the swaps are deterministic, the waiting time equals the time until all elementary links in the network have been prepared.
Thus, the nested structure does not exist and the only relevant parameters are the number of elementary links and the elementary-link success probability $\pgen$.

The waiting time in the deterministic-swap case can be used as an approximation to the case where $\pswap$ is slightly lower than 1 and bounds from below the waiting time for general $\pswap$.
The quality of the approximation is shown in Fig.~\ref{fig:error of approxiamtion}.

\subsubsection{Methods for analyzing distillation-based repeater schemes with memory-decoherence}
\label{sec:methods for distillation protocols}
In contrast to entanglement swapping, distillation has a fidelity-dependent success probability.
In the absence of decoherence, the fidelity of a pair does not decrease while it is waiting for other components to succeed.
Hence, the success probability $\pdist$ is a constant for each level and distillation can be studied in the same way as entanglement swapping.
However, in the presence of decoherence, fidelity and success probability become correlated, which complicates the analysis.

We finish by mentioning two tools for bounding the fidelity and generation rate of distillation-based repeater schemes in the presence of memory decoherence.
First, upper bounds on the achievable fidelity can be derived using fixed-point analysis~\cite{briegel1998quantum, deutsch1996quantum}.
In this approach, one makes use of the fact that entanglement distillation does not improve the fidelity when the quality of the input links is sufficiently high.
Such a fidelity is thus a fixed point of the entanglement distillation procedure and it depends on the quality of the local operations~\cite{briegel1998quantum} and memories~\cite{hartmann2007role}.
If the fixed-point is an attractor and the input links have fidelity lower than the fixed-point, repeated application of entanglement distillation cannot boost fidelity beyond the fixed-point and it thus forms an upper bound.
Next, lower bounds on the fidelity can be trivially obtained for protocols that impose a fidelity cut-off, i.e. protocols that discard the entanglement if the fidelity is lower than a certain threshold~\cite{vinay2019statistical}.
Because the distillation success probability is a monotonic function of the fidelity of the input states, a lower bound on the fidelity by a cut-off also directly yields a lower bound on the success probability.

\subsection{Numerical tools for evaluating analytical expressions}
\label{sec:numerical study on the rate and fidelity}

Above, in Section~\ref{sec:analytical study on the rate and fidelity}, we reviewed analytical tools for computing the probability distribution of the waiting time for generating remote entanglement and of the entanglement quality.
For models that do not include memory decoherence, distillation, or cutoffs, these tools are sufficient.
For more complex models and for the analysis of many-node networks, the tools presented above are often still applicable but analytically evaluating the resulting expressions to compact, closed-form expressions is unfeasible.
An example of such a case is a nested repeater chain with cut-offs and non-deterministic swapping.
In this case, no concise analytical expression for the waiting time is known. A priory, it is possible to write down a recursive analytical expression for the waiting time using a similar reasoning to Section~\ref{sec:single repeater swap protocol}, where the single-repeater case was treated. However, the recursion relation has so far not been solved for general repeater chains.
Fortunately, numerical tools enable the evaluation of such expressions.
In this section, we treat three classes of such tools: Markov Chain algorithms, probability-tracking algorithms and Monte Carlo methods for abstract models.
These numerical tools differ from the simulation tools in Section~\ref{sec:simulation} in that they are all directly applied to expressions that can in principle be evaluated analytically.

\subsubsection{Markov Chain methods}
\label{sec:markov cahin method}
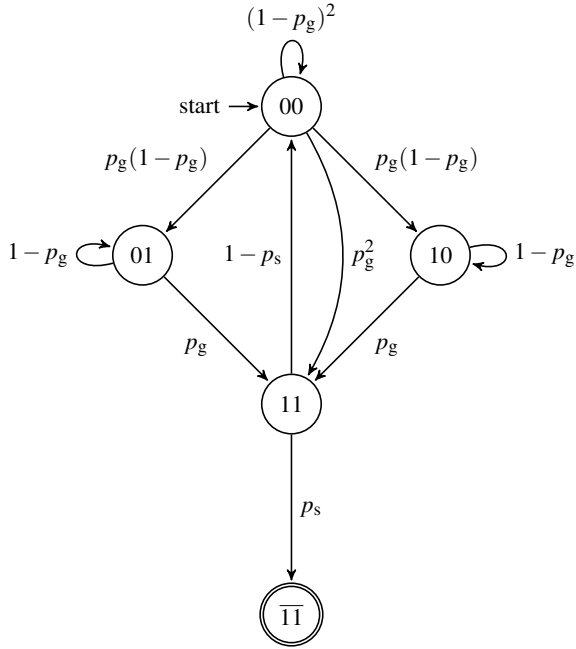
\begin{figure}
\begin{tikzpicture}[->,>=stealth',shorten >=1pt,auto,node distance=2.8cm,
    semithick]
\tikzstyle{every state}=[]

\node[initial, state]         (A)                    {00};
\node[state]         (B) [below left of=A]  {01};
\node[state]         (C) [below right of=A] {10};
\node[state]         (D) [below right of=B] {11};
\node[accepting, state]         (E) [below of=D]       {$\overline{11}$};

\path (A) edge [loop above] node  {$(1-\pgen)^2$}   (A)
          edge              node[midway,above left] {$\pgen(1-\pgen)$} (B)
          edge              node {$\pgen(1-\pgen)$} (C)
          edge [bend left]  node {$\pgen^2$}        (D)
      (B) edge [loop left]  node {$1-\pgen$}        (B)
          edge              node[midway,below left] {$\pgen$}          (D)
      (C) edge              node {$\pgen$}          (D)
          edge [loop right] node {$1-\pgen$}        (C)
      (D) edge              node[midway] {$1-\pswap$}          (A)
          edge              node {$\pswap$}         (E)
    ;

\end{tikzpicture}
	\caption{The directed graph for a Markov chain of the single repeater swapping protocol, which consists of two end nodes with a repeater node positioned in between.
	A vertex in the graph corresponds to a state of the network, while the labeled edges represent possible transitions between states with corresponding transition probabilities.
	The Markov state $00$ represents the initial state with no entanglement; state $01$ ($10$) is the state with one elementary link on the right (left) segment; state $11$ is the state with an elementary link on both segments; state $\overline{11}$ denotes the state after the successful entanglement swapping by the repeater node, yielding entanglement between the end nodes.
	The double cycle indicates that this Markov state is absorbing, i.e.\ has only incoming transitions.
Such an absorbing state corresponds to the protocol being finished.
Reprinted with permission from E. Shchukin, F. Schmidt, and P. van Loock, Phys. Rev. A, vol. 100, p. 032322, (2019)\cite{shchukin2019waiting}. Copyright 2019 by the American Physical Society.
}
\label{fig:markov chain}
\end{figure}
In many repeater protocols, the change of the entanglement in the network in the next time step only depends on the existing entanglement.
Shchukin \textit{et al.} used this idea to model the network as a discrete Markov chain~\cite{shchukin2019waiting}, which can be visually depicted as a directed graph, an example of which is shown in Fig.~\ref{fig:markov chain}.
A vertex in this graph is a state of the network, i.e. the collection of entanglement that exists at a given point in time.
The network transitions from one state to the other with a fixed probability, which is visualized by directed edges of the labeled graph.
At each time step, a network randomly transitions from its current state to the next state according to the transition probabilities over outgoing edges.
For example, in the single-repeater protocol depicted in Fig.~\ref{fig:markov chain}, the transition from the `an entangled pair exists on each of the two segments' state ($11$) to `entanglement exists between end nodes' ($\overline{11}$) occurs with the entanglement swapping success probability $\pswap$ (entanglement swapping succeeded), whereas the transition to `no entanglement' (state $00$) has probability $1 - \pswap$ (entanglement swapping failed and the two involved links are lost).

An equivalent representation of a Markov chain is the transition probability matrix (TPM), where entry $(j,k)$ represents the transition probability from state $j$ to state $k$.
Since a single transition corresponds to a single time step, the waiting time distribution equals the distribution of the number of edges traversed before reaching a predefined target state, such as `entanglement between the end nodes of the repeater chain' ($\overline{11}$ in Fig.~\ref{fig:markov chain}).
Shchukin {\it et al.} used this equivalence to compute the average waiting time, as well as its variance, by solving a linear equation system that has the same size as the number of states in the Markov chain.

The original proposal by Shchukin {\it et al.\ }included an analysis of the waiting time.
Later, the idea was refined to include memory decoherence by Vinay {\it et al.\ }\cite{vinay2019statistical}
They computed the fidelity distribution by assigning a noise parameter to certain transition edges and calculated how many times the edges are traversed given that the entanglement distribution is completed at time $t$.
With this noise model, the Markov chain method was used to study the BDCZ protocol \cite{briegel1998quantum}, which includes entanglement distillation.
Due to the assumption of the Markov process, i.e.\ the system has no memory of the past, this method cannot handle fidelity-dependent success probability without assigning each possible fidelity a state representation.
As an alternative, Vinay {\it et al.\ } provided a lower bound to the final fidelity using fidelity cutoffs (see Section~\ref{sec:methods for distillation protocols}).

Apart from repeater chain protocols, Markov chains have also been used to study more general network protocols, such as a quantum switch by Vardoyan {\it et al.}~\cite{vardoyan2019capacity,vardoyan2019stochastic}, who used the continuous-time Markov chains as an approximation to discrete-time Markov chains.
In this model, the transition probability is replaced by the transition rate.
Compared to their discrete counterparts, continuous-time Markov chains simplify the analysis in various aspects.
For instance, Vardoyan {\it et al.} included a model of decoherence where the states are lost at a fixed rate by adding an additional transition edge indicating the loss of one entangled pair.
Moreover, Vardoyan {\it et al.} show that the quality of the approximation is high in many scenarios~\cite{vardoyan2019capacity}.

The Markov chain method is rather general and flexible:
in principle, the waiting time of any entanglement distribution protocol with predefined transition probabilities can be calculated, regardless of the topology or entanglement swapping policy (such as swapping as soon as two links are available, regardless of the segments on which this entanglement has been produced, or swapping only between predefined segments).
However, this method is computationally expensive.
The size of the TPM is the same as the number of possible Markov states and, in general, grows exponentially with the number of nodes.

This rapid growth of the size of TPM can be partially mitigated. For instance, by grouping equivalent Markov states and treating them as one state, the complexity can be drastically reduced.
With this technique, Shchukin {\it et al.\ }gave examples for the BDCZ protocol with analytical expressions for up to 4 nodes, while numerically they reached 32 nodes~\cite{shchukin2019waiting}.
Vinay {\it et al.} reduced the computational complexity of this approach using probability generating functions and complex analysis, but the scaling remains exponential \cite{vinay2019statistical}.
To process a larger number of nodes, Shchukin {\it et al.\ }proposed to use the average waiting time to replace the random variable for low-level sub-protocols.
This idea is similar to approximating the waiting time distribution at every nesting level of a repeater protocol with the bottom-level distribution (see Section \ref{sec:approximation with geometric distribution}).

Finally, in a recent development, Khatri \cite{khatri2020policies} introduced a method for describing network protocols based on quantum partially observable Markov decision processes \cite{barry2014quantum}. A quantum partially observable Markov decision process is a reinforcement-learning based framework for protocol optimization. In this framework, the protocol obtains feedback from its actions in the form of classical information about the quantum state that the network holds, which the protocol uses to choose its next action. As an application of the method, Khatri found analytical solutions for optimizing a cut-off for elementary link generation under different constraints. It is an interesting open question whether this approach can be extended for efficiently characterizing and optimizing protocols over large repeater chains and networks.

\subsubsection{Probability-tracking algorithm}
\label{sec:probability-tracking algorithms}
The next numerical tool tracks the full waiting time probability distribution and the average fidelity of the delivered quantum state.
We explain this method via a concrete example, a symmetric nested repeater protocol with $2^n$ segments and no entanglement distillation (depicted in Fig.~\ref{fig:bdcz}(b) for $n=2$).
In Section~\ref{sec:single repeater swap protocol}, we treated the case for $n=1$, which resulted in an expression for the waiting time random variable consisting of the maximum of two copies of the waiting time of the bottom level (Eq.~\eqref{eq:max of two T}) and a compound sum (Eq.~\eqref{eq:compound sum}).
The first element, the maximum, corresponds to the fact that an entanglement swap acts on two links that need both be generated, so one needs to wait until the latest of the two links has been prepared.
The second element is the sum of the waiting time until the first successful swap attempt.
Since the number of attempts is probabilistic, the result is a compound sum.
The analysis for the $n=1$ case can be generalized to an arbitrary number of nesting levels $n$ and yields a recursive expression of the waiting time $T_n$ which alternates between compound sums and maximums of two copies of the waiting time $T_{n-1}$ on the previous level.
Unfortunately, as discussed in Section~\ref{sec:analytical study on the rate and fidelity}, to our knowledge, this recursive expression of $T_n$ has not been analytically evaluated for $n>1$.
Hence, various approximation methods were introduced in that section.
The exact evaluation can, however, be achieved with numerical tools.
By truncating the waiting time at a prespecified time $t_{\rm trunc}$, the waiting time distribution becomes finite.
The evaluation with numerical tools leads to an algorithm that tracks both the truncated probability distribution of $T_n$ and the associated fidelity \cite{brand2020efficient,li2020efficient}.

On the $2^n$-segment nested repeater protocol, the algorithm computes the waiting time distribution as follows.
If $n=0$, i.e. if there is no repeater and the two end nodes obtain entanglement by direct generation, the waiting time $T_0$ follows a geometric distribution (Eq.~\eqref{eq:geometric-distribution}).
If $n=1$, i.e. there is a single repeater and two segments, then the algorithm evaluates Eq.~\eqref{eq:compound sum}, which describes how the probability distribution of the waiting time $T_1$ can be obtained from the distribution of $T_0$.
Although the two elements in Eq.~\eqref{eq:compound sum}, the maximum and the geometric compound sum, can in principle be evaluated sequentially\cite{brand2020efficient}, a computationally faster approach is to separate the probability distributions of failed and successful swap attempts\cite{li2020efficient}.
For $n>1$, the algorithm is applied iteratively over the nesting levels until level $n$ has been reached.
The algorithm can be extended in polynomial time to also track the average fidelity of the produced quantum state.
This fidelity is a function of the delivery time and it can include the effect of memory decoherence.

Although the example protocol above only consists of entanglement swapping, the algorithm is applicable to any protocol which is composed of entanglement distillation and cut-offs, in addition to entanglement swaps\cite{li2020efficient}.
The algorithm presupposes that the protocol is composed of these three operations in a predefined order, e.g.\ which entangled pairs are swapped must be known in advance.
The algorithm scales polynomially in the number of nodes and in the truncation time $t_{\rm trunc}$ and has been used to track over 1000 nodes for some parameter regimes~\cite{brand2020efficient}.

A related approach to the probability-tracking algorithm explained above is taken by Kuzmin {\it et al.} \cite{kuzmin2019scalable, kuzmin2020diagrammatic}.
This method assumes that the waiting time of an elementary link is exponentially distributed (Eq.~\eqref{eq:exponential-distribution}), after which the mean waiting time for the next level is computed by evaluating a continuous integral, as well as the quantum state in the presence of memory decay.
These are then used as input to the next nesting level, by assuming that at that level, the waiting time follows an exponential distribution also.
With this approximation, the calculation of the maximum in Eq.~(\ref{eq:max of two T}) is simplified.

\subsubsection{Sampling the analytical expressions with Monte Carlo method}
So far in this section, we have discussed two methods that compute the statistical distribution of the waiting time and produced quantum state.
For a given model, both of them evaluate the analytical expression exactly up to machine precision.
An alternative to this semi-analytical computation is to sample the expressions on random variables for the waiting time and the delivered state using a Monte Carlo approach~\cite{brand2020efficient}.
Instead of tracking the whole distribution, this approach samples a pair of waiting time and the produced state between the end nodes.
By sampling many times, the probability distribution of the waiting time and the quantum state can be reconstructed.

Again, let us take the $2^n$-segment nested repeater protocol as an example to explain the algorithm.
The individual sample pairs are produced by iterating over the different components of the repeater protocol, following its nested structure.
At each component, a pair is sampled recursively, following the expressions on random variables, which thus become expressions on individual events.
For instance, Eq.~(\ref{eq:max of two T}) requires sampling two instances of $T_0$ for entanglement generation and then taking the maximum of both to produce a sample of $M_0$.
Also, memory decoherence can be calculated from the time difference of two events.
Similarly, the method can handle cut-offs.
Furthermore, distillation can also be included in the protocol, since the input states to a distillation attempt, which determine its success probability, are also sampled.
For nested protocols, the Monte Carlo algorithm can be defined as a recursive function, following the nested structure of the protocols.

\subsection{Second and third generation repeater protocols}
\label{sec:second- and third generation repeater protocols}
So far, we have only treated first-generation quantum repeater protocols, i.e. protocols for which the building blocks -- fresh entanglement generation, entanglement swapping, and entanglement distillation -- are probabilistic operations.
The quantum repeater proposals that do not fall into this category make explicit use of quantum error correction codes and are referred to as second-generation (probabilistic entanglement generation, deterministic entanglement swapping, and one-way quantum error correction) and third-generation repeaters (loss-tolerant entanglement generation) \cite{munro2015inside,muralidharan2016optimal}.

In first-generation repeaters, entanglement generation and swapping are probabilistic, and once it has succeeded, the entanglement is kept in quantum memories and needs to wait until other components performed in parallel have succeeded.
Consequently, the waiting time probability distribution and state quality are a complex function of the success probabilities of components in the repeater chain.

In contrast, for second-generation repeater protocols, such as~\cite{jiang2009quantum,munro2010quantum,li2013long,mazurek2014long}, entanglement swapping and (one-way) quantum error correction are no longer probabilistic. (We note that although entanglement generation is also probabilistic, it may be parallelized to achieve near-unit generation success probability). As a result, the time at which the entire repeater chain finishes with a single attempt at generating end-to-end entanglement is simply a sum of the (constant) times that the individual components take.
Not only there is no waiting for other components, but there are also no feedback loops here, i.e.\ components that need to restart in case others have failed.
The unit time step at which such repeater chains operate is an attempt at end-to-end entanglement generation (i.e.\ the sum of the individual component times); the probability that such an attempt succeeds is the product of all individual steps succeeding.
For this reason, the distribution of the waiting time is a geometric distribution. 

A similar reasoning applies to third generation repeaters~\cite{knill1996concatenated,fowler2010surface,munro2012quantumcommunication,azuma2015all,muralidharan2014ultrafast,muralidharan2016optimal,glaudell2016serialized,ewert2016ultrafastPRL,ewert2016ultrafastPRA,pant2017rate,lee2019fundamental,borregaard2020one}, where entanglement between adjacent nodes is established almost deterministically, rather than probabilistically, by encoding part of locally-generated entanglement into a large state of photons, followed by transmission of the encoded state.
Commonly, the analysis of the propagation of operational errors (for 2nd and 3rd generation) and the propagation of physical loss errors into logical errors (for 3rd generation) is based on work on quantum error correction codes (combined with explicit counting of the combinations of losses of photons which yield errors beyond recovery) and optical quantum computation.
We consider such tools out of scope for this review.

 \section{Simulation tools}
\label{sec:simulation}
In this section, we sketch the methodology for evaluating the network performance based on simulation and then discuss the particularities of quantum networks together with a selection of existing tools for quantum networks. For an in-depth discussion about network simulation, we refer to the books \cite{wehrle2010modeling,fernandes2017performance,burbank2011introduction}.

Let us first compare simulation tools against analytical tools. 
Analytical tools are well suited for predicting the performance of scenarios which are simple in terms of streamlined models, but also in terms of the communication protocol, network topology, and network usage. 
These scenarios typically are simplified versions of the scenarios of interest and the output of analytical tools can be understood as a rough statement about feasibility (Section \ref{sec:analytical}) or of the optimal achievable performance (Section \ref{sec:fundamental}). 
Simulation finds its role in use cases that move beyond simple scenarios.

\subsection{Simulation methodology}
In the following, we sketch a methodology for network performance evaluation via simulation. The methodology consists of four steps: problem formulation, theoretical modeling, implementation, and simulation. It is a streamlined version of the best practice methodologies discussed in \cite{wehrle2010modeling,law2019build}.

\textit{Problem formulation. }
The first step is to define the goal of the performance evaluation. Some examples are the validation of protocol designs, the comparison between several alternative solutions, the study of network stability, the sensitivity analysis of performance with respect to hardware parameters, parameter optimization, etc. After the problem has been formulated, the desired performance metrics should be identified; different metrics impose different constraints to the modeling step. Some examples are: average rate, fidelity, latency, throughput, etc. 

\textit{Theoretical modeling. }
The second step is modeling. Best practices \cite{wehrle2010modeling,law2019build} dictate a separation between the theoretical characterization of a network element and its implementation. The theoretical characterization begins with an identification of boundary conditions and assumptions. An important design consideration is the level of detail of the models and protocols. This is a difficult choice that depends on several factors in addition to the problem goal and the performance metric: the availability or lack of data can condition the modeling of some elements, and the computation time might limit the level of detail as well as more general constraints. 

Let us sketch the qualitative trade-offs between simple and detailed modeling.  Simple models have several advantages: the modeling effort is reduced, simulations are comparatively faster and they can be easily modified. Conversely, the output should be taken as an estimation of the real behavior of the network. In some cases, the output of the simulations might be similar to that of analytical methods. As the models increase in detail, the simulations become slower, it becomes more difficult to modify the network and the output becomes representative of the output expected from the real devices. As the level of detail increases, the cost also increases. 

The output of the theoretical modeling process is a full specification of the behavior of all elements of interest together with their interactions. The relevant question at this stage is whether the level of detail is both necessary and sufficient. In particular, it is whether either the model does not allow to evaluate the desired metric or, on the contrary, a simpler model would be sufficient. This can be quantified by estimating whether additional or smaller amounts of detail will impact the simulation outcome. Finally, the model is documented.

\textit{Implementation. }
The documented model is then translated into software. The first decision is whether to develop an ad-hoc solution or to opt for an existing framework. The advantage of an ad-hoc solution is that it can be tailored to the project at hand, which in turn can benefit both the accuracy of the implementation and its computational efficiency. On the other hand, tailored solutions have several drawbacks: they might be difficult to expand at a later stage, they can be costly to develop and it is difficult to compare the outcome with other simulation studies.
Most projects build on an existing framework, and the choice can depend on many factors including the existence of models, the validation means, the capability of quantifying the desired metric, the cost, the flexibility for further investigation, and the accuracy.  

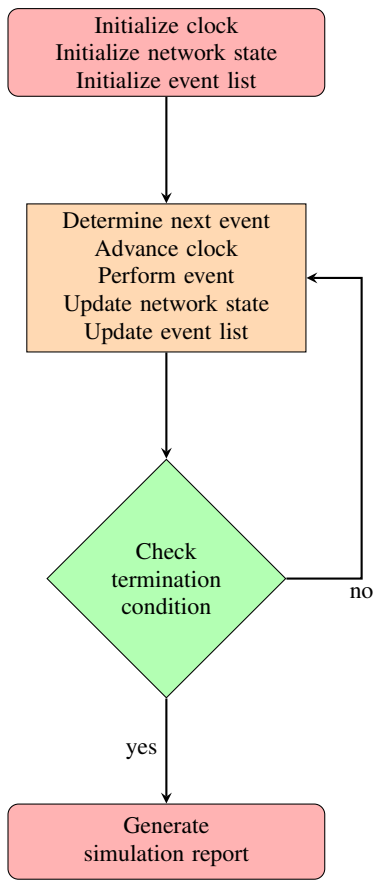
\begin{figure}
\begin{tikzpicture}[node distance=2cm]
\tikzstyle{startstop} = [rectangle, rounded corners, minimum width=3cm, text width=4cm,minimum height=1cm,text centered, draw=black, fill=red!30]
\tikzstyle{io} = [trapezium, trapezium left angle=70, trapezium right angle=110, minimum width=3cm, minimum height=1cm, text centered, draw=black, fill=blue!30]
\tikzstyle{process} = [rectangle, minimum width=3cm, minimum height=1cm, text width=3.5cm, text centered, draw=black, fill=orange!30]
\tikzstyle{decision} = [diamond, minimum width=1cm, minimum height=1cm, text width=1.8cm, text centered,draw=black, fill=green!30]
\tikzstyle{arrow} = [thick,->,>=stealth]
\node (start) [startstop] {Initialize clock\\
Initialize network state\\
Initialize event list};
\node (pro1) [process, below of=start, yshift=-1cm] {Determine next event\\
Advance clock\\
Perform event\\
Update network state\\
Update event list};
\node (dec1) [decision, below of=pro1, yshift=-2cm] {Check\\
termination\\
condition};
\node (stop) [startstop, below of=dec1, yshift=-1.5cm] {Generate\\
simulation report};
\draw [arrow] (start) -- (pro1);
\draw [arrow] (pro1) -- (dec1);
\draw [arrow] (dec1) -- node[anchor=east] {yes} (stop);
\draw [arrow] (dec1.east) -| node[anchor=north] {no} ($(dec1.east) + (10mm,0mm)$) |-  (pro1.east);
\end{tikzpicture}
\caption{
\label{fig:des}
    Flowchart of a discrete event simulator. Discrete event simulators have a global state, a global agenda of events, and global clock. At the beginning of the simulation the global variables are initialized. Then the main loop begins, the clock is moved to the next event in the agenda, the event is executed and the global state and agenda are updated. The loop is iterated until a termination condition is met.}
\end{figure}

Simulation accuracy has a particular implication for the choice of the simulation engine. If accurate physical modeling is a requirement, accurate characterization of timing between actions is necessary.
Because of this reason, it is already the case that most classical network simulators are based on the discrete-event paradigm \cite{wehrle2010modeling}, a paradigm for simulation that guarantees reliable timing behavior (see Figure \ref{fig:des}).  The requirement of accurate tracking of time is particularly strong in the case of quantum networks, for instance, if an accurate characterization of the average fidelity is a goal; the reason is that the amount of noise a quantum memory undergoes is highly dependent on time.
In consequence, inaccurate time handling can yield strong variations in the performance evaluation.
A side benefit of discrete event simulators is that simulations are inherently repeatable for a fixed randomness seed. This is more difficult for simulators where the ordering of events is not guaranteed, such as in multi-threaded or distributed implementations.

Once the models have been translated to software, they should first be verified against the theoretical models and, if possible, validated against experimental data. 

\textit{Simulation. }
The final step is the performance of the simulations and the analysis of the results.

\subsection{The challenge of simulating quantum networks} 
The main difference between simulating quantum and classical networks is that in full generality the state of the network can be an entangled state. This implies that the state can not be described as an aggregate of the local description of the states at each network node. Moreover, the representation of the state grows exponentially with the number of qubits in the network. For this reason, classical simulators of quantum computation can not simulate computations involving more than tens of qubits \cite{wu2019full,huang2020classical}. 

Fortunately, many scenarios of interest in quantum networks involve only Clifford gates and Pauli noise, yielding classical simulation tractable \cite{van2014quantum}. Other scenarios involve many small entangled states which dynamically interact, shrink via measurement or are fused but never grow larger than a few qubits. For these scenarios, classical simulation is also possible. 

\subsection{Existing simulation platforms}
We now introduce several existing platforms for simulating quantum networks. We limit our discussion to platforms that can simulate arbitrary quantum network topologies and protocols. There are a number of specialized platforms that fall beyond the scope of this review \cite{mailloux2015modeling,mehic2017implementation,buhari2012efficient,zhao2008event,zhang2007object,niemiec2011quantum,pereszlenyi2005simulation,chatterjee2019qkdsim,dasari2016programmable,dasari2019openflow}.

The first two platforms are SimulaQron~\cite{dahlberg2018simulaqron} and the QUantum NETwork SIMulator (QuNetSim)~\cite{diadamo2020qunetsim}. Both of them aim at facilitating quantum network application development. QuNetSim focuses on ease of implementation. It posits a concrete architecture of quantum networks inspired by the OSI model \cite{zimmermann1980osi} and a convenient feature of the architecture is the joint arrival of the quantum payload together with a corresponding classical header. This simplifies the process of synchronization and the logic at the nodes, because a node's action after message reception can now be based on the classical header, irrespective of whether the payload is classical or quantum. QuNetSim has been used to explore routing schemes \cite{sawerwain2020quantum}. 
SimulaQron can be run, distributed over a classical network, i.e. on physically-distinct machines, to simulate a network of quantum computers. The back-end can be accessed via a quantum hardware interface, thus acting as a substitute for the quantum device. SimulaQron  has been used to explore entanglement verification schemes \cite{murray2020implementing} and several multipartite protocols \cite{amoretti2020entanglement,amoretti2019enhancing}; moreover, within the context of the quantum protocol zoo \cite{quantumprotocolzoo}, SimulaQron has been used to implement other protocols such as a quantum cheque scheme \cite{moulick2016quantum}, a protocol for leader election \cite{ganz2009quantum}, universal blind quantum computation \cite{dunjko2014composable} and coin flipping \cite{pappa2011practical}.
 
The next platform is the Simulator for Quantum Networks and CHannels (SQUANCH)~\cite{bartlett2018distributed}. The goal of SQUANCH is to simulate quantum information protocols with realistic noise models over large networks. The simulator builds on the notion of the agent, a party in the quantum network connected to other agents via classical and quantum channels. An agent can manipulate in runtime a classical and a quantum memory. The quantum memory can be a subset of a larger distributed quantum state, and the underlying registers are dynamically resized as they get measured or fused. The simulation mirrors the network configuration by assigning a different process to each agent, and performance is optimized with a NumPy backend. 
It has been used to evaluate quantum channel discrimination schemes \cite{qiu2019solving} and to investigate a photonic lattice architecture for a universal quantum programmable gate array \cite{bartlett2020universal}. 

While noise can be added to these first three simulators, and noise itself can sometimes be accurately modeled, the noise parameter depends on the time elapsed which is challenging to capture. 
Next, we introduce three discrete event simulators. 

The QUantum Internet Simulator Package (QuISP) \cite{githubquisp, matsuo2019simulation} aims to capture complex quantum network behavior. In particular, one of the target use cases is understanding emergent behavior with complex networks composed of large quantum networks interconnected. QuISP has been designed as a module for OMNeT++, an open-source simulator of classical networks with a long history \cite{varga2008overview}. This implies that most of OMNeT++ functionality is potentially available. This includes visualization, logging, and data analysis tools but also the models and protocols of classical networks.
The main mode of operation of QuISP is based on tracking Pauli errors.
While this is not accurate enough for link-layer protocols, it enables very efficient simulation.
QuISP has already been used to study the capability of first-generation repeater links including realistic parameters compatible with Pauli tracking and to evaluate the RuleSet-based communication protocol \cite{matsuo2019quantum}. 

The two last discrete event simulators are the Simulator of QUantum Network Communication (SeQUeNCE) \cite{wu2020sequence} and the NETwork Simulator for QUantum Information with Discrete events (NetSquid) \cite{netsquidpaper}.
Both projects aim at accurate physical simulation.
SeQUenCE is designed according to a concrete network stack architecture, consisting of reprogrammable modules for given functionalities.
SeQUeNCE has been used for a full simulation of the BB84 quantum key distribution protocol and quantum teleportation~\cite{suchara2020designing,Wu2019}, as well as for analysis of a nine-node quantum network with quantum memories modeled after single erbium ions in solids~\cite{wu2020sequence}.
Regarding NetSquid, some particular features are a dynamical handling of the quantum registers similar to SQUANCH and a fully parallelizable architecture.
NetSquid has shown to enable physically-accurate modeling through simulation of a link layer protocol with node models based on NV centers in diamond~\cite{dahlberg2019link}.
Its modular design was showcased by comparing two types of atomic-ensemble based quantum memories through only replacing the hardware model component, while its quantum engine is powerful enough to simulate up to one thousand nodes for some models~\cite{netsquidpaper}.
NetSquid has also been used to evaluate a quantum router architecture \cite{lee2020quantum} and to go beyond the analytical regime in the analysis of a quantum switch \cite{vardoyan2019capacity,netsquidpaper}.
 \section{Outlook}
\label{sec:outlook}
Quantum networks will enable the implementation of communication tasks with qualitative advantages with respect to the communication networks we know today. To compare different solutions, optimize over parameter space and inform experiments, it is necessary to evaluate the performance of concrete quantum network scenarios. Here, we discussed the existing tools for evaluating the performance of quantum networks from three different angles: information-theoretic benchmarks, analytical tools, and simulation.

First, we presented information-theoretic tools, which allow comparing the performance that a protocol with realistic parameters achieves against the optimal performance of a protocol with similar resources. In particular, we discussed a number of recent tools that allow us to bound the capacity of the network for entanglement and key distribution in multi-user scenarios with a complexity that grows polynomially with the number of nodes in the network.

Then we discussed analytical tools, mostly for estimating the average waiting time and fidelity of entanglement distribution over repeater chains. We described tools that can analyze moderately complicated protocols composed of entanglement generation, distillation, swapping, and cut-offs, and discussed the trade-offs between different approximations used in the literature.

Simulation tools take the limelight when one moves away from simple scenarios.  They can accurately model complicated scenarios and be used for problems such as validating protocols, evaluating alternative network solutions, studying the stability of a network, etc. Here, we reviewed the methodology for network simulation in the context of quantum networks and introduced six simulation frameworks with different goals and target audiences.

Our goal in this review has been to cover the tools for evaluating the performance of quantum networks. This is a rapidly developing topic. For instance, most of the benchmarks for quantum networks have been developed in the past five years and all the six simulation frameworks discussed have been released or presented to the scientific community in the past two years. We hope that this review raises awareness of the wealth of existing tools and stimulates the development of new ones.

 \section*{Data availability statement}
The data that support the findings of this study are available from the corresponding authors upon request.
\section*{Acknowledgements}
This work was supported by the QIA project (funded by European Union's Horizon 2020, Grant Agreement No. 820445) and by the Netherlands Organization for Scientific Research (NWO/OCW), as part of the Quantum Software Consortium program (project number 024.003.037 / 3368). 
K.A. thanks support, in part, from PREST, JST JP-MJPR1861 and from CREST, JST JP-MJCR1671.
S.B. acknowledges support from the European Union's Horizon 2020 research and innovation programme, grant agreement No. 820466 (project CiViQ), as well asn from the the Government of Spain (FIS2020-TRANQI and Severo Ochoa CEX2019-000910-S), Fundaci\'{o} Cellex, Fundaci\'{o} Mir-Puig, Generalitat de Catalunya (CERCA, AGAUR SGR 1381).
B.L. is supported by the IDEA League student grant programme.

\bibliographystyle{ieeetr}
\bibliography{bibliography}

\begin{thebibliography}{100}

\bibitem{law2019build}
A.~M. Law, ``How to build valid and credible simulation models,'' in {\em 2019
  Winter Simulation Conference (WSC)}, pp.~1402--1414, IEEE, 2019.

\bibitem{bennett1984quantum}
C.~H. Bennett and G.~Brassard, ``{Q}uantum cryptography: {P}ublic key
  distribution and coin tossing,'' {\em Proceedings of IEEE International
  Conference on Computers, Systems and Signal Processing}, vol.~175, 1984.

\bibitem{ekert1991quantum}
A.~K. Ekert, ``{Q}uantum cryptography based on {B}ell's theorem,'' {\em Phys.
  Rev. Lett.}, vol.~67, pp.~661--663, Aug 1991.

\bibitem{komar2014quantum}
P.~K{\'o}m{\'a}r, E.~M. Kessler, M.~Bishof, L.~Jiang, A.~S. S{\o}rensen, J.~Ye,
  and M.~D. Lukin, ``A quantum network of clocks,'' {\em Nature Physics},
  vol.~10, no.~8, pp.~582--587, 2014.

\bibitem{gottesman2012longer}
D.~Gottesman, T.~Jennewein, and S.~Croke, ``Longer-baseline telescopes using
  quantum repeaters,'' {\em Phyisical Review Letters}, vol.~109, no.~7,
  p.~070503, 2012.

\bibitem{wehner2018quantum}
S.~Wehner, D.~Elkouss, and R.~Hanson, ``Quantum internet: A vision for the road
  ahead,'' {\em Science}, vol.~362, no.~6412, 2018.

\bibitem{aparicio2011protocol}
L.~Aparicio, R.~Van~Meter, and H.~Esaki, ``Protocol design for quantum repeater
  networks,'' in {\em Proceedings of the 7th Asian Internet Engineering
  Conference}, pp.~73--80, 2011.

\bibitem{dahlberg2019link}
A.~Dahlberg, M.~Skrzypczyk, T.~Coopmans, L.~Wubben, F.~Rozp\k{e}dek,
  M.~Pompili, A.~Stolk, P.~Pawe\l{}czak, R.~Knegjens, J.~de~Oliveira~Filho,
  R.~Hanson, and S.~Wehner, ``A link layer protocol for quantum networks,'' in
  {\em Proceedings of the ACM Special Interest Group on Data Communication},
  SIGCOMM ’19, (New York, NY, USA), p.~159–173, Association for Computing
  Machinery, 2019.

\bibitem{zhou2019security}
H.~Zhou, K.~Lv, L.~Huang, and X.~Ma, ``Security assessment and key management
  in a quantum network,'' {\em arXiv preprint arXiv:1907.08963}, 2019.

\bibitem{kozlowski2020designing}
W.~Kozlowski, A.~Dahlberg, and S.~Wehner, ``Designing a quantum network
  protocol,'' in {\em Proceedings of the 16th International Conference on
  Emerging Networking EXperiments and Technologies}, CoNEXT '20, (New York, NY,
  USA), p.~1–16, Association for Computing Machinery, 2020.

\bibitem{huberman2020quantum}
B.~A. Huberman and B.~Lund, ``A quantum router for the entangled web,'' {\em
  Information Systems Frontiers}, vol.~22, no.~1, pp.~37--43, 2020.

\bibitem{yu2019protocols}
N.~Yu, C.-Y. Lai, and L.~Zhou, ``Protocols for packet quantum network
  intercommunication,'' {\em arXiv preprint arXiv:1903.10685}, 2019.

\bibitem{matsuo2019quantum}
T.~Matsuo, C.~Durand, and R.~Van~Meter, ``Quantum link bootstrapping using a
  ruleset-based communication protocol,'' {\em Physical Review A}, vol.~100,
  no.~5, p.~052320, 2019.

\bibitem{van2012quantum}
R.~Van~Meter, ``Quantum networking and internetworking,'' {\em IEEE Network},
  vol.~26, no.~4, pp.~59--64, 2012.

\bibitem{van2013designing}
R.~Van~Meter and J.~Touch, ``Designing quantum repeater networks,'' {\em IEEE
  Communications Magazine}, vol.~51, no.~8, pp.~64--71, 2013.

\bibitem{huberman2020quantumB}
B.~Huberman, B.~Lund, and J.~Wang, ``Quantum secured internet transport,'' {\em
  arXiv preprint arXiv:2007.05522}, 2020.

\bibitem{pirker2018modular}
A.~Pirker, J.~Walln{\"o}fer, and W.~D{\"u}r, ``Modular architectures for
  quantum networks,'' {\em New Journal of Physics}, vol.~20, no.~5, p.~053054,
  2018.

\bibitem{rabbie2020designing}
J.~Rabbie, K.~Chakraborty, G.~Avis, and S.~Wehner, ``Designing quantum networks
  using preexisting infrastructure,'' {\em arXiv preprint arXiv:2005.14715},
  2020.

\bibitem{da2020optimizing}
F.~F. da~Silva, A.~Torres-Knoop, T.~Coopmans, D.~Maier, and S.~Wehner,
  ``Optimizing entanglement generation and distribution using genetic
  algorithms,'' {\em arXiv preprint arXiv:2010.16373}, 2020.

\bibitem{chakraborty2019distributed}
K.~Chakraborty, F.~Rozp\ifmmode~\mbox{\k{e}}\else \k{e}\fi{}dek, A.~Dahlberg,
  and S.~Wehner, ``Distributed routing in a quantum internet,'' {\em arXiv
  preprint arXiv:1907.11630}, 2019.

\bibitem{aaberg2020semidefinite}
J.~\AA{}berg, R.~Nery, C.~Duarte, and R.~Chaves, ``Semidefinite tests for
  quantum network topologies,'' {\em Phys. Rev. Lett.}, vol.~125, p.~110505,
  Sep 2020.

\bibitem{kraft2020characterizing}
T.~Kraft, C.~Spee, X.-D. Yu, and O.~G{\"u}hne, ``Characterizing quantum
  networks: Insights from coherence theory,'' {\em arXiv preprint
  arXiv:2006.06693}, 2020.

\bibitem{wolfe2019quantum}
E.~Wolfe, A.~Pozas-Kerstjens, M.~Grinberg, D.~Rosset, A.~Ac{\'\i}n, and
  M.~Navascu{\'e}s, ``Quantum inflation: a general approach to quantum causal
  compatibility,'' {\em arXiv preprint arXiv:1909.10519}, 2019.

\bibitem{renou2019genuine}
M.-O. Renou, E.~B{\"a}umer, S.~Boreiri, N.~Brunner, N.~Gisin, and S.~Beigi,
  ``Genuine quantum nonlocality in the triangle network,'' {\em Phyisical
  Review Letters}, vol.~123, no.~14, p.~140401, 2019.

\bibitem{renou2019limits}
M.-O. Renou, Y.~Wang, S.~Boreiri, S.~Beigi, N.~Gisin, and N.~Brunner, ``Limits
  on correlations in networks for quantum and no-signaling resources,'' {\em
  Phyisical Review Letters}, vol.~123, no.~7, p.~070403, 2019.

\bibitem{brito2020statistical}
S.~Brito, A.~Canabarro, R.~Chaves, and D.~Cavalcanti, ``Statistical properties
  of the quantum internet,'' {\em Physical Review Letters}, vol.~124, no.~21,
  p.~210501, 2020.

\bibitem{perseguers2013distribution}
S.~Perseguers, G.~Lapeyre~Jr, D.~Cavalcanti, M.~Lewenstein, and A.~Ac{\'\i}n,
  ``Distribution of entanglement in large-scale quantum networks,'' {\em
  Reports on Progress in Physics}, vol.~76, no.~9, p.~096001, 2013.

\bibitem{duer2007entanglement}
W.~D\"ur and H.~Briegel, ``Entanglement purification and quantum error
  correction,'' {\em Reports on Progress in Physics}, vol.~70, no.~8, p.~1381,
  2007.

\bibitem{holevo2012quantum}
A.~S. Holevo, {\em Quantum systems, channels, information: a mathematical
  introduction}, vol.~16.
\newblock Walter de Gruyter, 2012.

\bibitem{wilde2013quantum}
M.~M. Wilde, {\em Quantum information theory}.
\newblock Cambridge University Press, 2013.

\bibitem{arimoto1972algorithm}
S.~Arimoto, ``An algorithm for computing the capacity of arbitrary discrete
  memoryless channels,'' {\em IEEE Transactions on Information Theory},
  vol.~18, no.~1, pp.~14--20, 1972.

\bibitem{blahut1972computation}
R.~Blahut, ``Computation of channel capacity and rate-distortion functions,''
  {\em IEEE transactions on Information Theory}, vol.~18, no.~4, pp.~460--473,
  1972.

\bibitem{pirandola2017fundamental}
S.~Pirandola, R.~Laurenza, C.~Ottaviani, and L.~Banchi, ``Fundamental limits of
  repeaterless quantum communications,'' {\em Nature Communications}, vol.~8,
  no.~15043, pp.~1--15, 2017.

\bibitem{nielsen2002quantum}
M.~A. Nielsen and I.~Chuang, {\em Quantum computation and quantum information}.
\newblock Cambridge University Press, 2000.

\bibitem{AML16}
K.~Azuma, A.~Mizutani, and H.-K. Lo, ``Fundamental rate-loss trade-off for the
  quantum internet,'' {\em Nature Communications}, vol.~7, no.~13523, pp.~1 --
  8, 2016.

\bibitem{bennett1993teleporting}
C.~H. Bennett, G.~Brassard, C.~Cr\'epeau, R.~Jozsa, A.~Peres, and W.~K.
  Wootters, ``Teleporting an unknown quantum state via dual classical and
  {E}instein-{P}odolsky-{R}osen channels,'' {\em Phys. Rev. Lett.}, vol.~70,
  pp.~1895--1899, Mar 1993.

\bibitem{renner2005universally}
R.~Renner and R.~K{\"o}nig, ``Universally composable privacy amplification
  against quantum adversaries,'' in {\em Theory of Cryptography Conference},
  pp.~407--425, Springer, 2005.

\bibitem{horodecki2005secure}
K.~Horodecki, M.~Horodecki, P.~Horodecki, and J.~Oppenheim, ``Secure key from
  bound entanglement,'' {\em Phyisical Review Letters}, vol.~94, no.~16,
  p.~160502, 2005.

\bibitem{greenberger1989going}
D.~M. Greenberger, M.~A. Horne, and A.~Zeilinger, ``Going beyond bell’s
  theorem,'' in {\em Bell’s theorem, quantum theory and conceptions of the
  universe}, pp.~69--72, Springer, 1989.

\bibitem{hillery1999quantum}
M.~Hillery, V.~Bu{\v{z}}ek, and A.~Berthiaume, ``Quantum secret sharing,'' {\em
  Physical Review A}, vol.~59, no.~3, p.~1829, 1999.

\bibitem{augusiak2009multipartite}
R.~Augusiak and P.~Horodecki, ``Multipartite secret key distillation and bound
  entanglement,'' {\em Physical Review A}, vol.~80, no.~4, p.~042307, 2009.

\bibitem{horodecki2009quantum}
R.~Horodecki, P.~Horodecki, M.~Horodecki, and K.~Horodecki, ``Quantum
  entanglement,'' {\em Reviews of modern physics}, vol.~81, no.~2, p.~865,
  2009.

\bibitem{fuchs1999cryptographic}
C.~A. Fuchs and J.~Van De~Graaf, ``Cryptographic distinguishability measures
  for quantum-mechanical states,'' {\em IEEE Transactions on Information
  Theory}, vol.~45, no.~4, pp.~1216--1227, 1999.

\bibitem{ben2005universal}
M.~Ben-Or, M.~Horodecki, D.~W. Leung, D.~Mayers, and J.~Oppenheim, ``The
  universal composable security of quantum key distribution,'' in {\em Theory
  of Cryptography Conference}, pp.~386--406, Springer, 2005.

\bibitem{P16}
S.~Pirandola, ``Capacities of repeater-assisted quantum communications,'' {\em
  arXiv preprint arXiv:1601.00966}, 2016.

\bibitem{p19}
S.~Pirandola, ``End-to-end capacities of a quantum communication network,''
  {\em Communications Physics}, vol.~2, no.~51, pp.~1--10, 2019.

\bibitem{takeoka2014fundamental}
M.~Takeoka, S.~Guha, and M.~M. Wilde, ``Fundamental rate-loss tradeoff for
  optical quantum key distribution,'' {\em Nature Communications}, vol.~5,
  no.~5235, 2014.

\bibitem{rigovacca2018versatile}
L.~Rigovacca, G.~Kato, S.~B{\"a}uml, M.~Kim, W.~J. Munro, and K.~Azuma,
  ``Versatile relative entropy bounds for quantum networks,'' {\em New Journal
  of Physics}, vol.~20, no.~1, p.~013033, 2018.

\bibitem{christandl2017relative}
M.~Christandl and A.~M{\"u}ller-Hermes, ``Relative entropy bounds on quantum,
  private and repeater capacities,'' {\em Communications in Mathematical
  Physics}, vol.~353, no.~2, pp.~821--852, 2017.

\bibitem{wilde2017converse}
M.~M. Wilde, M.~Tomamichel, and M.~Berta, ``Converse bounds for private
  communication over quantum channels,'' {\em IEEE Transactions on Information
  Theory}, vol.~63, no.~3, pp.~1792--1817, 2017.

\bibitem{wilde2016squashed}
M.~M. Wilde, ``Squashed entanglement and approximate private states,'' {\em
  Quantum Information Processing}, vol.~15, no.~11, pp.~4563--4580, 2016.

\bibitem{takeoka2014squashed}
M.~Takeoka, S.~Guha, and M.~M. Wilde, ``The squashed entanglement of a quantum
  channel,'' {\em IEEE Transactions on Information Theory}, vol.~60, no.~8,
  pp.~4987--4998, 2014.

\bibitem{tucci2002entanglement}
R.~Tucci, ``Entanglement of distillation and conditional mutual information,''
  {\em arXiv preprint arXiv: quant-ph/0202144}, 2002.

\bibitem{christandl2004squashed}
M.~Christandl and A.~Winter, ``“squashed entanglement”: an additive
  entanglement measure,'' {\em Journal of mathematical physics}, vol.~45,
  no.~3, pp.~829--840, 2004.

\bibitem{datta2009min}
N.~Datta, ``Min-and max-relative entropies and a new entanglement monotone,''
  {\em IEEE Transactions on Information Theory}, vol.~55, no.~6,
  pp.~2816--2826, 2009.

\bibitem{vedral1997quantifying}
V.~Vedral, M.~B. Plenio, M.~A. Rippin, and P.~L. Knight, ``Quantifying
  entanglement,'' {\em Physical Review Letters}, vol.~78, no.~12, p.~2275,
  1997.

\bibitem{vedral1998entanglement}
V.~Vedral and M.~B. Plenio, ``Entanglement measures and purification
  procedures,'' {\em Physical Review A}, vol.~57, no.~3, p.~1619, 1998.

\bibitem{horodecki2009general}
K.~Horodecki, M.~Horodecki, P.~Horodecki, and J.~Oppenheim, ``General paradigm
  for distilling classical key from quantum states,'' {\em IEEE Transactions on
  Information Theory}, vol.~55, no.~4, pp.~1898--1929, 2009.

\bibitem{bennett1996mixed}
C.~H. Bennett, D.~P. DiVincenzo, J.~A. Smolin, and W.~K. Wootters,
  ``Mixed-state entanglement and quantum error correction,'' {\em Physical
  Review A}, vol.~54, no.~5, p.~3824, 1996.

\bibitem{gottesman1999demonstrating}
D.~Gottesman and I.~L. Chuang, ``Demonstrating the viability of universal
  quantum computation using teleportation and single-qubit operations,'' {\em
  Nature}, vol.~402, no.~6760, pp.~390--393, 1999.

\bibitem{horodecki1999general}
M.~Horodecki, P.~Horodecki, and R.~Horodecki, ``General teleportation channel,
  singlet fraction, and quasidistillation,'' {\em Physical Review A}, vol.~60,
  no.~3, p.~1888, 1999.

\bibitem{knill2001scheme}
E.~Knill, R.~Laflamme, and G.~J. Milburn, ``A scheme for efficient quantum
  computation with linear optics,'' {\em nature}, vol.~409, no.~6816,
  pp.~46--52, 2001.

\bibitem{wolf2007quantum}
M.~M. Wolf, D.~P{\'e}rez-Garc{\'\i}a, and G.~Giedke, ``Quantum capacities of
  bosonic channels,'' {\em Phyisical Review Letters}, vol.~98, no.~13,
  p.~130501, 2007.

\bibitem{niset2009no}
J.~Niset, J.~Fiur{\'a}{\v{s}}ek, and N.~J. Cerf, ``No-go theorem for gaussian
  quantum error correction,'' {\em Phyisical Review Letters}, vol.~102, no.~12,
  p.~120501, 2009.

\bibitem{muller2012transposition}
A.~M{\"u}ller-Hermes, ``Transposition in quantum information theory,'' {\em
  Master's thesis, Technical University of Munich}, 2012.

\bibitem{AK17}
K.~Azuma and G.~Kato, ``Aggregating quantum repeaters for the quantum
  internet,'' {\em Physical Review A}, vol.~96, no.~3, p.~032332, 2017.

\bibitem{BAKE20}
S.~B{\"a}uml, K.~Azuma, G.~Kato, and D.~Elkouss, ``Linear programs for
  entanglement and key distribution in the quantum internet,'' {\em
  Communications Physics}, vol.~3, no.~55, pp.~1--12, 2020.

\bibitem{pirandola2009direct}
S.~Pirandola, R.~Garc{\'\i}a-Patr{\'o}n, S.~L. Braunstein, and S.~Lloyd,
  ``Direct and reverse secret-key capacities of a quantum channel,'' {\em
  Physical review letters}, vol.~102, no.~5, p.~050503, 2009.

\bibitem{pirandola2019advances}
S.~Pirandola, U.~L. Andersen, L.~Banchi, M.~Berta, D.~Bunandar, R.~Colbeck,
  D.~Englund, T.~Gehring, C.~Lupo, C.~Ottaviani, {\em et~al.}, ``Advances in
  quantum cryptography,'' {\em Advances in Optics and Photonics}, vol.~12,
  no.~4, pp.~1012--1236, 2020.

\bibitem{menger1927allgemeinen}
K.~Menger, ``Zur allgemeinen {K}urventheorie,'' {\em Fundamenta Mathematicae},
  vol.~10, no.~1, pp.~96--115, 1927.

\bibitem{bondy2008graph}
J.~A. Bondy and U.~S.~R. Murty, ``Graph theory, volume 244 of,'' {\em Graduate
  texts in Mathematics}, p.~81, 2008.

\bibitem{P19b}
S.~Pirandola, ``Bounds for multi-end communication over quantum networks,''
  {\em Quantum Science and Technology}, vol.~4, p.~045006, sep 2019.

\bibitem{BA17}
S.~B{\"a}uml and K.~Azuma, ``Fundamental limitation on quantum broadcast
  networks,'' {\em Quantum Science and Technology}, vol.~2, no.~2, p.~024004,
  2017.

\bibitem{yang2009squashed}
D.~Yang, K.~Horodecki, M.~Horodecki, P.~Horodecki, J.~Oppenheim, and W.~Song,
  ``Squashed entanglement for multipartite states and entanglement measures
  based on the mixed convex roof,'' {\em IEEE Transactions on Information
  Theory}, vol.~55, no.~7, pp.~3375--3387, 2009.

\bibitem{avis2008distributed}
D.~Avis, P.~Hayden, and I.~Savov, ``Distributed compression and multiparty
  squashed entanglement,'' {\em Journal of Physics A: Mathematical and
  Theoretical}, vol.~41, no.~11, p.~115301, 2008.

\bibitem{seshadreesan2016bounds}
K.~P. Seshadreesan, M.~Takeoka, and M.~M. Wilde, ``Bounds on entanglement
  distillation and secret key agreement for quantum broadcast channels,'' {\em
  IEEE Transactions on Information Theory}, vol.~62, no.~5, pp.~2849--2866,
  2016.

\bibitem{wallnofer20162d}
J.~Walln{\"o}fer, M.~Zwerger, C.~Muschik, N.~Sangouard, and W.~D{\"u}r,
  ``Two-dimensional quantum repeaters,'' {\em Physical Review A}, vol.~94,
  no.~5, p.~052307, 2016.

\bibitem{kaski2004packing}
P.~Kaski, ``Packing {S}teiner trees with identical terminal sets,'' {\em
  Information Processing Letters}, vol.~91, no.~1, pp.~1--5, 2004.

\bibitem{das2019universal}
S.~Das, S.~B{\"a}uml, M.~Winczewski, and K.~Horodecki, ``Universal limitations
  on quantum key distribution over a network,'' {\em arXiv preprint
  arXiv:1912.03646}, 2019.

\bibitem{ye1991n3l}
Y.~Ye, ``An {O}(n$^3${L}) potential reduction algorithm for linear
  programming,'' {\em Mathematical programming}, vol.~50, no.~1-3,
  pp.~239--258, 1991.

\bibitem{wright1997primal}
S.~J. Wright, {\em Primal-dual interior-point methods}.
\newblock Siam, 1997.

\bibitem{ford1956maximal}
L.~R. Ford and D.~R. Fulkerson, ``Maximal flow through a network,'' {\em
  Canadian journal of Mathematics}, vol.~8, no.~3, pp.~399--404, 1956.

\bibitem{EFS96}
P.~Elias, A.~Feinstein, and C.~Shannon, ``A note on the maximum flow through a
  network,'' {\em IRE Transactions on Information Theory}, vol.~2, no.~4,
  pp.~117--119, 1956.

\bibitem{karp1972reducibility}
R.~M. Karp, ``Reducibility among combinatorial problems,'' in {\em Complexity
  of computer computations}, pp.~85--103, Springer, 1972.

\bibitem{hu1963multi}
T.~C. Hu, ``Multi-commodity network flows,'' {\em Operations research},
  vol.~11, no.~3, pp.~344--360, 1963.

\bibitem{garg1997primal}
N.~Garg, V.~V. Vazirani, and M.~Yannakakis, ``Primal-dual approximation
  algorithms for integral flow and multicut in trees,'' {\em Algorithmica},
  vol.~18, no.~1, pp.~3--20, 1997.

\bibitem{aumann1998log}
Y.~Aumann and Y.~Rabani, ``An {O}(log k) approximate min-cut max-flow theorem
  and approximation algorithm,'' {\em SIAM Journal on Computing}, vol.~27,
  no.~1, pp.~291--301, 1998.

\bibitem{garg1996approximate}
N.~Garg, V.~V. Vazirani, and M.~Yannakakis, ``Approximate max-flow min-(multi)
  cut theorems and their applications,'' {\em SIAM Journal on Computing},
  vol.~25, no.~2, pp.~235--251, 1996.

\bibitem{cheriyan2006hardness}
J.~Cheriyan and M.~R. Salavatipour, ``Hardness and approximation results for
  packing steiner trees,'' {\em Algorithmica}, vol.~45, no.~1, pp.~21--43,
  2006.

\bibitem{kriesell2003edge}
M.~Kriesell, ``Edge-disjoint trees containing some given vertices in a graph,''
  {\em Journal of Combinatorial Theory, Series B}, vol.~88, no.~1, pp.~53--65,
  2003.

\bibitem{lau2004approximate}
L.~C. Lau, ``An approximate max-{S}teiner-tree-packing min-{S}teiner-cut
  theorem,'' in {\em Proceedings of the 45th Annual IEEE Symposium on
  Foundations of Computer Science}, pp.~61--70, IEEE, 2004.

\bibitem{petingi2009packing}
L.~Petingi and M.~Talafha, ``Packing the steiner trees of a graph,'' {\em
  Networks}, vol.~54, no.~2, pp.~90--94, 2009.

\bibitem{chakraborty2020entanglement}
K.~Chakraborty, D.~Elkouss, B.~Rijsman, and S.~Wehner, ``Entanglement
  distribution in a quantum network, a multi-commodity flow-based approach,''
  {\em arXiv preprint arXiv:2005.14304}, 2020.

\bibitem{yamasaki2017graph}
H.~Yamasaki, A.~Soeda, and M.~Murao, ``Graph-associated entanglement cost of a
  multipartite state in exact and finite-block-length approximate
  constructions,'' {\em Physical Review A}, vol.~96, no.~3, p.~032330, 2017.

\bibitem{pant2019routing}
M.~Pant, H.~Krovi, D.~Towsley, L.~Tassiulas, L.~Jiang, P.~Basu, D.~Englund, and
  S.~Guha, ``Routing entanglement in the quantum internet,'' {\em npj Quantum
  Information}, vol.~5, no.~25, pp.~1--9, 2019.

\bibitem{schoute2016shortcuts}
E.~Schoute, L.~Mancinska, T.~Islam, I.~Kerenidis, and S.~Wehner, ``Shortcuts to
  quantum network routing,'' {\em arXiv preprint arXiv:1610.05238}, 2016.

\bibitem{ahlswede2000network}
R.~Ahlswede, N.~Cai, S.-Y. Li, and R.~W. Yeung, ``Network information flow,''
  {\em IEEE Transactions on information theory}, vol.~46, no.~4,
  pp.~1204--1216, 2000.

\bibitem{li2004network}
Z.~Li and B.~Li, ``Network coding: The case of multiple unicast sessions,'' in
  {\em Allerton Conference on Communications}, vol.~16, p.~8, IEEE, 2004.

\bibitem{el2011network}
A.~El~Gamal and Y.-H. Kim, {\em Network information theory}.
\newblock Cambridge university press, 2011.

\bibitem{kobayashi2010perfect}
H.~Kobayashi, F.~Le~Gall, H.~Nishimura, and M.~R{\"o}tteler, ``Perfect quantum
  network communication protocol based on classical network coding,'' in {\em
  2010 IEEE International Symposium on Information Theory}, pp.~2686--2690,
  IEEE, 2010.

\bibitem{kobayashi2011constructing}
H.~Kobayashi, F.~Le~Gall, H.~Nishimura, and M.~R{\"o}tteler, ``Constructing
  quantum network coding schemes from classical nonlinear protocols,'' in {\em
  2011 IEEE International Symposium on Information Theory Proceedings},
  pp.~109--113, IEEE, 2011.

\bibitem{epping2017multi}
M.~Epping, H.~Kampermann, D.~Bru{\ss}, {\em et~al.}, ``Multi-partite
  entanglement can speed up quantum key distribution in networks,'' {\em New
  Journal of Physics}, vol.~19, no.~9, p.~093012, 2017.

\bibitem{leung2010quantum}
D.~Leung, J.~Oppenheim, and A.~Winter, ``Quantum network communication—the
  butterfly and beyond,'' {\em IEEE Transactions on Information Theory},
  vol.~56, no.~7, pp.~3478--3490, 2010.

\bibitem{epping2016large}
M.~Epping, H.~Kampermann, and D.~Bru{\ss}, ``Large-scale quantum networks based
  on graphs,'' {\em New Journal of Physics}, vol.~18, no.~5, p.~053036, 2016.

\bibitem{epping2016robust}
M.~Epping, H.~Kampermann, and D.~Bruss, ``Robust entanglement distribution via
  quantum network coding,'' {\em New Journal of Physics}, vol.~18, no.~10,
  p.~103052, 2016.

\bibitem{hahn2019quantum}
F.~Hahn, A.~Pappa, and J.~Eisert, ``Quantum network routing and local
  complementation,'' {\em npj Quantum Information}, vol.~5, no.~76, pp.~1--7,
  2019.

\bibitem{bouchet1988graphic}
A.~Bouchet, ``Graphic presentations of isotropic systems,'' {\em Journal of
  Combinatorial Theory, Series B}, vol.~45, no.~1, pp.~58--76, 1988.

\bibitem{van2004graphical}
M.~Van~den Nest, J.~Dehaene, and B.~De~Moor, ``Graphical description of the
  action of local clifford transformations on graph states,'' {\em Physical
  Review A}, vol.~69, no.~2, p.~022316, 2004.

\bibitem{dahlberg2018transforming}
A.~Dahlberg and S.~Wehner, ``Transforming graph states using single-qubit
  operations,'' {\em Philosophical Transactions of the Royal Society A:
  Mathematical, Physical and Engineering Sciences}, vol.~376, no.~2123,
  p.~20170325, 2018.

\bibitem{dahlberg2018transform}
A.~Dahlberg, J.~Helsen, and S.~Wehner, ``How to transform graph states using
  single-qubit operations: computational complexity and algorithms,'' {\em
  arXiv preprint arXiv:1805.05306}, 2018.

\bibitem{chiribella2013quantum}
G.~Chiribella, G.~M. D’Ariano, P.~Perinotti, and B.~Valiron, ``Quantum
  computations without definite causal structure,'' {\em Physical Review A},
  vol.~88, no.~2, p.~022318, 2013.

\bibitem{ebler2018enhanced}
D.~Ebler, S.~Salek, and G.~Chiribella, ``Enhanced communication with the
  assistance of indefinite causal order,'' {\em Phyisical Review Letters},
  vol.~120, no.~12, p.~120502, 2018.

\bibitem{chiribella2018indefinite}
G.~Chiribella, M.~Banik, S.~S. Bhattacharya, T.~Guha, M.~Alimuddin, A.~Roy,
  S.~Saha, S.~Agrawal, and G.~Kar, ``Indefinite causal order enables perfect
  quantum communication with zero capacity channel,'' {\em arXiv preprint
  arXiv:1810.10457}, 2018.

\bibitem{chiribella2019quantum}
G.~Chiribella and H.~Kristj{\'a}nsson, ``Quantum shannon theory with
  superpositions of trajectories,'' {\em Proceedings of the Royal Society A},
  vol.~475, no.~2225, p.~20180903, 2019.

\bibitem{abbott2018communication}
A.~A. Abbott, J.~Wechs, D.~Horsman, M.~Mhalla, and C.~Branciard,
  ``Communication through coherent control of quantum channels,'' {\em
  Quantum}, vol.~4, p.~333, 2020.

\bibitem{miguel2020genuine}
J.~Miguel-Ramiro, A.~Pirker, and W.~D{\"u}r, ``Genuine quantum networks:
  superposed tasks and addressing,'' {\em arXiv preprint arXiv:2005.00020},
  2020.

\bibitem{munro2015inside}
W.~J. Munro, K.~Azuma, K.~Tamaki, and K.~Nemoto, ``Inside quantum repeaters,''
  {\em {IEEE} Journal of Selected Topics in Quantum Electronics}, vol.~21,
  pp.~78--90, may 2015.

\bibitem{zukowski1993}
M.~\ifmmode~\dot{Z}\else \.{Z}\fi{}ukowski, A.~Zeilinger, M.~A. Horne, and
  A.~K. Ekert, ````{E}vent-ready-detectors'' {Bell} experiment via entanglement
  swapping,'' {\em Phys. Rev. Lett.}, vol.~71, pp.~4287--4290, Dec 1993.

\bibitem{calsamiglia2001maximum}
J.~Calsamiglia and N.~L{\"u}tkenhaus, ``Maximum efficiency of a linear-optical
  bell-state analyzer,'' {\em Applied Physics B}, vol.~72, no.~1, pp.~67--71,
  2001.

\bibitem{grice2011arbitrarily}
W.~P. Grice, ``Arbitrarily complete bell-state measurement using only linear
  optical elements,'' {\em Physical Review A}, vol.~84, no.~4, p.~042331, 2011.

\bibitem{olivo2018ancilla}
A.~Olivo and F.~Grosshans, ``Ancilla-assisted linear optical bell measurements
  and their optimality,'' {\em Physical Review A}, vol.~98, no.~4, p.~042323,
  2018.

\bibitem{ewert20143}
F.~Ewert and P.~van Loock, ``3/4-efficient bell measurement with passive linear
  optics and unentangled ancillae,'' {\em Phyisical Review Letters}, vol.~113,
  no.~14, p.~140403, 2014.

\bibitem{wu2020nearterm}
Y.~Wu, J.~Liu, and C.~Simon, ``Near-term performance of quantum repeaters with
  imperfect ensemble-based quantum memories,'' {\em Phys. Rev. A}, vol.~101,
  p.~042301, Apr 2020.

\bibitem{bennett1996purification}
C.~H. Bennett, G.~Brassard, S.~Popescu, B.~Schumacher, J.~A. Smolin, and W.~K.
  Wootters, ``Purification of noisy entanglement and faithful teleportation via
  noisy channels,'' {\em Phys. Rev. Lett.}, vol.~76, pp.~722--725, Jan 1996.

\bibitem{deutsch1996quantum}
D.~Deutsch, A.~Ekert, R.~Jozsa, C.~Macchiavello, S.~Popescu, and A.~Sanpera,
  ``Quantum privacy amplification and the security of quantum cryptography over
  noisy channels,'' {\em Phys. Rev. Lett.}, vol.~77, pp.~2818--2821, Sep 1996.

\bibitem{werner1989quantum}
R.~F. Werner, ``Quantum states with {E}instein-{P}odolsky-{R}osen correlations
  admitting a hidden-variable model,'' {\em Phys. Rev. A}, vol.~40,
  pp.~4277--4281, Oct 1989.

\bibitem{briegel1999quantumrepeaters}
W.~D\"ur, H.-J. Briegel, J.~I. Cirac, and P.~Zoller, ``Quantum repeaters based
  on entanglement purification,'' {\em Phys. Rev. A}, vol.~59, pp.~169--181,
  Jan 1999.

\bibitem{collins2007multiplexed}
O.~A. Collins, S.~D. Jenkins, A.~Kuzmich, and T.~A.~B. Kennedy, ``Multiplexed
  memory-insensitive quantum repeaters,'' {\em Phys. Rev. Lett.}, vol.~98,
  p.~060502, Feb 2007.

\bibitem{abruzzo2014measurement}
S.~Abruzzo, H.~Kampermann, and D.~Bru\ss{}, ``Measurement-device-independent
  quantum key distribution with quantum memories,'' {\em Phys. Rev. A},
  vol.~89, p.~012301, Jan 2014.

\bibitem{briegel1998quantum}
H.-J. Briegel, W.~D\"ur, J.~I. Cirac, and P.~Zoller, ``Quantum repeaters: The
  role of imperfect local operations in quantum communication,'' {\em Phys.
  Rev. Lett.}, vol.~81, pp.~5932--5935, Dec 1998.

\bibitem{duan2001long}
L.-M. Duan, M.~D. Lukin, J.~I. Cirac, and P.~Zoller, ``Long-distance quantum
  communication with atomic ensembles and linear optics,'' {\em Nature},
  vol.~414, pp.~413 EP --, Nov 2001.
\newblock Article.

\bibitem{kok2003construction}
P.~Kok, C.~P. Williams, and J.~P. Dowling, ``Construction of a quantum repeater
  with linear optics,'' {\em Physical Review A}, vol.~68, no.~2, p.~022301,
  2003.

\bibitem{childress2005fault}
L.~Childress, J.~M. Taylor, A.~S. S\o{}rensen, and M.~D. Lukin,
  ``Fault-tolerant quantum repeaters with minimal physical resources and
  implementations based on single-photon emitters,'' {\em Phys. Rev. A},
  vol.~72, p.~052330, Nov 2005.

\bibitem{van2006hybrid}
P.~Van~Loock, T.~Ladd, K.~Sanaka, F.~Yamaguchi, K.~Nemoto, W.~Munro, and
  Y.~Yamamoto, ``Hybrid quantum repeater using bright coherent light,'' {\em
  Phyisical Review Letters}, vol.~96, no.~24, p.~240501, 2006.

\bibitem{munro2008high}
W.~Munro, R.~Van~Meter, S.~G. Louis, and K.~Nemoto, ``High-bandwidth hybrid
  quantum repeater,'' {\em Phyisical Review Letters}, vol.~101, no.~4,
  p.~040502, 2008.

\bibitem{azuma2012quantum}
K.~Azuma, H.~Takeda, M.~Koashi, and N.~Imoto, ``Quantum repeaters and
  computation by a single module: Remote nondestructive parity measurement,''
  {\em Physical Review A}, vol.~85, no.~6, p.~062309, 2012.

\bibitem{zwerger2012measurement}
M.~Zwerger, W.~D{\"u}r, and H.~Briegel, ``Measurement-based quantum
  repeaters,'' {\em Physical Review A}, vol.~85, no.~6, p.~062326, 2012.

\bibitem{jiang2007fast}
L.~Jiang, J.~M. Taylor, and M.~D. Lukin, ``Fast and robust approach to
  long-distance quantum communication with atomic ensembles,'' {\em Phys. Rev.
  A}, vol.~76, p.~012301, Jul 2007.

\bibitem{brask2008memory}
J.~B. Brask and A.~S. S\o{}rensen, ``Memory imperfections in
  atomic-ensemble-based quantum repeaters,'' {\em Phys. Rev. A}, vol.~78,
  p.~012350, Jul 2008.

\bibitem{sangouard2011quantum}
N.~Sangouard, C.~Simon, H.~de~Riedmatten, and N.~Gisin, ``Quantum repeaters
  based on atomic ensembles and linear optics,'' {\em Rev. Mod. Phys.},
  vol.~83, pp.~33--80, Mar 2011.

\bibitem{loock2019extending}
P.~van Loock, W.~Alt, C.~Becher, O.~Benson, H.~Boche, C.~Deppe, J.~Eschner,
  S.~H{\"o}fling, D.~Meschede, P.~Michler, {\em et~al.}, ``Extending quantum
  links: Modules for fiber-and memory-based quantum repeaters,'' {\em arXiv
  preprint arXiv:1912.10123}, 2019.

\bibitem{abruzzo2013quantum}
S.~Abruzzo, S.~Bratzik, N.~K. Bernardes, H.~Kampermann, P.~van Loock, and
  D.~Bru\ss{}, ``Quantum repeaters and quantum key distribution: Analysis of
  secret-key rates,'' {\em Phys. Rev. A}, vol.~87, p.~052315, May 2013.

\bibitem{asadi2018quantum}
F.~K. Asadi, N.~Lauk, S.~Wein, N.~Sinclair, C.~O'Brien, and C.~Simon, ``Quantum
  repeaters with individual rare-earth ions at telecommunication wavelengths,''
  {\em Quantum}, vol.~2, p.~93, 2018.

\bibitem{hartmann2007role}
L.~Hartmann, B.~Kraus, H.-J. Briegel, and W.~D\"ur, ``Role of memory errors in
  quantum repeaters,'' {\em Phys. Rev. A}, vol.~75, p.~032310, Mar 2007.

\bibitem{schmidt2019memory}
F.~Schmidt and P.~van Loock, ``Memory-assisted long-distance phase-matching
  quantum key distribution,'' {\em Physical Review A}, vol.~102, no.~4,
  p.~042614, 2020.

\bibitem{rozpedek2018parameters}
F.~Rozp\ifmmode~\mbox{\k{e}}\else \k{e}\fi{}dek, K.~Goodenough, J.~Ribeiro,
  N.~Kalb, V.~C. Vivoli, A.~Reiserer, R.~Hanson, S.~Wehner, and D.~Elkouss,
  ``Parameter regimes for a single sequential quantum repeater,'' {\em Quantum
  Science and Technology}, 2018.

\bibitem{shchukin2019waiting}
E.~Shchukin, F.~Schmidt, and P.~van Loock, ``Waiting time in quantum repeaters
  with probabilistic entanglement swapping,'' {\em Phys. Rev. A}, vol.~100,
  p.~032322, Sep 2019.

\bibitem{kuzmin2019scalable}
V.~Kuzmin, D.~Vasilyev, N.~Sangouard, W.~D{\"u}r, and C.~Muschik, ``Scalable
  repeater architectures for multi-party states,'' {\em npj Quantum
  Information}, vol.~5, no.~115, pp.~1--6, 2019.

\bibitem{kuzmin2020diagrammatic}
V.~V. Kuzmin and D.~V. Vasilyev, ``Diagrammatic technique for simulation of
  large-scale quantum repeater networks with dissipating quantum memories,''
  {\em arXiv preprint arXiv:2009.10415}, 2020.

\bibitem{santra2018quantum}
S.~Santra, L.~Jiang, and V.~S. Malinovsky, ``Quantum repeater architecture with
  hierarchically optimized memory buffer times,'' {\em Quantum Science and
  Technology}, vol.~4, p.~025010, mar 2019.

\bibitem{goodenough2020}
K.~Goodenough, D.~Elkouss, and S.~Wehner, ``Optimising repeater schemes for the
  quantum internet,'' {\em arXiv preprint arXiv:2006.12221}, 2020.

\bibitem{vinay2017practical}
S.~E. Vinay and P.~Kok, ``Practical repeaters for ultralong-distance quantum
  communication,'' {\em Phys. Rev. A}, vol.~95, p.~052336, May 2017.

\bibitem{bernardes2011rate}
N.~K. Bernardes, L.~Praxmeyer, and P.~van Loock, ``Rate analysis for a hybrid
  quantum repeater,'' {\em Phys. Rev. A}, vol.~83, p.~012323, Jan 2011.

\bibitem{praxmeyer}
L.~Praxmeyer, ``Reposition time in probabilistic imperfect memories,'' {\em
  arXiv preprint arXiv:1309.3407}, 2013.

\bibitem{khatri2019practicalPRR}
S.~Khatri, C.~T. Matyas, A.~U. Siddiqui, and J.~P. Dowling, ``Practical figures
  of merit and thresholds for entanglement distribution in quantum networks,''
  {\em Phys. Rev. Research}, vol.~1, p.~023032, Sep 2019.

\bibitem{vinay2019statistical}
S.~E. Vinay and P.~Kok, ``Statistical analysis of quantum-entangled-network
  generation,'' {\em Phys. Rev. A}, vol.~99, p.~042313, Apr 2019.

\bibitem{vardoyan2019capacity}
G.~Vardoyan, S.~Guha, P.~Nain, and D.~Towsley, ``On the capacity region of
  bipartite and tripartite entanglement switching,'' {\em arXiv preprint
  arXiv:1901.06786}, 2019.

\bibitem{vardoyan2019stochastic}
G.~Vardoyan, S.~Guha, P.~Nain, and D.~Towsley, ``On the stochastic analysis of
  a quantum entanglement switch,'' {\em SIGMETRICS Perform. Eval. Rev.},
  vol.~47, p.~27–29, Dec. 2019.

\bibitem{khatri2020policies}
S.~Khatri, ``Policies for elementary link generation in quantum networks,''
  {\em arXiv preprint arXiv:2007.03193}, 2020.

\bibitem{barry2014quantum}
J.~Barry, D.~T. Barry, and S.~Aaronson, ``Quantum partially observable markov
  decision processes,'' {\em Physical Review A}, vol.~90, no.~3, p.~032311,
  2014.

\bibitem{brand2020efficient}
S.~{Brand}, T.~{Coopmans}, and D.~{Elkouss}, ``Efficient computation of the
  waiting time and fidelity in quantum repeater chains,'' {\em IEEE Journal on
  Selected Areas in Communications}, vol.~38, no.~3, pp.~619--639, 2020.

\bibitem{li2020efficient}
B.~Li, T.~Coopmans, and D.~Elkouss, ``Efficient optimization of cut-offs in
  quantum repeater chains,'' {\em arXiv preprint arXiv:2005.04946}, 2020.

\bibitem{muralidharan2016optimal}
S.~Muralidharan, L.~Li, J.~Kim, N.~L{\"u}tkenhaus, M.~D. Lukin, and L.~Jiang,
  ``Optimal architectures for long distance quantum communication,'' {\em
  Scientific reports}, vol.~6, p.~20463, 2016.

\bibitem{jiang2009quantum}
L.~Jiang, J.~M. Taylor, K.~Nemoto, W.~J. Munro, R.~Van~Meter, and M.~D. Lukin,
  ``Quantum repeater with encoding,'' {\em Physical Review A}, vol.~79, no.~3,
  p.~032325, 2009.

\bibitem{munro2010quantum}
W.~Munro, K.~Harrison, A.~Stephens, S.~Devitt, and K.~Nemoto, ``From quantum
  multiplexing to high-performance quantum networking,'' {\em Nature
  Photonics}, vol.~4, no.~11, pp.~792--796, 2010.

\bibitem{li2013long}
Y.~Li, S.~D. Barrett, T.~M. Stace, and S.~C. Benjamin, ``Long range
  failure-tolerant entanglement distribution,'' {\em New Journal of Physics},
  vol.~15, no.~2, p.~023012, 2013.

\bibitem{mazurek2014long}
P.~Mazurek, A.~Grudka, M.~Horodecki, P.~Horodecki, J.~{\L}odyga,
  {\L}.~Pankowski, and A.~Przysi{\k{e}}{\.z}na, ``Long-distance quantum
  communication over noisy networks without long-time quantum memory,'' {\em
  Physical Review A}, vol.~90, no.~6, p.~062311, 2014.

\bibitem{knill1996concatenated}
E.~Knill and R.~Laflamme, ``Concatenated quantum codes,'' {\em arXiv preprint
  quant-ph/9608012}, 1996.

\bibitem{fowler2010surface}
A.~G. Fowler, D.~S. Wang, C.~D. Hill, T.~D. Ladd, R.~Van~Meter, and L.~C.
  Hollenberg, ``Surface code quantum communication,'' {\em Phyisical Review
  Letters}, vol.~104, no.~18, p.~180503, 2010.

\bibitem{munro2012quantumcommunication}
W.~J. Munro, A.~M. Stephens, S.~J. Devitt, K.~A. Harrison, and K.~Nemoto,
  ``Quantum communication without the necessity of quantum memories,'' {\em
  Nature Photonics}, vol.~6, pp.~777--781, Nov 2012.

\bibitem{azuma2015all}
K.~Azuma, K.~Tamaki, and H.-K. Lo, ``All-photonic quantum repeaters,'' {\em
  Nature Communications}, vol.~6, p.~6787, 2015.

\bibitem{muralidharan2014ultrafast}
S.~Muralidharan, J.~Kim, N.~L{\"u}tkenhaus, M.~D. Lukin, and L.~Jiang,
  ``Ultrafast and fault-tolerant quantum communication across long distances,''
  {\em Phyisical Review Letters}, vol.~112, no.~25, p.~250501, 2014.

\bibitem{glaudell2016serialized}
A.~N. Glaudell, E.~Waks, and J.~M. Taylor, ``Serialized quantum error
  correction protocol for high-bandwidth quantum repeaters,'' {\em New Journal
  of Physics}, vol.~18, p.~093008, sep 2016.

\bibitem{ewert2016ultrafastPRL}
F.~Ewert, M.~Bergmann, and P.~van Loock, ``Ultrafast long-distance quantum
  communication with static linear optics,'' {\em Phys. Rev. Lett.}, vol.~117,
  p.~210501, Nov 2016.

\bibitem{ewert2016ultrafastPRA}
F.~Ewert and P.~van Loock, ``Ultrafast fault-tolerant long-distance quantum
  communication with static linear optics,'' {\em Phys. Rev. A}, vol.~95,
  p.~012327, Jan 2017.

\bibitem{pant2017rate}
M.~Pant, H.~Krovi, D.~Englund, and S.~Guha, ``Rate-distance tradeoff and
  resource costs for all-optical quantum repeaters,'' {\em Physical Review A},
  vol.~95, no.~1, p.~012304, 2017.

\bibitem{lee2019fundamental}
S.-W. Lee, T.~C. Ralph, and H.~Jeong, ``Fundamental building block for
  all-optical scalable quantum networks,'' {\em Phys. Rev. A}, vol.~100,
  p.~052303, Nov 2019.

\bibitem{borregaard2020one}
J.~Borregaard, H.~Pichler, T.~Schr{\"o}der, M.~D. Lukin, P.~Lodahl, and A.~S.
  S{\o}rensen, ``One-way quantum repeater based on near-deterministic
  photon-emitter interfaces,'' {\em Physical Review X}, vol.~10, no.~2,
  p.~021071, 2020.

\bibitem{wehrle2010modeling}
K.~Wehrle, M.~G{\"u}nes, and J.~Gross, {\em Modeling and tools for network
  simulation}.
\newblock Springer Science \& Business Media, 2010.

\bibitem{fernandes2017performance}
S.~Fernandes, {\em Performance Evaluation for Network Services, Systems and
  Protocols}.
\newblock Springer, 2017.

\bibitem{burbank2011introduction}
J.~L. Burbank, W.~Kasch, and J.~Ward, {\em An introduction to network modeling
  and simulation for the practicing engineer}, vol.~5.
\newblock John Wiley \& Sons, 2011.

\bibitem{wu2019full}
X.-C. Wu, S.~Di, E.~M. Dasgupta, F.~Cappello, H.~Finkel, Y.~Alexeev, and F.~T.
  Chong, ``Full-state quantum circuit simulation by using data compression,''
  in {\em Proceedings of the International Conference for High Performance
  Computing, Networking, Storage and Analysis}, pp.~1--24, 2019.

\bibitem{huang2020classical}
C.~Huang, F.~Zhang, M.~Newman, J.~Cai, X.~Gao, Z.~Tian, J.~Wu, H.~Xu, H.~Yu,
  B.~Yuan, {\em et~al.}, ``Classical simulation of quantum supremacy
  circuits,'' {\em arXiv preprint arXiv:2005.06787}, 2020.

\bibitem{van2014quantum}
R.~Van~Meter, {\em Quantum networking}.
\newblock John Wiley \& Sons, 2014.

\bibitem{mailloux2015modeling}
L.~O. Mailloux, J.~D. Morris, M.~R. Grimaila, D.~D. Hodson, D.~R. Jacques,
  J.~M. Colombi, C.~V. Mclaughlin, and J.~A. Holes, ``A modeling framework for
  studying quantum key distribution system implementation nonidealities,'' {\em
  IEEE Access}, vol.~3, pp.~110--130, 2015.

\bibitem{mehic2017implementation}
M.~Mehic, O.~Maurhart, S.~Rass, and M.~Voznak, ``Implementation of quantum key
  distribution network simulation module in the network simulator ns-3,'' {\em
  Quantum Information Processing}, vol.~16, no.~10, p.~253, 2017.

\bibitem{buhari2012efficient}
A.~Buhari, Z.~A. Zukarnain, S.~K. Subramaniam, H.~Zainuddin, and S.~Saharudin,
  ``An efficient modeling and simulation of quantum key distribution protocols
  using optisystem™,'' in {\em 2012 IEEE Symposium on Industrial Electronics
  and Applications}, pp.~84--89, IEEE, 2012.

\bibitem{zhao2008event}
S.~Zhao and H.~De~Raedt, ``Event-by-event simulation of quantum cryptography
  protocols,'' {\em Journal of Computational and Theoretical Nanoscience},
  vol.~5, no.~4, pp.~490--504, 2008.

\bibitem{zhang2007object}
X.~Zhang, Q.~Wen, and F.~Zhu, ``Object-oriented quantum cryptography simulation
  model,'' in {\em Third International Conference on Natural Computation (ICNC
  2007)}, vol.~4, pp.~599--602, IEEE, 2007.

\bibitem{niemiec2011quantum}
M.~Niemiec, {\L}.~Roma{\'n}ski, and M.~{\'S}wi{\k{e}}ty, ``Quantum cryptography
  protocol simulator,'' in {\em International Conference on Multimedia
  Communications, Services and Security}, pp.~286--292, Springer, 2011.

\bibitem{pereszlenyi2005simulation}
A.~Pereszlenyi, ``Simulation of quantum key distribution with noisy channels,''
  in {\em Proceedings of the 8th International Conference on
  Telecommunications, 2005. ConTEL 2005.}, vol.~1, pp.~203--210, IEEE, 2005.

\bibitem{chatterjee2019qkdsim}
R.~Chatterjee, K.~Joarder, S.~Chatterjee, B.~C. Sanders, and U.~Sinha,
  ``qkdsim, a simulation toolkit for quantum key distribution including
  imperfections: Performance analysis and demonstration of the b92 protocol
  using heralded photons,'' {\em Phys. Rev. Applied}, vol.~14, p.~024036, Aug
  2020.

\bibitem{dasari2016programmable}
V.~R. Dasari, R.~J. Sadlier, R.~Prout, B.~P. Williams, and T.~S. Humble,
  ``Programmable multi-node quantum network design and simulation,'' in {\em
  Quantum Information and Computation IX}, vol.~9873, p.~98730B, International
  Society for Optics and Photonics, 2016.

\bibitem{dasari2019openflow}
V.~R. Dasari and T.~S. Humble, ``Openflow arbitrated programmable network
  channels for managing quantum metadata,'' {\em The Journal of Defense
  Modeling and Simulation}, vol.~16, no.~1, pp.~67--77, 2019.

\bibitem{dahlberg2018simulaqron}
A.~Dahlberg and S.~Wehner, ``Simula{Q}ron—a simulator for developing quantum
  internet software,'' {\em Quantum Science and Technology}, vol.~4, no.~1,
  p.~015001, 2018.

\bibitem{diadamo2020qunetsim}
S.~DiAdamo, J.~N{\"o}zel, B.~Zanger, and M.~M. Be{\c{s}}e, ``Qu{N}et{S}im: A
  software framework for quantum networks,'' {\em arXiv preprint
  arXiv:2003.06397}, 2020.

\bibitem{zimmermann1980osi}
H.~Zimmermann, ``Osi reference model-the iso model of architecture for open
  systems interconnection,'' {\em IEEE Transactions on communications},
  vol.~28, no.~4, pp.~425--432, 1980.

\bibitem{sawerwain2020quantum}
M.~Sawerwain and J.~Wi{\'s}niewska, ``Quantum router for qutrit networks,'' in
  {\em International Conference on Computer Networks}, pp.~41--51, Springer,
  2020.

\bibitem{murray2020implementing}
H.~Murray, J.~Horgan, J.~F. Santos, D.~Malone, and H.~Siljak, ``Implementing a
  quantum coin scheme,'' {\em arXiv preprint arXiv:2006.02149}, 2020.

\bibitem{amoretti2020entanglement}
M.~Amoretti and S.~Carretta, ``Entanglement verification in quantum networks
  with tampered nodes,'' {\em IEEE Journal on Selected Areas in
  Communications}, vol.~38, no.~3, pp.~598--604, 2020.

\bibitem{amoretti2019enhancing}
M.~Amoretti, M.~Pizzoni, and S.~Carretta, ``Enhancing distributed functional
  monitoring with quantum protocols,'' {\em Quantum Information Processing},
  vol.~18, no.~12, p.~371, 2019.

\bibitem{quantumprotocolzoo}
``The quantum protocol zoo.'' \url{https://wiki.veriqloud.fr/}, 2019.

\bibitem{moulick2016quantum}
S.~R. Moulick and P.~K. Panigrahi, ``Quantum cheques,'' {\em Quantum
  Information Processing}, vol.~15, no.~6, pp.~2475--2486, 2016.

\bibitem{ganz2009quantum}
M.~Ganz, ``Quantum leader election,'' {\em arXiv preprint arXiv:0910.4952},
  2009.

\bibitem{dunjko2014composable}
V.~Dunjko, J.~F. Fitzsimons, C.~Portmann, and R.~Renner, ``Composable security
  of delegated quantum computation,'' in {\em International Conference on the
  Theory and Application of Cryptology and Information Security}, pp.~406--425,
  Springer, 2014.

\bibitem{pappa2011practical}
A.~Pappa, A.~Chailloux, E.~Diamanti, and I.~Kerenidis, ``Practical quantum coin
  flipping,'' {\em Physical Review A}, vol.~84, no.~5, p.~052305, 2011.

\bibitem{bartlett2018distributed}
B.~Bartlett, ``A distributed simulation framework for quantum networks and
  channels,'' {\em arXiv preprint arXiv:1808.07047}, 2018.

\bibitem{qiu2019solving}
P.-H. Qiu, X.-G. Chen, and Y.-W. Shi, ``Solving quantum channel discrimination
  problem with quantum networks and quantum neural networks,'' {\em IEEE
  Access}, vol.~7, pp.~50214--50222, 2019.

\bibitem{bartlett2020universal}
B.~Bartlett and S.~Fan, ``Universal programmable photonic architecture for
  quantum information processing,'' {\em Physical Review A}, vol.~101, no.~4,
  p.~042319, 2020.

\bibitem{githubquisp}
``Quantum {I}nternet {S}imulation {P}ackage ({Q}u{I}{S}{P}).''
  \url{https://github.com/sfc-aqua/quisp}, 2020.

\bibitem{matsuo2019simulation}
T.~Matsuo, ``Simulation of a dynamic, {R}ule{S}et-based quantum network,'' {\em
  arXiv:1908.10758}, 2020.

\bibitem{varga2008overview}
A.~Varga and R.~Hornig, ``An overview of the {O}{M}{N}e{T}++ simulation
  environment,'' in {\em Proceedings of the 1st international conference on
  Simulation tools and techniques for communications, networks and systems \&
  workshops}, p.~60, ICST, 2008.

\bibitem{wu2020sequence}
X.~Wu, A.~Kolar, J.~Chung, D.~Jin, T.~Zhong, R.~Kettimuthu, and M.~Suchara,
  ``{S}e{Q}{U}e{N}{C}e: A customizable discrete-event simulator of quantum
  networks,'' {\em arXiv preprint arXiv:2009.12000}, 2020.

\bibitem{netsquidpaper}
T.~Coopmans, R.~Knegjens, A.~Dahlberg, D.~Maier, L.~Nijsten, J.~Oliveira,
  M.~Papendrecht, J.~Rabbie, F.~Rozpędek, M.~Skrzypczyk, L.~Wubben,
  W.~de~Jong, D.~Podareanu, A.~T. Knoop, D.~Elkouss, and S.~Wehner, ``Netsquid,
  a discrete-event simulation platform for quantum networks,'' {\em arXiv
  preprint arXiv:2010.12535}, 2020.

\bibitem{suchara2020designing}
M.~Suchara, J.~Chung~Miranda, R.~Kettimuthu, A.~Kolar, X.~Wu, and T.~Zhong,
  ``Designing scalable quantum network architectures,'' {\em Bulletin of the
  American Physical Society}, 2020.

\bibitem{Wu2019}
X.~{Wu}, J.~{Chung}, A.~{Kolar}, E.~{Wang}, T.~{Zhong}, R.~{Kettimuthu}, and
  M.~{Suchara}, ``Simulations of photonic quantum networks for performance
  analysis and experiment design,'' in {\em 2019 IEEE/ACM Workshop on
  Photonics-Optics Technology Oriented Networking, Information and Computing
  Systems (PHOTONICS)}, pp.~28--35, 2019.

\bibitem{lee2020quantum}
Y.~Lee, E.~Bersin, A.~Dahlberg, S.~Wehner, and D.~Englund, ``A quantum router
  architecture for high-fidelity entanglement flows in multi-user quantum
  networks,'' {\em arXiv preprint arXiv:2005.01852}, 2020.

\end{thebibliography}

\end{document}